\documentclass[aps,pra,twocolumn,showpacs]{revtex4}
\usepackage{graphicx}
\usepackage{amsmath}
\usepackage{amssymb,wasysym}
\usepackage{amsfonts}
\usepackage{bm,verbatim}

\newcommand{\bea}{\begin{eqnarray}}
\newcommand{\eea}{\end{eqnarray}}
\newcommand{\be}{\begin{equation}}
\newcommand{\ee}{\end{equation}}
\newcommand{\bmu}{\begin{multline}}
\newcommand{\emu}{\end{multline}}

\newcommand{\rr}{\mathbf{r}}
\newcommand{\kk}{\mathbf{k}}

\newcommand{\KK}{\mathbf{K}}

\newcommand{\uu}{\mathbf{u}}
\newcommand{\yy}{\mathbf{y}}
\newcommand{\xx}{\mathbf{x}}
\newcommand{\argsh}{\mathrm{argsh}}

\begin{document}
\title{Trimers in the resonant $2+1$ fermionic problem on a narrow Feshbach resonance: Crossover from Efimovian
to Hydrogenoid spectrum}

\author{Yvan Castin}
\affiliation{
Laboratoire Kastler Brossel,
\'{E}cole Normale Sup\'{e}rieure and CNRS, UPMC, 24 rue Lhomond, 75231 Paris, France}
\author{Edoardo Tignone}
\affiliation{
Laboratoire Kastler Brossel,
\'{E}cole Normale Sup\'{e}rieure and CNRS, UPMC, 24 rue Lhomond, 75231 Paris, France}

\begin{abstract}
We study the quantum three-body free space 
problem of two same-spin-state fermions
of mass $m$ interacting with a different particle of mass $M$,
on an infinitely narrow Feshbach resonance with infinite $s$-wave scattering length.
This problem is made interesting by the existence of a 
tunable parameter, the mass ratio $\alpha=m/M$.
By a combination of analytical and numerical techniques, we obtain
a detailed picture of the spectrum of three-body bound states,
within {\sl each} sector of fixed total angular momentum $l$.
For $\alpha$ increasing from $0$,
we find that the trimer states first appear at the $l$-dependent Efimovian 
threshold $\alpha_c^{(l)}$, where the Efimov exponent $s$ vanishes,
and that the {\sl entire} trimer spectrum (starting from the ground
trimer state) is geometric 
for $\alpha$ tending to $\alpha_c^{(l)}$ from above, with a global energy scale
that has a finite and non-zero limit.
For further increasing values of $\alpha$, the least bound trimer states
still form a geometric spectrum, with an energy ratio
$\exp(2\pi/|s|)$ that becomes closer and closer to unity,
but the most bound trimer states deviate more and more 
from that geometric spectrum and eventually form a hydrogenoid
spectrum.
\end{abstract}



\maketitle

\section{Introduction}

The quantum few-body problem is the subject of a renewed interest 
\cite{revue_Efimov}, 
thanks to the possibility of experimentally studying this problem 
in a resonant regime with cold atoms
close to a Feshbach resonance \cite{revue_feshbach}.
In this resonant regime, the $s$-wave scattering length $a$
associated to the interaction among the particles can be made
much larger in absolute value than the interaction range.
This has in particular allowed to study the Efimov effect in the lab,
up to now for three bosons and for three distinguishable particles
\cite{manips_Efimov}, that is the emergence for $1/a=0$
of an infinite number of trimer states with an accumulation point at
zero energy in the vicinity of which the spectrum forms
a geometric sequence. Whereas the existence of an infinite
number of bound states is common for long range interactions
(vanishing for diverging interparticle distance $r_{12}$
as $1/r_{12}^2$ or more slowly), this is quite intriguing
for short range interactions.  Initially predicted by Efimov for
three bosons, this effect can actually take place in more general
situations \cite{Efimov}, in particular in the so-called $2+1$
fermionic problem if the extra particle is light enough \cite{Petrov_fermions}.

What we call here the $2+1$ fermionic problem consists in the system 
of two same spin state fermions
of mass $m$ interacting with a particle of mass $M$ of another species.
It is assumed that there is no direct interaction among the fermions, whereas there is a resonant
interaction between each fermion and the extra particle, that is with an infinite $s$-wave scattering
length, $1/a=0$. Furthermore, it is assumed that this resonant interaction is due to an infinitely
narrow Feshbach resonance, that is of vanishing van der Waals range $b\to 0$ and finite effective range $r_e$.

This concept is most easily understood in a two-channel model. In the open channel,
the particles exist in the form of atoms, and have a weak, non-resonant direct van der Waals interaction
corresponding to the background scattering length $a_{\rm bg}\approx b$ and the interaction range $b$.
In the closed channel, the particles exist in the form of a bound state of a fermion with the other-species-atom, the so-called 
closed-channel molecule with radius $\approx b$.
Due to a coupling $\Lambda$ between the two channels, that we shall precisely define later,
the closed-channel molecule is coherently
converted into a pair of atoms in the open channel, and {\sl vice-versa}. For an appropriate Zeeman tuning
(with a magnetic field) of the bare energy $E_{\rm mol}$ of the closed-channel molecule with respect to the dissociation
limit of the open channel, the $s$-wave scattering length $a$ between a fermion and the other-species-atom
is infinite. In this case, in simple models, the effective range $r_e$ is the sum of two contributions \cite{Tarruell}. The first one is, 
as expected, of the order of the van der Waals length $b$. The second one is induced by the interchannel coupling;
it is expressed as $-2 R_*$ \cite{Petrov_rstar}, where the Feshbach length $R_*$ is positive and scales
as $1/\Lambda^2$:
\be
R_* = \frac{\pi \hbar^4}{\Lambda^2 \mu^2}
\ee
where $\mu$ is the reduced mass of a fermion and the other-species particle.
When the interchannel coupling $\Lambda$ is very weak,
this second contribution dominates over the first one, $R_*\gg b$, and this is the narrow Feshbach resonance
regime.  
An example under current theoretical and experimental investigation is the case of the interspecies Feshbach resonances
of the fermionic ${}^{6}$Li and the fermionic ${}^{40}$K, which are narrow: The Feshbach length $R_*$
exceeds $100$ nm, whereas the van der Waals length is a few nanometers \cite{Levinsen,Wille,Tiecke}.

To obtain the infinitely narrow Feshbach resonance model, here for $1/a=0$, one takes the mathematical limit 
of a vanishing van der Waals range $b$ with a fixed non-zero interchannel coupling $\Lambda$.
The $s$-wave scattering amplitude between one fermion and the extra particle for a relative
wave vector $\kk$ is then  \cite{Petrov_rstar}:
\be
f_k = \frac{-1}{ik+ k^2 R_*}.
\ee
This implies the absence of two-body bound states, since $f_k$ can not have a pole for $k=iq$,
$q>0$. The model can however certainly supports trimer states since the $2+1$ fermionic problem  is subjected
to the Efimov effect for a large enough mass ratio $m/M$ \cite{Efimov,Petrov_fermions}.

The main motivation of the present work is to study the spectrum of trimers for this problem in free space
for $1/a=0$, in particular to determine analytically the global energy scale of the Efimovian part of the spectrum,
related to the so-called three-body parameter.
This global energy scale is out of reach of Efimov's zero range theory \cite{Efimov} but it was determined 
analytically for three bosons for a narrow Feshbach resonance in \cite{Mora_3corps,Mora_CRAS}. Here we shall generalize
this calculation to the present $2+1$ fermionic problem.
A second motivation is to determine the low-lying states of each Efimov trimer series, which
in principle are not accurately described by Efimov theory, and to look for possible trimer states
that are not related to the Efimov effect and may thus appear for lower mass ratios.

The paper is organized as follows. After a presentation of the model and the derivation of an integral
equation {\sl \`a la} Skorniakov-Ter-Martirosian \cite{STM} for the three-body problem in momentum space in 
section~\ref{sec:model_equa}, analytical solutions of this
integral equation are obtained in limiting cases in section~\ref{sec:analy}. 
In the central section \ref{sec:arots} of the paper, we analytically explore the physics
of the trimers: Of particular interest
are the exact results on the global energy scale in the Efimovian part of the spectrum, see
subsection~\ref{subsec:aftts},
and the study of an hydrogenoid part of the spectrum in the Born-Oppenheimer regime,
see subsection~\ref{subsec:BO}.
An efficient numerical solution of the integral equation is used in 
section~\ref{sec:numerics} to explore intermediate
regimes not covered by the analytics. We conclude in section~\ref{sec:conclusion}.

\section{The Model and the General Momentum Space Equation}
\label{sec:model_equa}

A tractable though realistic description of a Feshbach resonance is obtained with
the so-called two-channel models, where the particles exist either in the form of atoms in the open channel
or in the form of molecules in the closed channel 
\cite{revue_feshbach,Holland,KoehlerBurnett,Koehler,Lee,Gurarie,ondep,Kokkelmans}.
We use here the same free space two-channel model Hamiltonian $H$ as in \cite{Castin_CRAS} written in momentum space
in second quantized form in terms of the fermionic annihilation operators $c_\kk$, the extra-particle 
annihilation operators $a_\kk$ and the closed-channel molecule annihilation operators $b_\kk$:
\be
H = H_{\rm at} + H_{\rm mol} + H_{\rm at-mol} + H_{\rm open},
\ee
with 
\bea
H_{\rm at}\! &=&\! \int \frac{d^3k}{(2\pi)^3} \left[E_\kk c_\kk^\dagger c_\kk + \alpha E_\kk a_\kk^\dagger a_\kk\right] \\
H_{\rm mol}\! &=&\! \int \frac{d^3k}{(2\pi)^3} \left(E_{\rm mol} +\frac{\alpha}{1+\alpha} E_\kk \right) b_\kk^\dagger b_\kk \\
H_{\rm at-mol}\! &=&\! \Lambda\! \int\! \frac{d^3k_1 d^3 k_2}{[(2\pi)^3]^2}
\chi(\kk_{12})
[b_{\kk_1+\kk_2}^\dagger a_{\kk_1} c_{\kk_2} \!
+\! \mbox{h.c.}] \\
H_{\rm open} &=&\! g_0 \int \frac{d^3k_1 d^3 k_2 d^3 k_3 d^3 k_4}{[(2\pi)^3]^4}
\chi(\kk_{12}) \chi(\kk_{43}) \nonumber \\
&\times&  (2\pi)^3 \delta(\kk_1+\kk_2-\kk_3-\kk_4)
a_{\kk_4}^\dagger c_{\kk_3}^\dagger c_{\kk_2} a_{\kk_1}.
\eea
Whereas the $c_\kk$ obey the usual free space anticommutation relations
\be
\{ c_\kk, c_{\kk'}^\dagger\} = (2\pi)^3 \delta(\kk-\kk'),
\ee
the statistical nature (fermionic or bosonic) of the extra particle and of the closed-channel molecule does not need
to be specified here, since there will be at most one of the particles in the state vector. 
Simply, the $c_\kk$ and $c_\kk^\dagger$ commute with $a_\kk$ and $b_\kk$.
In the kinetic energy terms of the atoms $H_{\rm at}$ and of the closed-channel molecule $H_{\rm mol}$,
we have introduced the free fermion dispersion relation $E_\kk=\hbar^2 k^2/(2m)$ and the mass
ratio of a fermion to the extra particle:
\be
\alpha \equiv \frac{m}{M}.
\ee
The internal energy $E_{\rm mol}$ of the closed-channel molecule is counted with respect to the dissociation
limit of the open channel and is experimentally adjusted thanks to the Zeeman effect by tuning of the external magnetic field.
$H_{\rm at-mol}$ represents the coherent interconversion of a closed-channel molecule into one fermionic atom
and the extra particle, due to the coupling between the closed channel and the open channel.
It involves the interchannel coupling constant $\Lambda$ and is regularized by the momentum space cut-off function
$\chi$, assumed to be real and rotationally invariant, that tends to one a zero momentum,
and that rapidly tends to zero at large momenta with a
width $1/b$, where the interaction range $b$ is of the order of the van der Waals length.
Note that the argument of the cut-off function $\chi$ is the relative wave vector of a fermion (of momentum $\hbar\kk_2$) with respect to
the extra particle (of momentum $\hbar \kk_1$):
\be
\kk_{12} \equiv \mu \left(\frac{\kk_2}{m}-\frac{\kk_1}{M}\right)= \frac{\kk_2-\alpha \kk_1}{1+\alpha},
\ee
where 
\be
\mu=\frac{m M}{m+M} 
\ee
is the reduced mass, so as to preserve Galilean invariance. 
Finally, $H_{\rm open}$ models the direct interaction between atoms in the open channel, in the form of a separable
potential with bare coupling constant $g_0$ and the same cut-off function $\chi$ as in $H_{\rm at-mol}$.
This direct interaction is characterized by the so-called background scattering length $a_{\rm bg}$.
The inclusion of both $H_{\rm at-mol}$ and $H_{\rm open}$ allows to recover
the usual expression for the scattering length
as a function of the magnetic field $B$ \cite{revue_feshbach},
\be
a(B)=a_{\rm bg}\times \left(1-\frac{\Delta B}{B-B_0}\right)
\label{eq:adb}
\ee
if $E_{\rm mol}$ is taken to be an affine function of $B$.
The quantity $\Delta B$ is the so-called magnetic width of the Feshbach resonance.

We now derive from the model Hamiltonian a momentum space integral equation {\sl \`a la} Skorniakov-Ter-Martirosian \cite{STM}
for the three-body problem of two fermions and one extra particle,
closely following reference \cite{Castin_CRAS}
downgraded from the $3+1$ to the $2+1$ case. In the search for bound states, we take a negative
eigenenergy,  $E < 0 $, and we express Schr\"odinger's equation $0=(H-E)|\Psi\rangle$ for a ket
of zero total momentum and 
being the sum of general ansatz with zero or one closed-channel molecule: 
$|\Psi\rangle = |\psi_{3\,\rm at}\rangle + |\psi_{1\,{\rm at}+1\,{\rm mol}}\rangle$, with
\bea
\label{eq:psi3at}
|\psi_{3\,\rm at}\rangle &= &
\int \frac{d^3k_1 d^3k_2 d^3k_3  }{[(2\pi)^3]^3} 
(2\pi)^3 \delta(\kk_1+\kk_2+\kk_3) \nonumber \\
&\times& A(\kk_1,\kk_2,\kk_3) a_{\kk_1}^\dagger c_{\kk_2}^\dagger c_{\kk_3}^\dagger |0\rangle \\
|\psi_{1\,\rm at+1\,mol}\rangle &=&  \int \frac{d^3k}{(2\pi)^3} 
B(\kk) b_{-\kk}^\dagger c_\kk^\dagger |0\rangle.
\eea
Thanks to the fermionic antisymmetry we can impose that $A(\kk_1,\kk_2,\kk_3)$ is an antisymmetric
function of $\kk_2$ and $\kk_3$. On the contrary, a fermionic atom and a closed-channel molecule
are distinguishable objects and there is no exchange symmetry constraint on the function $B(\kk)$.
Projecting Schr\"odinger's equation on the subspace with three atoms, and using $E<0$, we are
able to express $A$ in terms of $B$ and of an auxiliary unknown function $\tilde{B}$ 
obtained by a partial contraction of $A$:
\bea
\tilde{B}(\kk_3) &=& \int \frac{d^3k_1d^3k_2}{[(2\pi)^3]^2}\chi(\kk_{12}) (2\pi)^3 \delta(\kk_1+\kk_2+\kk_3) \nonumber \\
&& \times A(\kk_1,\kk_2,\kk_3).
\label{eq:defBt}
\eea
More precisely,
\begin{multline}
A(\kk_1,\kk_2,\kk_3) = \frac{\Lambda/2}{E-(\alpha E_{\kk_1}+E_{\kk_2}+E_{\kk_3})} \\
\times [\chi(\kk_{12}) D(\kk_3) - \chi(\kk_{13}) D(\kk_2)]
\label{eq:expA}
\end{multline}
where the convenient unknown function is actually $D$ such that
\be
\Lambda D(\kk) = \Lambda B(\kk) + 2 g_0 \tilde{B}(\kk).
\label{eq:defD}
\ee
Plugging the expression (\ref{eq:expA}) of $A$ into the definition (\ref{eq:defBt}) of $\tilde{B}$
gives a first important equation
\begin{multline}
\frac{2}{\Lambda} \tilde{B}(\kk_3) = \int \frac{d^3k_1d^3k_2}{[(2\pi)^3]^2} \chi(\kk_{12})
\frac{(2\pi)^3 \delta(\kk_1+\kk_2+\kk_3)}{E-(\alpha E_{\kk_1}+E_{\kk_2}+E_{\kk_3})} \\
\times [\chi(\kk_{12}) D(\kk_3) - \chi(\kk_{13}) D(\kk_2)].
\label{eq:imp1}
\end{multline}

The second important equation is obtained by projecting Schr\"odinger's equation on the subspace with one
atom and one closed-channel molecule (in which case the direct open-channel interaction can not contribute):
\be
2 \Lambda \tilde{B}(\kk) = [E_{\rm rel}(\kk)-E_{\rm mol}] B(\kk)
\label{eq:molat}
\ee
where we have introduced what we call the relative energy 
\be
E_{\rm rel}(\kk) = E - \left(E_{\kk} + \frac{\alpha}{1+\alpha} E_{\kk}\right).
\label{eq:erel}
\ee
This is indeed the relative energy of one of the fermions and of the extra particle, knowing that the second
fermion has a wave vector $\kk$, since one subtracts in (\ref{eq:erel}) from the total energy $E$ the
kinetic energy $E_\kk$ of the second fermion and the center of mass kinetic energy of the first-fermion-plus-extra-particle.
One expresses $\tilde{B}$ in terms of $D$ by elimination of $B$ between (\ref{eq:defD})
and (\ref{eq:molat}). One then eliminates $\tilde{B}$ between the resulting equation and (\ref{eq:imp1})
to finally obtain a closed equation for $D$:
\begin{multline}
0=\frac{\mu D(\kk_3)}{2\pi\hbar^2 f[E_{\rm rel}(\kk_3)]}  - 
\int \frac{d^3k_1d^3k_2}{[(2\pi)^3]^2} 
(2\pi)^3 \delta(\kk_1+\kk_2+\kk_3) \\
\times \frac{\chi(\kk_{12}) \chi(\kk_{13})  D(\kk_2)}{E-(\alpha E_{\kk_1}+E_{\kk_2}+E_{\kk_3})}.
\end{multline}
The function $f$ is related to the two-body $T$ matrix for the scattering of a fermion
and of the extra particle \cite{Castin_CRAS},
\be
\langle \kk_f | T(\epsilon + i0^+) |\kk_i\rangle = -\frac{2\pi\hbar^2}{\mu} \chi(\kk_f)
\chi(\kk_i) f(\epsilon+i0^+)
\ee
where the relative wave vectors $\kk_i$, $\kk_f$ and the energy $\epsilon$ are arbitrary (the $T$ matrix
is not necessarily on shell in that expression).

The general expression of $f$ is given in \cite{Castin_CRAS}.
Here however, we shall concentrate on the limit of an infinitely narrow Feshbach resonance.
We thus take the zero-range limit $b\to 0$, 
in which case the cut-off function $\chi$ (of width $\propto 1/b$) tends to unity.
It is assumed that there is no resonant interaction in the open channel, so that the corresponding background
scattering length $a_{\rm bg}$ is $O(b)$ and also tends to zero. On the contrary, the interchannel coupling
$\Lambda$ is kept fixed, so as to keep a non-zero effective range, and $E_{\rm mol}$ is adjusted
to keep a fixed value of the scattering length $a$.
In this case, the function $f$ for a positive and fixed energy $\epsilon > 0$ (so that $\mu b^2 \epsilon/\hbar^2\to 0$) simply tends
to the $s$-wave scattering amplitude $f_k$ on an infinitely narrow Feshbach resonance \cite{Petrov_rstar}:
\be
f(\epsilon+i0^+)=f_k
\label{eq:valf}
\ee
with
\be
f_k = -\frac{1}{a^{-1} + ik + k^2 R_*}.
\label{eq:fk}
\ee
The Feshbach length $R_*$ is expressed in terms of the width $\Delta B$ of the Feshbach
resonance as \cite{Petrov_rstar}:
\be
R_* =\frac{\hbar^2}{2\mu  a_{\rm bg} \mu_b \Delta B},
\ee
where the differential magnetic moment between the closed and open channels
is $\mu_b = d E_{\rm mol}/dB$ taken for $B=B_0$.
In (\ref{eq:fk}), the relative wavenumber $k=(2\mu \epsilon)^{1/2}/\hbar$ since
the energy is positive.  For a negative energy, the function $f$ has the same expression (\ref{eq:valf})
if one uses the analytic continuation $k=i (-2\mu \epsilon)^{1/2}/\hbar$ in $f_k$
\cite{footnote_clear}.

It is convenient to represent the eigenenergy $E$ in terms of a wavenumber, setting
\be
E = -\frac{\hbar^2 q^2}{2\mu}
\label{eq:defq}
\ee
where we recall that $\mu=m M/(m+M)$ is the reduced mass and $q \geq 0$. 
Similarly the relative energy is represented
by a wavenumber, $E_{\rm rel}(\kk) = -\hbar^2 q^2_{\rm rel}(k)/(2\mu)$ leading to
\be
q_{\rm rel}(k) = \left[q^2+ \frac{1+2\alpha}{(1+\alpha)^2}\, k^2\right]^{1/2}.
\ee
For an infinitely narrow $s$-wave Feshbach resonance with a scattering length $a$,
the integral equation resulting from Schr\"odinger's equation, the equivalent of the Skorniakov-Ter-Martirosian
equation \cite{STM} for our problem,
is then
\begin{multline}
0 = \left[-a^{-1}+q_{\rm rel}(k)+q_{\rm rel}^2(k) R_*\right] D(\kk) \\
+\int \frac{d^3k'}{2\pi^2} \frac{D(\kk')}{q^2+k^2+k^{'2}+\frac{2\alpha}{1+\alpha} \kk\cdot \kk'}.
\label{eq:int_gen}
\end{multline}
In that equation, for the sake of generality, we have kept an arbitrary value of the scattering length $a$.
In what follows, we shall restrict to the exact resonance location where $1/a=0$.

We shall also take advantage of rotational invariance to reduce the integral
equation to an unknown function of a single variable only, as done in \cite{STM}.
The eigenstates of the Hamiltonian $H$ may be assumed of a fixed total angular momentum
of angular quantum number $l$. Without loss of generality one can also assume that the angular momentum
along the quantization axis $z$ is zero. The corresponding ansatz for $D$ is thus
\be
D(\kk) = Y_l^0(\kk) f^{(l)}(k)
\label{eq:Dansatz}
\ee
where the notation $Y_l^{m_l}(\kk)$ stands for the spherical harmonics $Y_l^{m_l}(\theta,\phi)$
where $\theta$ and $\phi$ are respectively the polar and azimuthal angles of the vector
$\kk$ in a system of spherical coordinates of polar axis $z$. The unknown function $f^{(l)}$
then depends only on the modulus $k$ of $\kk$. We see that this also fixes the parity of
the eigenstate to the value $(-1)^l$.
In the integral over $\kk'$ in (\ref{eq:int_gen}), we then perform the change of
variable of unit Jacobian,
\be
\kk'= \mathcal{R} \KK
\ee
where $\mathcal{R}$ is the rotation in $\mathbb{R}^3$  defined by the Euler decomposition
of its inverse:
\be
\mathcal{R}^{-1} = \mathcal{R}_z(-\phi) \mathcal{R}_y(\theta) \mathcal{R}_z(\pi-\phi)
\ee
where the notation $\mathcal{R}_i(\alpha)$ stands for the rotation of an angle $\alpha$
around the axis $i\in \{x,y,z\}$. This choice ensures that
$\kk/k = \mathcal{R} \mathbf{e}_z$, where $\mathbf{e}_z$ is the unit vector defining
the $z$ axis. In the denominator of the integrand of (\ref{eq:int_gen}), the scalar
product $\kk\cdot \kk'$ is then transformed as $k \mathbf{e}_z \cdot \KK$, so that
the denominator is invariant by rotation of $\KK$ around $z$.
In the numerator we use the transformation of spherical harmonics under rotation,
see (8.6-2) and (8.6-1) in \cite{Wu}, and the relation $\left[Y_l^{m_l}(\theta,-\phi)\right]^*=
Y_l^{m_l}(\theta,\phi)$:
\be
Y_l^0(\mathcal{R}\KK) = 
\left(\frac{4\pi}{2l+1}\right)^{1/2} \sum_{m_l=-l}^{l} Y_l^{m_l}(\kk)
Y_l^{m_l}(\KK),
\label{eq:transf}
\ee
which in particular allows to pull out the factors $Y_l^{m_l}(\kk)$
expected from rotational invariance.
The integration over $\KK$ is then conveniently performed in spherical coordinates
of polar axis $z$. All the terms with $m_l\neq 0$ in (\ref{eq:transf}) vanish
in the integration over the azimuthal angle of $\KK$. 
From the expression of the spherical harmonics in terms of the Legendre polynomial
of degree $l$ \cite{Wu},
\be
Y_l^0(\KK) = \left(\frac{4\pi}{2l+1}\right)^{-1/2} P_l(u=\KK\cdot\mathbf{e}_z/K),
\ee
we finally obtain the reduced integral equation to be solved in the sector of angular momentum $l$,
for $1/a=0$:
\begin{multline}
0 = \left[q_{\rm rel}(k)+q_{\rm rel}^2(k) R_*\right] f^{(l)}(k) \\
+\int_0^{+\infty}\!\frac{dK}{\pi}  f^{(l)}(K) \int_{-1}^{1} du\,
\frac{P_l(u)K^2}{q^2+k^2+K^2+\frac{2\alpha}{1+\alpha} k K u}.
\label{eq:intl}
\end{multline}

\section{Analytical solutions in particular cases}
\label{sec:analy}

Several remarkable analytical techniques are now available to solve the three-body
problem in some appropriate limiting cases \cite{Petrov_fermions,
Mora_3corps,Mora_CRAS,Macek,Petrov_bbo}.
Whereas we do not know how to solve (\ref{eq:intl}) analytically in general, 
it is possible
to find solutions when there is an extra symmetry available, that is scale invariance.
The most standard regime corresponds to the limit $R_*\to 0$, in which case our model
reduces to the so-called zero-range or Bethe-Peierls model, where the interactions
are included {\sl via} two-body contact conditions on the wavefunction \cite{Bethe_Peierls}.
Since $1/a=0$, these contact conditions are indeed scaling invariant, which allows
to fully solve the problem \cite{Efimov,WernerCastinsymetrie}. 
Because the zero-range model is usually solved in position space, it is interesting
here to briefly show the calculations in momentum space.
In the regime of interest,
where the Efimov effect takes place, the zero-range model is however not well defined,
and an extra three-body condition has to be introduced to make it self-adjoint
\cite{Danilov}, involving the three-body parameter.

So the relevant case here is $R_*>0$. The existence of such a finite length scale
characterizing the interactions breaks the scale invariance. Fortunately,
as shown in \cite{Mora_CRAS}, if one restricts to the zero energy case $E=0$, 
equations of the type (\ref{eq:intl}) can still be solved analytically.
This gives access to the three-body parameter, and thus to the characterization
of the Efimov spectrum of trimers.

\subsection{Zero-range case ($R_*=0$) at zero energy}
\label{subsec:zrze}

For $R_*=0$ and at zero energy $q=0$, the integral equation (\ref{eq:intl}) is manifestly
scaling invariant: If $f^{(l)}(k)$ is a solution, the function $f_\lambda^{(l)}(k)=
f^{(l)}(k/\lambda)$ is also a solution, $\forall \lambda>0$, and we expect that the two
functions $f^{(l)}$ and $f_\lambda^{(l)}$ are proportional. We thus seek a solution
in the form of a power-law, 
\be
f^{(l)}(k)=k^{-(s+2)}.
\label{eq:ansatz_puissance}
\ee
The form of the exponent results from the general theory, see section 3.3 in \cite{livre_unitaire}:
For a $N$-body problem, with here $N=3$, for $s$ to be a direct generalization of the exponent $s_0$
introduced by Efimov \cite{Efimov}, the exponent in (\ref{eq:ansatz_puissance}) should be
$-[s+(3N-5)/2]$.
For convergence issues, it is simpler to assume in explicit calculations that $s=iS$, where $S$ is real.
By analytic continuation, the result however extends to real $s$ also, as also shown
by the real space calculation \cite{Werner3corps}.
We inject the ansatz (\ref{eq:ansatz_puissance}) in (\ref{eq:intl}) with $q=0$, $R_*=0$,
and we perform the change of variable $K=k e^x$ to obtain an equation for $s$:
\be
0=\Lambda_l(s),
\label{eq:lam_l_de_s=0}
\ee
with the function
\be
\Lambda_l(s)\equiv  \frac{(1+2\alpha)^{1/2}}{1+\alpha} +
\int_{-1}^{1} \!\! du\, P_l(u) \! \int_{-\infty}^{+\infty} \!\! \frac{dx}{2\pi} \frac{e^{-iS x}}{\cosh x
+\frac{\alpha}{1+\alpha} u}.
\label{eq:def_Lambda}
\ee
Using contour integration and the Cauchy residue formula, the integral over $x$ may be calculated:
Setting $\theta=\arccos\left(\frac{\alpha}{1+\alpha} u\right)$, so that
$\theta\in [0,\pi]$, we obtain
\be
\int_{-\infty}^{+\infty}\frac{dx}{2\pi} \frac{e^{-iS x}}{\cosh x+\frac{\alpha}{1+\alpha} u}
= \frac{\sin(s\theta)}{\sin(s\pi) \sin\theta}.
\label{eq:int_contour}
\ee
Successive integration over $u$ is simplified by taking $\theta$ rather than $u$ as integration
variable, with a Jacobian that simplifies with the factor $\sin \theta$
in the denominator of (\ref{eq:int_contour}).
Following \cite{Petrov_fermions} we then parameterize the mass ratio by an angle
$\nu\in ]0,\pi/2[$:
\be
\nu = \arcsin \frac{\alpha}{1+\alpha}=\arcsin \frac{m}{m+M},
\label{eq:def_nu}
\ee
which leads to the remarkable property
\be
\cos \nu = \frac{(1+2\alpha)^{1/2}}{1+\alpha}.
\label{eq:cosnu}
\ee
Also using $\arccos[\alpha/(1+\alpha)]=\frac{\pi}{2}-\nu$, we obtain
\be
\Lambda_l(s)= \cos \nu + \frac{1}{\sin \nu} \int_{\frac{\pi}{2}-\nu}^{\frac{\pi}{2}+\nu}
d\theta\, P_l\left(\frac{\cos\theta}{\sin \nu}\right) \frac{\sin (s\theta)}{\sin(s\pi)}.
\ee
This can be further simplified taking advantage of the parity of the Legendre
polynomial, $P_l(-u)=(-1)^l P_l(u)$ for all $u$, by shifting the integration variable
$\theta$ by $\pi/2$. Depending on the even or odd parity of the angular momentum $l$
this reduces to
\bea
\!\!\!\!\Lambda_l(s) &\!\stackrel{l\, {\rm even}}{=}\!& \!\!\cos \nu + \frac{1}{\sin \nu}\!\! \int_0^{\nu} \!\! d\theta 
P_l\left(\frac{\sin\theta}{\sin \nu}\right) \frac{\cos (s\theta)}{\cos(s\pi/2)} \\
\!\!\!\!\Lambda_l(s) &\!\stackrel{l\, {\rm odd}}{=}\!&\!\! \cos \nu - \frac{1}{\sin \nu}\!\! \int_0^{\nu}\!\!  d\theta 
P_l\left(\frac{\sin\theta}{\sin \nu}\right) \frac{\sin (s\theta)}{\sin(s\pi/2)}.
\label{eq:rep_odd}
\eea
Eq.~(\ref{eq:rep_odd}) will be quite useful to obtain analytical results on the Efimovian trimer
spectrum in the large $\alpha$ limit, see subsections \ref{subsec:etae} and \ref{subsec:aftts}.

An explicit expression of $\Lambda_l(s)$ as a sum of a finite number of simple functions
of $s$ may be obtained by representing the function $P_l\left(\frac{\sin\theta}{\sin \nu}\right)$
as a Fourier sum, that is a sum of $\cos n\theta$, $0\leq n\leq l$, $n$ even,
for an even $l$, and a sum of $\sin n\theta$, $1\leq n\leq l$, $n$ odd, for an odd $l$.
The coefficients are polynomials of $1/\sin\nu$ that are simple to calculate
analytically from the known coefficients of the Legendre polynomials $P_l(u)$
\cite{Gradstein}.  The resulting integrals over $\theta$, {\sl e.g.}\ of $\sin(n\theta)\sin(s\theta)$,
are then straightforward to evaluate.
We finally obtain the explicit formula valid for arbitrary parity of $l$:
\begin{multline}
\Lambda_l(s) = \cos \nu + \frac{1}{\sin\nu \cos[(s+l)\pi/2]} \\
\times
\sum_{n=0}^{l} c_n \left\{\frac{\sin[(s+n)\nu]}{s+n}+(-1)^l
\frac{\sin[(s-n)\nu]}{s-n}\right\}.
\label{eq:lambda_expli}
\end{multline}
The coefficients $c_n$ are zero for $l-n$ odd. For $l-n$ even,
\begin{multline}
c_n = \left(1-\frac{1}{2}\,\delta_{n,0}\right) 
\frac{(-1)^{(l+n)/2}}{(4\sin\nu)^l} \\
\times \sum_{k=0}^{(l-n)/2} \frac{(-4\sin^2\nu)^k(2l-2k)!}{k!(l-k)!
\left(\frac{l-n}{2}-k\right)!\left(\frac{l+n}{2}-k\right)!}
\end{multline}
where $\delta_{ij}$ is the usual Kronecker delta.

The form (\ref{eq:lambda_expli}) has the interesting feature that the coefficients are $s$-independent, 
which makes the numerical evaluation of $\Lambda_l(s)$ as a function of $s$ 
particularly efficient. As a test, it is however interesting to compare to the transcendental
equation for $s$ obtained by the direct calculation {\sl \`a la} Efimov in position
space. As detailed in the Appendix \ref{app:real_space}, introducing the hypergeometric function
${}_2 F_1$ as in \cite{hypergeom} to solve some differential equation,
we generalize the formulas of \cite{hypergeom} to an arbitrary mass ratio,
a generalization that was done already with the adiabatic hyperspherical method in \cite{Rittenhouse}:
\begin{multline}
\Lambda_l(s) = \cos\nu +(-1)^l \sin^l\nu\, \frac{\Gamma(\frac{l+1+s}{2})\Gamma(\frac{l+1-s}{2})}
{2\pi^{1/2} \Gamma(l+\frac{3}{2})} \\
\times
{}_2 F_1\left(\frac{l+1+s}{2},\frac{l+1-s}{2},l+\frac{3}{2};\sin^2\nu\right)
\label{eq:trans_hyper}
\end{multline}
where $\Gamma$ is the Gamma function. A variant of Eq.~(\ref{eq:trans_hyper}) will be quite useful to obtain analytical
results on the Efimovian trimer spectrum in the large $l$ limit, see subsections \ref{subsec:etae} and \ref{subsec:aftts}.

\subsection{Zero range case ($R_*=0$) at negative energy}
\label{subsec:zrne}

Once the imaginary values of the Efimov exponent $s$ are determined by solution
of the transcendental equation $\Lambda_l(s)=0$, which is possible for $l$ odd and $\alpha$ larger
than a critical value $\alpha_c^{(l)}$, see subsection \ref{subsec:etae},
one can determine the corresponding Efimovian trimer solutions of the zero-range
theory at arbitrary energies, in free space and also in an isotropic harmonic trap, using the general
real space formalism relying on separability of the Bethe-Peierls problem in hyperspherical
coordinates \cite{WernerCastinsymetrie,livre_unitaire,Werner3corps}. Here we find it interesting to deduce
from the real space solution the explicit form of the momentum space solution,
restricting for simplicity to the values of the mass
ratio $\alpha$ and the (necessarily odd) angular momentum $l$ such that the Efimov effect takes place.
This may be useful for example to calculate the atomic momentum distribution of the Efimov
trimer states in the zero-range limit, as was done for three bosons in \cite{nkefim}.

The idea to obtain $D(\kk)$ is to take the limit of the position $\rr_1$ of the  extra particle and the position $\rr_2$ of a fermionic atom
converging to the same location, with a fixed value of their center of mass position and of the position 
$\rr_3$ of the second fermionic atom. According to the Bethe-Peierls framework, the atomic wavefunction
$\psi(\rr_1,\rr_2,\rr_3)$ shall then diverge as $1/r$, with $\rr=\rr_1-\rr_2$, with a factor depending on the Jacobi
coordinate $\xx=\rr_3-(M\rr_1+m\rr_2)/(M+m)$:
\be
\psi(\rr_1,\rr_2,\rr_3) \underset{r\to 0}{\stackrel{\xx\ {\rm fixed}}{\sim}} \frac{\mathcal{A}(\xx)}{r}.
\ee
Then one calculates $\mathcal{A}(\xx)$ in two different ways. First, one uses Efimov solution exposed in the Appendix
\ref{app:real_space}: From Efimov's ansatz (\ref{eq:ansatz_psi}) and from the solution $F(R)\propto
K_s[(\bar{m}/\mu)^{1/2}q R]$ of the hyperradial equation (\ref{eq:hyper_radial}), where $\bar{m}$ is an arbitrary mass unit 
and $K_s$ a Bessel function, one finds
for a vanishing angular momentum along $z$:
\be
\mathcal{A}(\xx) \propto Y_l^{0}(\xx) \frac{K_s(Q x)}{x},
\label{eq:expr_acalli}
\ee
where we have set
\be
Q\equiv q\left(\frac{\mu_{\rm am}}{\mu}\right)^{1/2} = \frac{1+\alpha}{(1+2\alpha)^{1/2}}\, q =\frac{q}{\cos \nu}
\label{eq:defQ}
\ee
where $\mu_{\rm am}$ is the reduced mass of one fermionic atom of mass $m$ and one pair of fermion-plus-extra-particle 
of mass $m+M$.
Second, after inspection of (\ref{eq:psi3at}), one calculates the atomic wavefunction $\psi$ by taking the Fourier transform 
of $A(\kk_1,\kk_2,\kk_3) (2\pi)^3\delta(\kk_1+\kk_2+\kk_3)$, where $A$ is given by
of (\ref{eq:expA}), taking the functions $\chi$ equal to unity. As expected, the contribution
involving $\chi(\kk_{12})D(\kk_3)$ is the one leading to a divergence of $\psi$ for $r\to 0$. 
Integration over $\kk_1$ is straightforward thanks to the factor $(2\pi)^3\delta(\kk_1+\kk_2+\kk_3)$.
Then integration over $\kk_2$ can be done after the change of variable $\kk_2=\kk_2'-\frac{\alpha}{1+\alpha} \kk_3$,
using 
\be
\int \frac{d^3k_2'}{(2\pi)^3} \frac{e^{-i\kk_2'\cdot\rr}}{k_2^{'2}+K^2} = \frac{e^{-K r}}{4\pi r}
\ee
where $K=[q^2+\frac{1+2\alpha}{(1+\alpha)^2}k_3^2]^{1/2}>0$. For $r\to 0$, one approximates $e^{-Kr}/r$ with $1/r$
and one obtains
\be
\mathcal{A}(\xx) \propto \int\frac{d^3k_3}{(2\pi)^3} D(\kk_3) e^{i \kk_3\cdot \xx}.
\ee
We thus set, for an arbitrary choice of normalization leading to a dimensionless function:
\be
D(\kk) = \int d^3x \, e^{-i\kk\cdot\xx} Y_l^{0}(\xx) \frac{Q^2 K_s(Q x)}{x}.
\label{eq:ddek}
\ee

A first technique to calculate the integral in (\ref{eq:ddek}) is to use the expansion of the
plane wave on spherical harmonics: According to the identity
(8.7-13) in \cite{Wu}, the amplitude of the function $\xx\to e^{i\kk\cdot\xx}$ on the spherical
harmonics $Y_l^{m_l}(\xx)$ is $4\pi i^l j_l(kx) [Y_l^{m_l}(\kk)]^*$, where the spherical Bessel
function is real and may be expressed in terms of the usual Bessel function $J$:
\be
j_l(kr) = \left(\frac{\pi}{2kr}\right)^{1/2} J_{l+1/2}(kr).
\ee
It turns out that the integral over $\mathbb{R}^+$ of the product of a power law and of two Bessel functions 
may be expressed exactly in terms of the hypergeometric function ${}_2 F_1$, see relation
6.576(3) in \cite{Gradstein}:
\begin{multline}
D(\kk) = 2\pi^{3/2}(-i)^l (k/Q)^l Y_l^0(\kk) \frac{\left|\Gamma\left(1+\frac{l+s}{2}\right)\right|^2}{\Gamma(l+3/2)} \\
\times {}_2 F_1\left(1+\frac{l+s}{2},1+\frac{l-s}{2},l+\frac{3}{2};-k^2/Q^2\right).
\label{eq:solun}
\end{multline}
This immediately shows that $D(\kk)$ vanishes as $k^l$ for $k\to 0$, which is generically the case
for a regular function of angular momentum $l$.
It also allows to obtain the large momentum behavior of $D(\kk)$ \cite{details}, 
\begin{multline}
D(\kk) \underset{k/Q\to +\infty}{=}  \frac{i \pi^2 Q^2}{k^2 \cos(\pi s/2)} Y_l^0(\kk) \\
\times \Bigg\{\left(\frac{2k}{Q}\right)^s \left[\prod_{n=0}^{(l-1)/2} \frac{s-(2n+1)}{s+2n}\right] \\
\times
\left[1+\frac{(l+s-1)(l-s+2)}{4(1-s)(k/Q)^2} + O(Q/k)^4\right]
 + \mbox{c.c} \Bigg\}
\label{eq:zr_grandk}
\end{multline}
which will play a crucial role
in what follows to obtain the three-body parameter at non-zero $R_*$.
In particular, Eq.~(\ref{eq:zr_grandk}) shows that $k^2 D(\kk)$ is asymptotically an oscillating function 
of $k/Q$ that is log-periodic: The same pattern is reproduced when $k$ is multiplied by 
$\exp(2\pi/|s|)$. For $k^2 D(\kk)$ to approach this asymptotic oscillating function, the very stringent
condition $k/Q \gg \exp(2\pi/|s|)$ is fortunately not required, it is simply sufficient that $k/Q \gg 1$
(for $l$ and $|s|$ not much larger than unity).

A second technique to calculate (\ref{eq:ddek}) is to use spherical coordinates of axis the quantization axis $z$.
The integral over the azimuthal angle $\phi$ is straightforward. The integral over the modulus $x$ 
can be performed if one uses the integral representation 8.432(1) of the Bessel function given in \cite{Gradstein},
$K_s(z)=\int_0^{+\infty} dt\, \exp(-z\cosh t) \cosh (st)$. In the integral over the polar angle $\theta$,
one replaces $Y_l^0(\xx)$ by its expression in terms of the Legendre polynomial $P_l(u)$ with the variable $u=\cos\theta$,
and one integrates by parts the factor $1/(\cosh t + i k u/Q)^2$ that appeared after integration over $x$. 
The integral over $t$ can then be performed with contour integration:
\be
\int_{-\infty}^{+\infty} \frac{dt}{2\pi} \frac{e^{-iSt}}{\cosh t + i \sinh\beta}
= \frac{\sin[S(\beta+i\pi/2)]}{i\sinh(S\pi)\cosh\beta}
\ee
where $S$ and $\beta$ are real quantities. Changing to the variable $\beta =\argsh (ku/Q)$ in the integral over $u$,
one obtains
\begin{multline}
D(\kk) = Y_l^0(\kk) \frac{2i\pi^2Q^2}{k\cosh(S\pi/2)} \left\{\frac{\cos[S\argsh(k/Q)]}{(k^2+Q^2)^{1/2}} \right.
\\
\left. -\frac{1}{k} \int_0^{\argsh(k/Q)} d\beta \, P_l'\left(\frac{Q\sinh \beta}{k}\right) \cos(S\beta)
\right\}
\end{multline}
where we have set $s=i S$. Generalizing the technique of subsection \ref{subsec:zrze} to the derivative
$P_l'$ of the Legendre polynomial allows a direct evaluation of $D(\kk)$ without the need of hypergeometric functions.

\subsection{Case $R_*>0$ at zero energy}
\label{subsec:rsze}

In this subsection, as in the previous one, we restrict to the case where an Efimov effect takes place:
Anticipating on results of subsection \ref{subsec:etae},
the angular momentum quantum number $l$ is odd and the mass ratio $\alpha$ is larger
than the corresponding critical value $\alpha_c^{(l)}$, so that the function $\Lambda_l(s)$ has a single purely imaginary
root $s_l=i S_l$ with positive imaginary part: 
\be
\Lambda_l(i S_l) = 0 \ \ \ {\rm with}\ \ \ S_l > 0.
\ee
As remarkably shown for three bosons in \cite{Mora_CRAS} the narrow Feshbach resonance model can be solved
analytically at zero energy, which gives access to the three-body parameter, that is to the energy scale
in the asymptotically geometric Efimovian trimer spectrum.

\begin{figure}[htb]
\includegraphics[width=8cm,clip=]{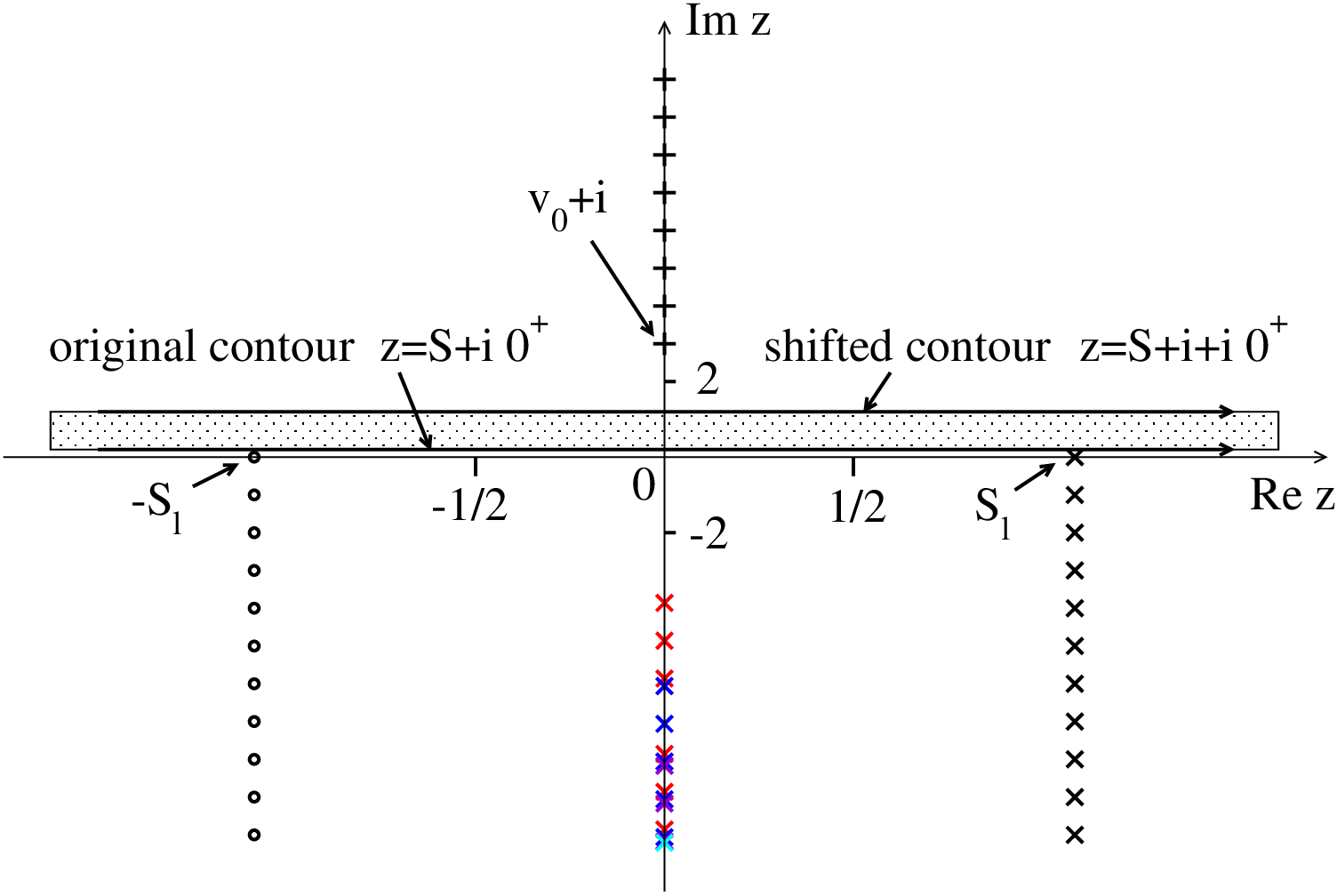}
\caption{(color online) In the complex plane, illustration of the procedure used to determine the Fourier transform
of the function $x\to e^x F^{(l)}(x)$: Under the {\sl a posteriori} checkable hypothesis (\ref{eq:hypothese}) 
(no pole of $\tilde{F}^{(l)}(z)$ in the grey area), one can shift the integration contour in (\ref{eq:a_decaler})
upwards by one unity along the vertical axis, to obtain the Fourier representation Eq.~(\ref{eq:shifted}) of
$x\to e^x F^{(l)}(x)$. Circles: Poles of $\tilde{F}^{(l)}(z)$ due to the function hyperbolic sine
in the denominator of (\ref{eq:introC}). Pluses: Poles of $\tilde{F}^{(l)}(z)$ due to the factors 
$\Gamma(1+iz-iv_n)$, $n\geq 0$, in the numerators of the infinite product (\ref{eq:weier_Ft}). 
Black crosses: Poles of $\tilde{F}^{(l)}(z)$ due to the factor $\Gamma(iS_l-iz)$ in (\ref{eq:weier_Ft}).
Colored crosses on the $\mbox{Im}\, z<0$ part of the imaginary axis: 
Poles of $\tilde{F}^{(l)}(z)$ due to the factors $\Gamma(-iu_n-iz)$ for $n>0$ in
(\ref{eq:weier_Ft}) [red: $n=1$; blue: $n=2$; violet: $n=3$; cyan: $n=4$];
from top to bottom, $n=1$ (three crosses), $n=2$ (two crosses), two triplets $n=1,2,3$ and
a quadruplet $n=1,2,3,4$.
The figure corresponds to $l=1$ and $\alpha=20$.
\label{fig:contour}}
\end{figure}

The underlying idea of the solution is that, at zero energy and after division of the overall Eq.~(\ref{eq:intl})
by a factor $k$, the integral part of the resulting equation is strictly scaling invariant. If one takes as variable
the logarithm of $k$ rather than $k$, this strict scaling invariance corresponds to a translation invariance,
which suggests that the integral part is simply a convolution product and leads one to perform a Fourier
transform with respect to $\ln k$.
More precisely, we use the ansatz:
\be
f^{(l)}(k) = e^{-2x}\, F^{(l)}(x)     \ \ \ \mbox{with}\ \ \ x=\ln\left(k R_*\cos \nu \right),
\label{eq:ansatz_fl}
\ee
where the factor $e^{-2x}$ shall lead to a convolution kernel with the desired even function,
and $\cos\nu$ is a function of the mass ratio given by Eq.~(\ref{eq:cosnu}).
Dividing Eq.~(\ref{eq:intl}) for $E=0$ by $k$, injecting the ansatz (\ref{eq:ansatz_fl}) in the resulting equation,
and finally multiplying by $e^{2x}$, we obtain
\be
0 = \left(1+e^x\right) F^{(l)}(x) \cos\nu+ \int_{-\infty}^{+\infty} dX\, \mathcal{K}_l(X-x) F^{(l)}(X)
\label{eq:convol}
\ee
with the kernel
\be
\mathcal{K}_l(x) = \frac{1}{2\pi} \int_{-1}^{1} du \frac{P_l(u)}{\cosh(x)+\frac{\alpha}{1+\alpha} u}.
\ee
As expected, (\ref{eq:convol}) is the convolution product.  We thus introduce the Fourier representation
of the function $F^{(l)}$ \cite{notation}:
\be
F^{(l)}(x) = \int_{-\infty+i0^+}^{+\infty+i0^+} \frac{dS}{2\pi} e^{iSx} \tilde{F}^{(l)}(S).
\label{eq:def_Fourier}
\ee
We expect that the function $F^{(l)}(x)$ oscillates periodically for $x\to -\infty$:
When $x\to -\infty$, the momentum $k$ tends to $0$, it becomes much smaller than $1/R_*$;
one enters a universal zero-energy, zero-range regime where solutions of the type (\ref{eq:ansatz_puissance})
are obtained, with $s$ purely imaginary for an Efimovian solution. According to (\ref{eq:ansatz_fl}),
this implies that $F^{(l)}(x)$ has plane wave oscillations at $x\to -\infty$, with a wavenumber $\pm S_l$
since both $s=iS_l$ and $s=-i S_l$ are roots of $\Lambda_l(s)=0$. More precisely, we expect that
there exist coefficients $A_{\pm}$ such that
\be
F^{(l)}(x) \underset{x\to-\infty}{=} A_+ e^{i S_l x} + A_- e^{-i S_l x} + o(1).
\label{eq:attendu}
\ee
This will be checked {\sl a posteriori}.  
As a consequence, the Fourier transform $\tilde{F}^{(l)}(S)$ has singularities on
the real axis. More precisely, Eq.~(\ref{eq:attendu}) leads to the natural conclusion that
\be
\tilde{F}^{(l)}(S) \ \mbox{has simple poles in}\ S=\pm S_l.
\label{eq:des_poles}
\ee
This is why the integration contour in (\ref{eq:def_Fourier}) is infinitesimally shifted
upwards in the complex plane. This is a standard procedure in physics, see for example the expression
of the unitary evolution operator as a Fourier transform of the resolvent for a system with time-independent
Hamiltonian \cite{Cohen-Tannoudji}.
In this Fourier representation, the convolution becomes a product, and one needs to calculate the Fourier
transform $\tilde{\mathcal{K}}_l(S)$ of the kernel function $K$. Exchanging the integration over $u$ and $x$, one then recovers exactly 
the integral in Eq.~(\ref{eq:def_Lambda}) knowing that $s=iS$ in that equation, so that
\be
\tilde{\mathcal{K}}_l(S) = \Lambda(iS) -\cos\nu.
\ee
The only subtle part is the determination of the Fourier representation of the function
$x\to e^x F^{(l)}(x)$. Multiplying (\ref{eq:def_Fourier}) by $e^x$ gives
\be
e^x  F^{(l)}(x) = \int_{-\infty+i0^+}^{+\infty+i0^+} \frac{dS}{2\pi} e^{i (S-i) x} \tilde{F}^{(l)}(S).
\label{eq:a_decaler}
\ee
This is not directly of the Fourier form (\ref{eq:def_Fourier}) because $S-i$ rather than $S$ appears
inside the exponential.
Now, if the integrand, that is here in practice the function $S\to \tilde{F}^{(l)}(S)$, is a meromorphic function
with no singularities
(no poles) in the band $0^+ \leq \mathrm{Im}\, z\leq 1+0^+$, that is $0< \mathrm{Im}\, z \leq 1$, where
$z\in \mathbb{C}$, we can shift the integration contour in (\ref{eq:a_decaler}) upwards by one unity along the vertical axis
in the complex plane, so as to replace $S-i$ with $S$.
Under the hypothesis
\be
\tilde{F}^{(l)}(z)\ \mbox{has no singularities for}\ 0<\mathrm{Im}\, z\leq 1,
\label{eq:hypothese}
\ee
we thus have
\be
e^x  F^{(l)}(x) = \int_{-\infty+i0^+}^{+\infty+i0^+} \frac{dS}{2\pi} e^{iSx} \tilde{F}^{(l)}(S+i),
\label{eq:shifted}
\ee
which is exactly of the Fourier functional form (\ref{eq:def_Fourier}), that is 
the function $S\to \tilde{F}^{(l)}(S+i)$ is the Fourier transform of the function $x\to e^x  F^{(l)}(x)$.
This procedure is summarized on Fig.\ref{fig:contour}, where the original (\ref{eq:a_decaler})
and the shifted (\ref{eq:shifted}) integration contours are plotted, and where the locations of the poles
of $\tilde{F}^{(l)}(z)$ are also indicated.
Eq.~(\ref{eq:convol}) then reduces to
\be
0 = \tilde{F}^{(l)}(S+i) \cos\nu  + \Lambda_l(iS)  \tilde{F}^{(l)}(S)
\label{eq:dif_finie}
\ee
for all real $S$. 

To solve (\ref{eq:dif_finie}) we follow \cite{Mora_CRAS}. We first introduce the ansatz
\be
\tilde{F}^{(l)}(S) = \frac{\pi}{\sinh[\pi(S+S_l)]} C_l(S)
\label{eq:introC}
\ee
where we recall that $iS_l$ is the positive-imaginary-part root of $\Lambda_l(s)$. The factor with the hyperbolic sine
is carefully chosen so as to give a minus sinus under translation $S\to S+i$, {\sl a priori} introducing poles in $z=-S_l$,
$z=-S_l\pm i$, $z=-S_l\pm 2 i$, etc,
for the function $\tilde{F}^{(l)}(z)$. This does not introduce singularities in the band $0<\mathrm{Im}\, z\leq 1$
provided that the function $C_l(z)$ tends to zero for $z\to -S_l+i$, a point to be checked {\sl a posteriori}.
The unknown function $C_l$ solves
\be
C_l(S+i)\cos\nu = \Lambda_l(iS) \, C_l(S)
\label{eq:eqC}
\ee
on the real axis. According to the expectation (\ref{eq:des_poles}), 
the function $C_l(S)$ has no pole in $S=-S_l$, since the $1/\sinh$ factor in (\ref{eq:introC}) already has a simple pole
in $S=-S_l$, and $C_l(S)$ has a simple pole in $S=S_l$, since the $1/\sinh$ factor has no pole there
\cite{Cp}. As in \cite{Mora_CRAS} one then uses the Weierstrass representation
\be
\Lambda_l(iS) = \cos\nu\, \prod_{n\in\mathbb{N}} \frac{S^2-u_n^2}{S^2-v_n^2}.
\label{eq:weier_Lambda}
\ee
where the overall factor $\cos\nu$ is the limit of $\Lambda_l(iS)$ for $S\to \infty$, see Eq.~(\ref{eq:lambda_expli}).
In Eq.~(\ref{eq:weier_Lambda}), $u_0=S_l$ is the real positive root of the function $S\to \Lambda_l(iS)$, $-u_0$ is the real
negative root, and the $\pm u_n$, $n\geq 1$, are the purely imaginary roots of $S\to \Lambda_l(iS)$, $\mathrm{Im}\, u_n>0$.
The $u_n$'s are sorted by ascending order of their imaginary parts, and for general values of the mass ratio $\alpha$,
they form an irregular, aperiodic sequence. They of course depend on the angular momentum $l$.
On the contrary, for general values of the mass ratio $\alpha$, the poles of $S\to \Lambda(iS)$ are found from (\ref{eq:rep_odd}) to
simply be $\pm v_n$, with $v_n=i(2n+l+1)$ for all integers $n\geq 0$ \cite{expli_poles}.
Finally, one can check as in \cite{Mora_CRAS} that the function $C_l$ is given by the infinite product
\begin{multline}
\label{eq:weier_Ft}
C_l(S) = \frac{\Gamma(iS_l-iS) \Gamma(1+iS-iv_0)}{\Gamma(1+iS+iS_l) \Gamma(-iS-iv_0)} \\
\times \prod_{n\in\mathbb{N}^*} \frac{\Gamma(-iS-iu_n)\Gamma(1+iS-i v_n)}{\Gamma(-iS-i v_n) \Gamma(1+iS-i u_n)},
\end{multline}
where we recall that the $u_n$ and $v_n$ depend on $l$.
In particular, one can check that this expression vanishes for $S\to -S_l+i$, as required
above Eq.~(\ref{eq:eqC}), and has no pole in the band $0< \mathrm{Im}\, z\leq 1$ 
of the complex plane \cite{pas_poles}.
Together with (\ref{eq:introC}), this constitutes the desired solution at zero energy for $R_*>0$.

An important application of this result is to calculate the previously mentioned low-$k$ or $x\to -\infty$ 
behavior (\ref{eq:attendu}) of the solution \cite{plus_inf},
which is a universal regime that has to match the zero-range model. 
Since $x<0$, in applying the usual contour integration technique to  (\ref{eq:def_Fourier}), we close the integration
contour following a half-circle in the lower part $\mbox{Im}\, z <0$ of the complex plane. According to the Cauchy residue formula,
one gets for $F^{(l)}(x)$ a sum of terms proportional to $e^{i z_n x}$, where the sum is taken
over all poles $z_n$ of the integrand in the lower half plane (see the pole locations in
Fig.\ref{fig:contour}). For $x\to -\infty$, the poles with a non-zero imaginary part have a contribution
that vanishes as $O(e^{x})$, and the sum is dominated
by the two poles $z=\pm S_l$ on the real axis, which are the only ones to give purely oscillating,
non-decaying contributions. The corresponding residues of $\tilde{F}^{(l)}(S)$ can be deduced from
\be
\tilde{F}^{(l)}(S) \underset{S\to -S_l}{\sim} \frac{C_l(-S_l)}{S+S_l} \ \ \mbox{and}\ \ \ 
\tilde{F}^{(l)}(S) \underset{S\to S_l}{\sim} \frac{[C_l(-S_l)]^*}{S-S_l}.
\label{eq:residues_Ft}
\ee
The value of the first residue directly results from the ansatz (\ref{eq:introC}) and the absence of pole of the function
$C_l(S)$ in $S=-S_l$. If one further uses (\ref{eq:eqC}) for $S\to S_l$, one finds for the second residue 
$[\pi/\sinh(2\pi S_l)]C_l(S_l+i)\cos\nu/i\Lambda'_l(iS_l)$. Properties of the explicit form (\ref{eq:weier_Ft})
and of the Gamma function, as in \cite{Mora_CRAS}, lead to (\ref{eq:residues_Ft}) \cite{longue_expli}.
Finally, turning back to the $k$ variable and to the function $D$:
\begin{multline}
D(\kk) \underset{k R_*\to 0^+}{=} -i \left(\frac{Q}{q R_* k}\right)^2 Y_l^0(\kk)  \\
\times \left\{[C_l(-S_l)]^* (qR_* k/Q)^{iS_l}+\mbox{c.c.} 
+ O(qR_*k/Q)\right\}
\label{eq:asympt_rs}
\end{multline}
where we recall that $S_l>0$ and $q/Q=\cos\nu$ as in (\ref{eq:defQ}).
The asymptotic form in the right-hand side of (\ref{eq:asympt_rs}) is satisfactory: 
It is indeed a superposition of solutions of the zero-range model at zero energy, see (\ref{eq:ansatz_puissance}).
A first important point is that it is actually a {\sl specific} linear combination of the solutions with exponents
$s=\pm i S_l$, with relative amplitudes depending on the Feshbach length $R_*$. This selection
of the right linear combination amounts to adjusting the three-body parameter to its right value
in the Danilov three-body contact conditions \cite{Danilov}. Eq.~(\ref{eq:asympt_rs}) thus constitutes
a microscopic derivation of this three-body parameter in the limit of an infinitely narrow
Feshbach resonance \cite{Mora_3corps,Mora_CRAS}.
A second important point is that the solution $D(\kk)$ starts approaching the log-periodic oscillatory asymptotic form
$(kR_*\cos\nu)^{\pm iS_l}$
as soon as $kR_*\cos \nu < 1$, there is no need to require that $kR_*\cos \nu < e^{-2\pi/S_l}$:
There is no need to require that the oscillatory form has performed at least one oscillation
to have $D(\kk)$ well approximated by it.

\section{Analytical results on trimer states}
\label{sec:arots}

In this central section of the paper, 
we develop a physical application of the particular analytical solutions of the previous section.
From the zero-energy and zero-range solution, we first determine, for each value of the angular momentum $l$,
the critical mass ratio $\alpha_c^{(l)}$ leading to the Efimov effect and the corresponding purely imaginary
Efimov exponent $s_l=iS_l$, with the convention $S_l>0$. Accurate asymptotic estimates for these quantities are obtained
and are compared to the Born-Oppenheimer approximation.
Second, from a matching of the $(E<0,R_*=0)$ solution to the $(E=0, R_*>0)$ solution,
we determine the Efimovian part of the spectrum, in particular the global energy scale $E_{\rm global}^{(l)}$
appearing in that geometric spectrum. The dependence of $E_{\rm global}^{(l)}$  on the mass ratio,
close to the Efimov threshold $\alpha\to \alpha_c^{(l)}$ and arbitrarily far from it ($\alpha\to +\infty$), is studied.
Also the variation of $E_{\rm global}^{(l)}$ with the angular momentum $l$, in particular
for $\alpha$ close to the critical mass ratio $\alpha_c^{(l)}$, is analyzed.
Third, using the Born-Oppenheimer approximation
expected to be asymptotically exact for a diverging mass ratio, we show that the hydrogenoid character
gradually takes over the Efimovian character in that limit, except in a vicinity of the $E=0$ accumulation
point (which remains Efimovian).

\subsection{Efimovian threshold $\alpha_c^{(l)}$ and exponent $s_l$}
\label{subsec:etae}

As a physical application of Eqs.~(\ref{eq:rep_odd},\ref{eq:lambda_expli},\ref{eq:trans_hyper}),
we determine the values of the mass ratio $\alpha$
and of the angular momentum $l$ such that the Efimov effect takes place, that is
the transcendental equation $\Lambda_l(s)=0$ admits some purely imaginary solutions.

A useful guide is the Born-Oppenheimer approximation \cite{Petrov_fermions}
that becomes exact in the limiting cases of vanishing or diverging mass ratio $\alpha$.
It indicates that the Efimov effect should take place for large enough values
of $\alpha$ and for odd values of $l$. In the limit $\alpha\to 0$,  the extra particle is indeed infinitely
massive, so that the fermions see a fixed point-like scatterer with infinite scattering length,
which does not support bound states. In the opposite limit $\alpha\to +\infty$,
the extra particle sees the very massive fermions as two fixed point-like scatterers of positions
$\rr_2$ and $\rr_3$, with which it forms a single bound state, of energy 
\be
\epsilon_0(r_{23})=-\frac{\hbar^2C^2}{2M r_{23}^2}
\label{eq:eps0}
\ee
in the zero-range Bethe-Peierls model,
and with a wavefunction that is symmetric under the exchange of $\rr_2$ and $\rr_3$.
Here the constant $C$ obeys
\be
C=\exp(-C) \ \ \mbox{so that}\ \ C=0.567\ 143\ 290\ 409 \ldots
\label{eq:C}
\ee
It can be related to the Lambert function by $C=W(1)$, and it is sometimes called the $\Omega$ constant.
Since the global state vector is fermionic, this extra particle wavefunction can be combined
in the Born-Oppenheimer factorized form with an odd orbital fermionic part only.
The effective potential seen by the fermions is thus the sum of the Born-Oppenheimer potential
$\epsilon_0(r_{23})$ and of the angular momentum centrifugal part $\hbar^2 l(l+1)/(mr_{23}^2)$. In 
three dimensions,  the zero-energy solution in that effective potential has to be written
as $r_{23}^{s-1/2}$ to match the usual definition of the Efimov exponent which is given
in two dimensions \cite{coherence}.
We then obtain
the Born-Oppenheimer approximation for the root $s$:
\be
s^2_{\rm BO} = \left(l+\frac{1}{2}\right)^2-\frac{1}{2} \alpha C^2 \ \ \ (l\ \mbox{odd})
\label{eq:BO}
\ee
to be used in the regime where $s^2_{\rm BO}\leq 0$. Interestingly, this Born-Oppenheimer approximation
can also be used for a non-zero $R_*$ to obtain exact results on the non-Efimovian low-energy trimers for $\alpha\to +\infty$,
as developed in subsection \ref{subsec:BO}.

\begin{figure}[htb]
\includegraphics[width=8cm,clip=]{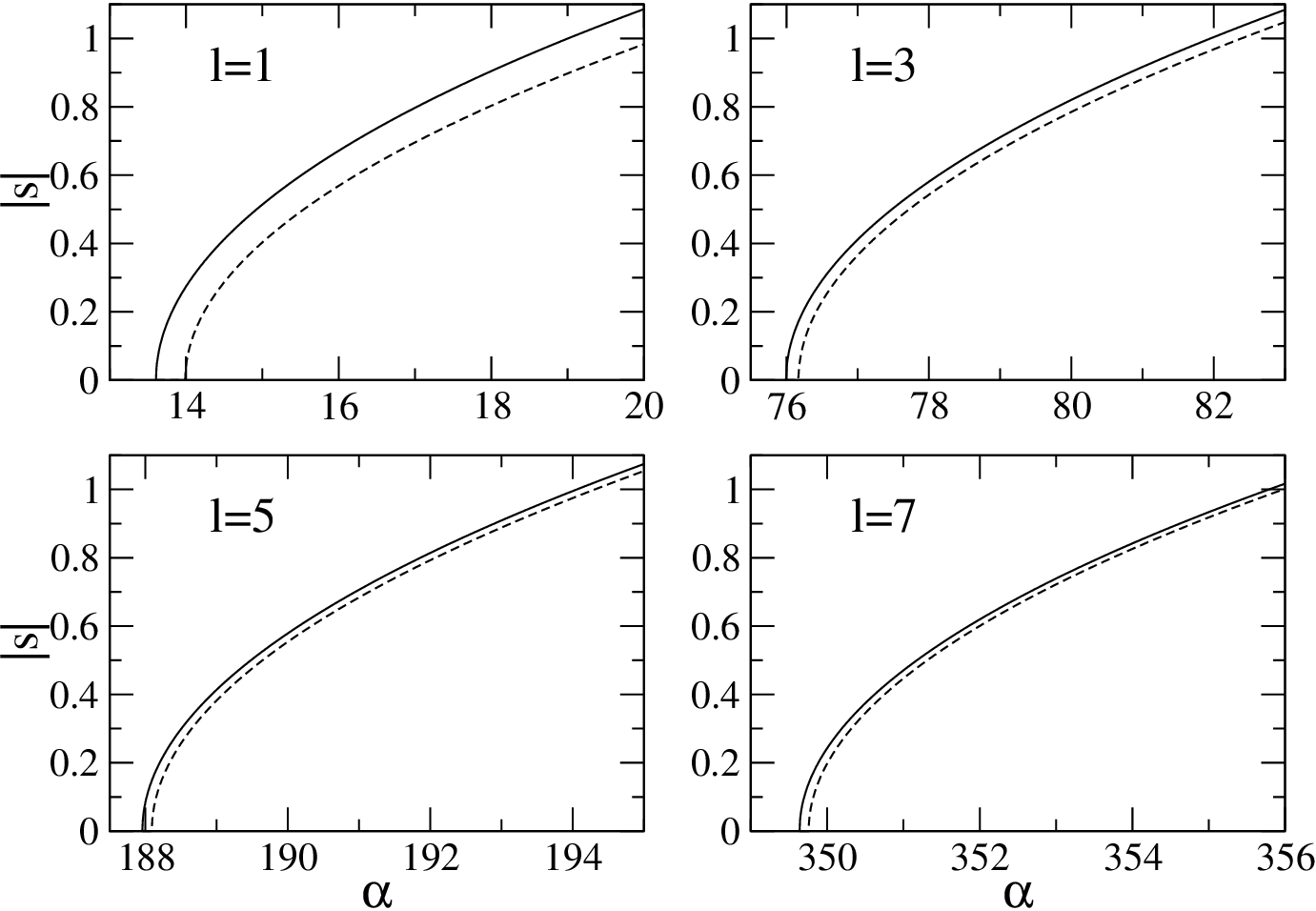}
\caption{Values of the Efimov exponent $s$ as a function of the fermion-to-extra-particle mass ratio $\alpha=m/M$,
for several values of the angular momentum $l$.
The modulus of $s$ is shown only over the interval of mass ratio
where $s$ is purely imaginary (that is where the Efimov effect takes place).
This was found to occur only for odd values of $l$, and for a single pair of $\pm s$ roots
of $\Lambda_l(s)=0$. Solid line: Numerical result from the exact function $\Lambda_l(s)$
as given by (\ref{eq:lambda_expli}). Dashed line: Born-Oppenheimer approximation
(\ref{eq:BO}).}
\label{fig:s}
\end{figure}

Turning back to the exact equation $\Lambda_l(s)=0$:
For a fixed value of $l$, we expect that the solutions $s$ of the transcendental equation
are continuous functions of the mass ratio $\alpha$. The critical values $\alpha_c^{(l)}$ of $\alpha$ 
for the emergence of the Efimov effect
are thus such that $\Lambda_l(0)=0$. From a numerical calculation of $\Lambda_l(s=0)$ as
a function of $\alpha$ ranging from $0$ to $+\infty$, we indeed find, for even $l$,
that $\Lambda_l(s=0)$ has a constant positive sign which means the absence of Efimov effect. 
For each odd $l$, we find that
$\Lambda_l(s=0)$ changes sign once (from positive to negative for increasing $\alpha$).
The resulting values of $\alpha_c^{(l)}$ are for example:
\bea
\alpha_c^{(l=1)}=13.60696\ldots, & \alpha_c^{(l=3)}=75.99449\ldots \nonumber \\
\alpha_c^{(l=5)}=187.9583\ldots, & \alpha_c^{(l=7)}=349.6384\ldots 
\eea
The critical mass ratio for $l=1$ was already given in \cite{Petrov_fermions}, and
in \cite{Kartavtsev} for $l=3$ and $l=5$.
For larger $l$ we have checked that $[\alpha_c^{(l)}C^2/2]^{1/2}$ is indeed very close to the approximation
$l+1/2$ resulting from (\ref{eq:BO}).
For these few odd values of $l$, we then solve numerically $\Lambda_l(iS)=0$, where $S>0$, to obtain
the Efimov exponent as a function of $\alpha$ for $\alpha>\alpha_c^{(l)}$. For each values
of $l$ and $\alpha>\alpha_c^{(l)}$, it is observed that $\Lambda_l(iS)$ for $S>0$
is an increasing function of $S$, so that a single pair of $\pm iS_l$ imaginary roots is obtained. 
The results are shown as solid lines in Fig.\ref{fig:s}.
For comparison, the Born-Oppenheimer approximation (\ref{eq:BO}) is plotted as dashed lines
in that figure. As expected intuitively, it approaches the exact result in the large $l$ (and thus
large $\alpha$) limit. 

One can however be more precise in the evaluation of the accuracy of Eq.~(\ref{eq:BO}).
For a {\sl fixed} value $s_l=i S_l$ of the Efimov exponent, one can perform a large $l$ expansion
of the mass ratio, as shown in Appendix \ref{app:qref_grandl}, to obtain:
\be
\frac{1}{2} C^2 (\alpha-\alpha_c^{(l)}) \underset{l\to \infty}{=} S_l^2[1 + O(1/l)]
\label{eq:justi_bo}
\ee
where the critical mass ratio $\alpha_c^{(l)}$ has the large $l$ expansion
\be
\label{eq:alphac_grandl}
\frac{1}{2} C^2 \alpha_c^{(l)} \underset{l\to \infty}{=}\left(l+\frac{1}{2}\right)^2 + \Delta +O(1/l).
\ee
$\Delta$ is thus the first correction to the critical mass ratio predicted by the Born-Oppenheimer
approximation (\ref{eq:BO}).  Accidentally it has a very small numerical value \cite{aide_epsilon}:
\be
\Delta = \frac{17-C^2}{12} -\frac{7}{6} \left(C+\frac{1}{C+1}\right)=-0.016\ 259\ 165\ldots
\label{eq:defDelta}
\ee
so that the Born-Oppenheimer approximation for $\alpha_c^{(l)}$ is in practice quite good
for large $l$.
For $\alpha\to +\infty$ for a {\sl fixed} angular momentum $l$, one can also obtain from the results
of Appendix \ref{app:qrefasympt} the following asymptotic expansion for the imaginary root $iS_l$ of $\Lambda_l(s)$:
\be
S_l^2 \underset{\alpha\to +\infty}{=} \frac{C^2}{2} \alpha -\left(l+\frac{1}{2}\right)^2 + \delta + O(1/\alpha)
\label{eq:Slasympt}
\ee
where the first two terms in the right-hand side constitute the Born-Oppenheimer approximation
and the angular-momentum independent constant $\delta$ is different from zero:
\be
\delta=\frac{5}{12} C^2 +\frac{4}{3} \frac{C^2}{1+C} + \frac{1}{4} =0.657\ 684\ldots
\label{eq:def_petit_delta}
\ee
Note that Eq.~(\ref{eq:Slasympt}) is not in contradiction with Eq.~(\ref{eq:justi_bo}) 
[combined with Eq.~(\ref{eq:alphac_grandl})] because the considered
limits are totally different, $S_l/l$ either diverges or vanishes.
As $s^2$ is a coefficient in the Born-Oppenheimer potential, a suggestive picture obtained from Eq.~(\ref{eq:Slasympt})
is that the Born-Oppenheimer approximation has an error $O(\alpha^0)$ on the effective potential seen by the fermions.
Mathematically, Eq.~(\ref{eq:Slasympt}) shows that the error on $S_l$ due to the Born-Oppenheimer approximation vanishes as $1/\alpha^{1/2}$,
and that the approximation gets the correct leading term $\alpha^{1/2}$ and the correct subleading term
$\alpha^0$ (which turns out to vanish) in an asymptotic expansion of $S_l$ in powers of $\alpha^{1/2}$.
These remarks are useful for subsection \ref{subsec:BO}.

\subsection{Efimovian part of the trimer spectrum}
\label{subsec:aftts}

In presence of the Efimov effect, we have obtained so far two solutions to the integral equation (\ref{eq:intl}) in limiting cases,
see subsection \ref{subsec:zrne} for $R_*=0, E<0$, and subsection \ref{subsec:rsze} for $R_*>0, E=0$.
How can we then obtain an approximation for the corresponding spectrum of trimers, which requires to
have both $R_*>0$ and $E<0$?

\begin{figure}[tb]
\includegraphics[width=8cm,clip=]{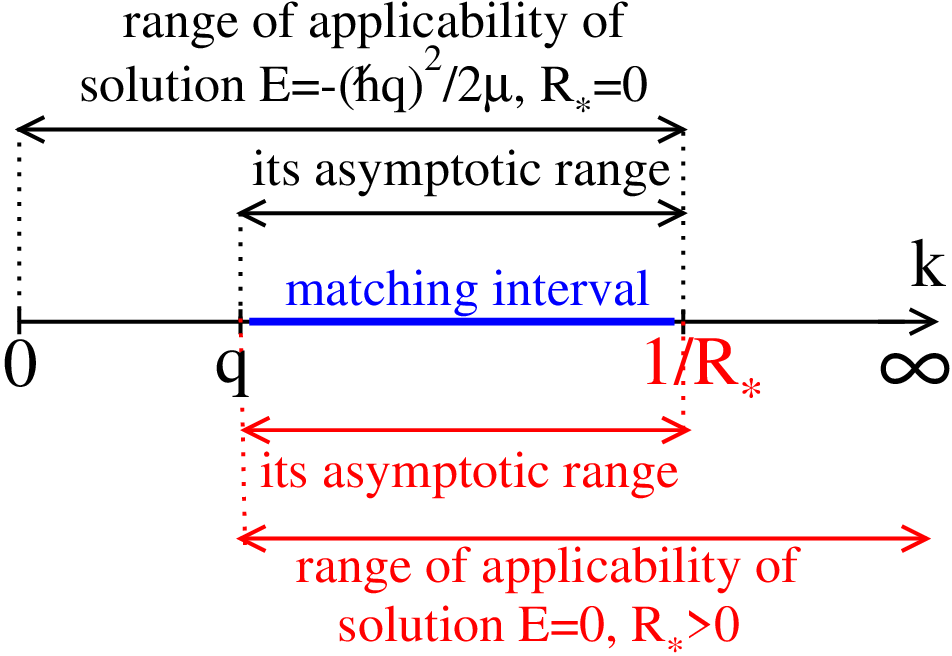}
\caption{(color online) Matching of two limiting solutions of the integral equation (\ref{eq:intl}), the solution 
($E=-\hbar^2 q^2/(2\mu)<0, R_*=0$) [see text in black, in the upper half of the figure] given by Eq.~(\ref{eq:solun}) and the solution
($E=0, R_*>0$) [see text in red, in the lower half of the figure] 
deducible from Eqs.~(\ref{eq:Dansatz},\ref{eq:ansatz_fl},\ref{eq:def_Fourier},\ref{eq:introC},\ref{eq:weier_Ft}), 
over a common interval of values of $k$ [the ``matching interval", blue segment on $k$ axis] where they are both in their
asymptotic regimes Eq.~(\ref{eq:zr_grandk}) and Eq.~(\ref{eq:asympt_rs}). This matching procedure
leads to the Efimovian spectrum formula (\ref{eq:spectrum_analy}) with a global scale
given by (\ref{eq:global}). This procedure makes sense when $q R_*\ll 1$, and it is expected to be exact
when $q R_*\to 0$, that is for the quantum number $n$  tending to infinity for a fixed (purely imaginary)
Efimov exponent $s_l$
(as in Efimov's historical solution) or (less usually) for $|s_l|$ tending to zero at fixed quantum number
$n \geq 1$. For simplicity of the figure, we have dropped factors slowly
depending on the mass ratio $\alpha$, such as $\cos \nu$, that is
we have assumed that the angular momentum $l$ and $|s_l|$ 
are not much larger than unity.
\label{fig:match}}
\end{figure}

Roughly speaking, dropping for simplicity a possible dependence on the mass ratio $\alpha$ of the various
bounds, the solution of subsection \ref{subsec:zrne} is expected to constitute an accurate approximation of the trimer solution
at interparticle distances much larger than $R_*$, that is at momenta smaller than $\hbar/R_*$.
In a symmetric manner, the solution of subsection \ref{subsec:rsze} is expected to well approximate the trimer solution 
at short enough interparticle distances, smaller than $1/q$, where the $E<0$ wavefunction only weakly departs from
the zero-energy one. This corresponds to momenta much larger than $\hbar q$.
There thus exists an interval of momentum $\hbar k$ over which both limiting solutions are close to the physical trimer solution,
$q \ll k \ll 1/R_*$,  if
\be
q R_* \ll 1.
\label{eq:contrainte}
\ee
In this case, we can match the two limiting solutions,
as summarized in Fig.~\ref{fig:match}. Over this matching interval of momentum, one has $k\gg q$ so that
the solution of subsection \ref{subsec:zrne} is in its large $k$ regime given by (\ref{eq:zr_grandk}) with $s=iS_l$.
Over this interval, one also has $k \ll 1/R_*$, 
so that the solution of subsection \ref{subsec:rsze} is in its low-$k$ regime given by (\ref{eq:asympt_rs}).
These two limiting regimes are compatible (within an arbitrary normalization factor) if $q$ is of the form
\cite{aide}
\be
q_n^{(l)} = q_{\rm global}^{(l)} e^{-\pi n/|s_l|}, \ \ \ n \ \mbox{integer},
\label{eq:spectrum_analy}
\ee
where we recall that the Efimov exponent is $s_l=iS_l$, $S_l>0$ and the trimer energy is
$E=-\hbar^2 q^2/(2\mu)$, with $\mu=mM/(m+M)$.  Eq.~(\ref{eq:spectrum_analy}) is the expected
geometric Efimovian spectrum, and we have the explicit expressions for 
the corresponding global wavenumber and energy scales:
\be
q_{\rm global}^{(l)} = \frac{2}{R_*}\, e^{\theta_l/|s_l|}
\ \mbox{and}\ E_{\rm global}^{(l)}=-\frac{2\hbar^2}{\mu R_*^2}\,
e^{2\theta_l/|s_l|}
\label{eq:global}
\ee
where $\theta_l$ is the phase of the complex number
\be
Z_l = |Z_l| e^{i\theta_l} \equiv \left[\prod_{n=1}^{l} (n-s_l)\right] s_l C_l(-S_l)
\label{eq:Z_l}
\ee
and the function $C_l(S)$ is given by Eq.~(\ref{eq:weier_Ft}).
At fixed angular momentum,
the phase $\theta_l$ depends on the mass
ratio $\alpha=m/M$. For $\alpha$ tending from above to the critical value $\alpha_c^{(l)}$, $S_l\to 0$, we show
in the Appendix \ref{app:qrefc} 
that $Z_l$ tends to a real and positive number. We thus choose the usual
determination $\theta_l=\mathrm{Arg}\, Z_l$ in that limit, and extend it by continuity to all larger
values of $\alpha$.

\begin{figure*}[htb]
\includegraphics[width=5.9cm,clip=]{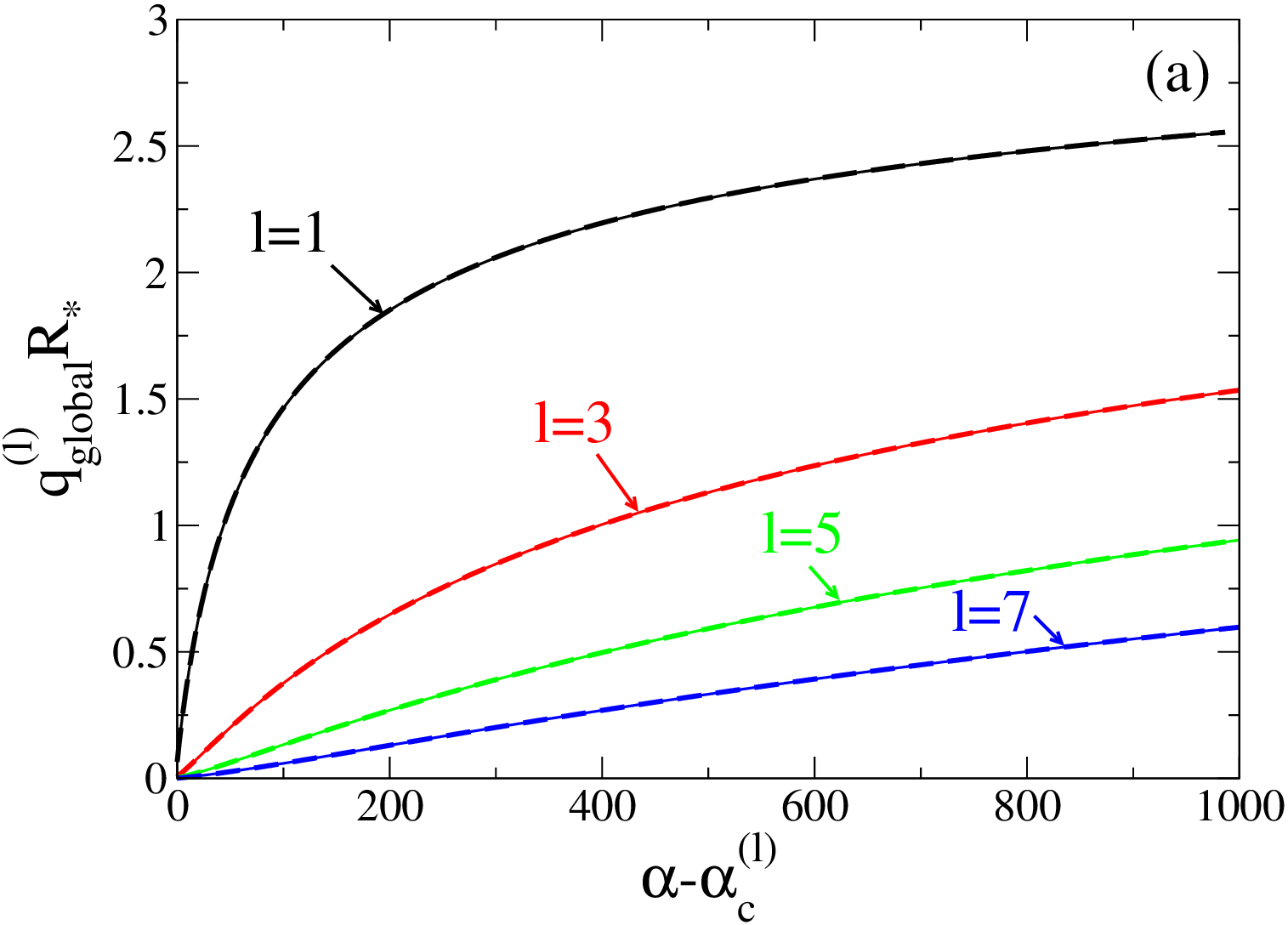}
\includegraphics[width=5.9cm,clip=]{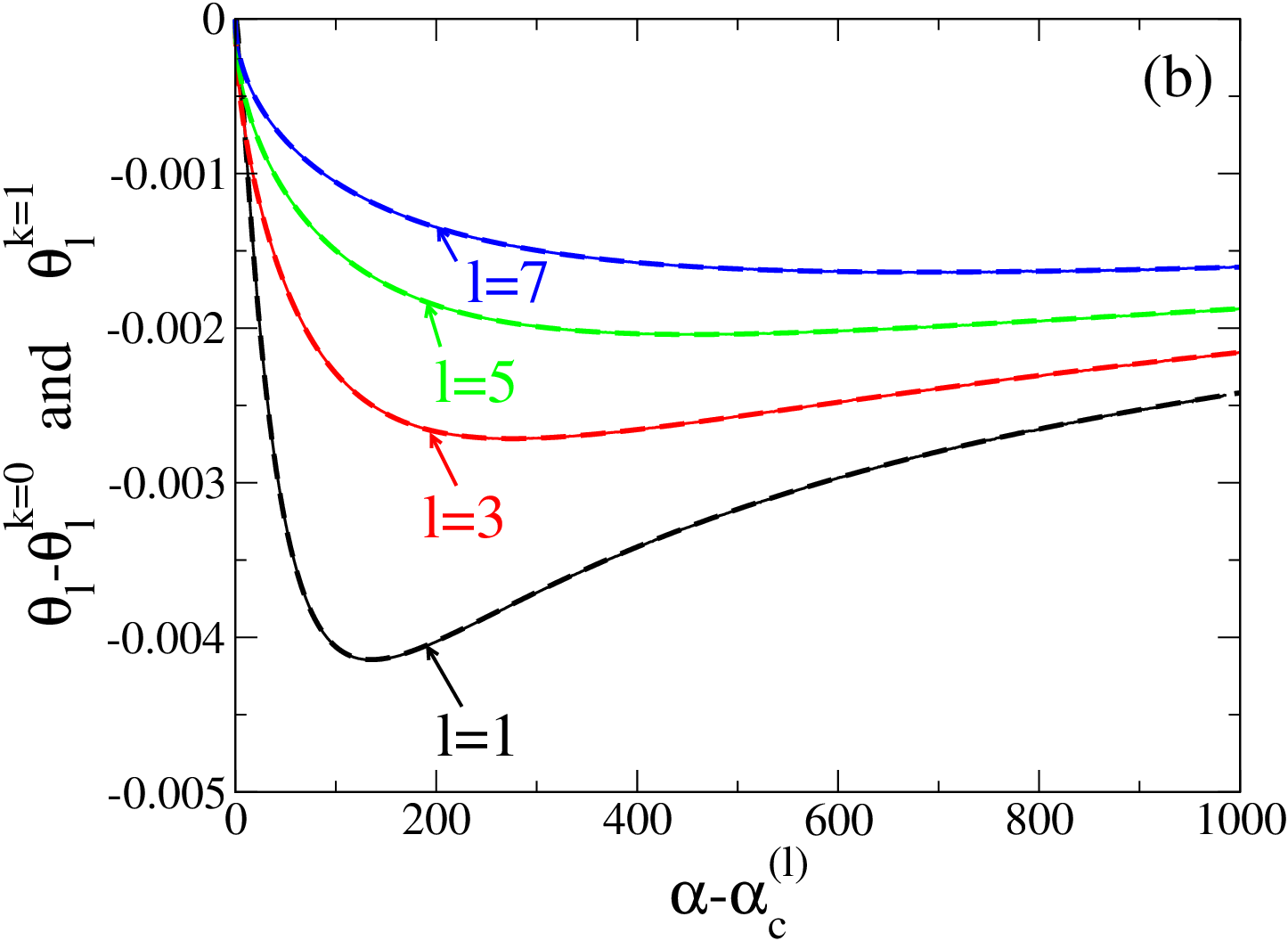}
\includegraphics[width=5.9cm,clip=]{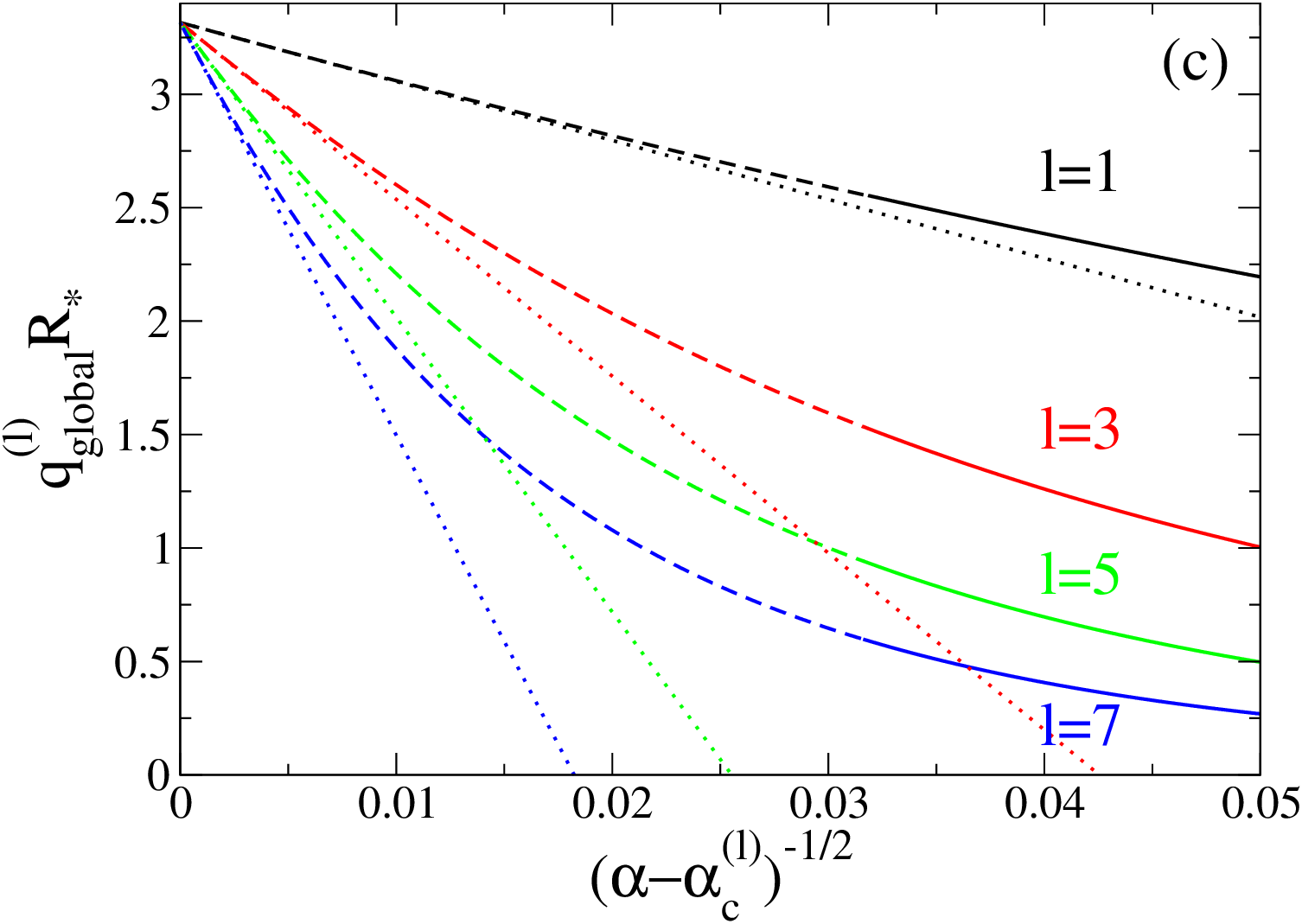}
\caption{(color online) In (a) value of the global wavenumber scale $q_{\rm global}^{(l)}$ of the Efimovian part of the spectrum,
see (\ref{eq:global}), as a function of the difference of the mass ratio $\alpha$ from its critical
value, for all odd angular momenta up to $l=7$. Thin solid lines: Exact result obtained 
from (\ref{eq:Z_l}) and from the infinite product representation (\ref{eq:weier_Ft}) [the more rapidly converging
form (\ref{eq:dls},\ref{eq:int_ims}) was used in the numerics]. Thick dashed lines: Zeroth-order approximation
omitting all the terms in the sum over $k$ in (\ref{eq:alter_thetal}). 
In (b)  the (small) difference between the exact $\theta_l$ and the zeroth-order approximation $\theta_l^{k=0}$ is
plotted as thin solid lines, and the value $\theta_l^{k=1}$ of the $k=1$ term of (\ref{eq:alter_thetal}) is plotted as thick dashed lines.
In (c) the values of $q_{\rm global}^{(l)}$ resulting from the Born-Oppenheimer plus semi-classical approximation
are given in dashed lines [see Eq.~(\ref{eq:bowkbspec})], as functions of $1/(\alpha-\alpha_c^{(l)})^{1/2}$,
showing how the $l$-independent $\alpha\to+\infty$ limit
is reached. The solid lines correspond to the exact numerical data of (a) and the dotted straight lines
corresponds to the asymptotic expansion Eq.~(\ref{eq:refesti}). 
\label{fig:qga}
}
\end{figure*}

For the analytical developments that follow, obtained for the global scale $q_{\rm global}^{(l)}$
in the limit of large angular momenta for $\alpha-\alpha_c^{(l)}$ fixed, or in the limit of a large mass
ratio at fixed angular momentum, there is an actually more operational expression for
$\theta_l$, that does not require the determination of all the roots and poles of
$\Lambda_l(iS)$ to evaluate $C_l(-S_l)$. As shown in Appendix \ref{app:alter} one has the series representation
of the phase $\theta_l$ of $Z_l$:
\begin{multline}
\label{eq:alter_thetal}
\theta_l = \mbox{Im}[\ln\Gamma(1+iS_l)+\ln \Gamma(1+2iS_l) +2 \ln \Gamma(l+1-iS_l)
\\  +\ln\Gamma(l+2-iS_l)]
+\int_0^{S_l}\!\! dS \ln\left[\frac{\Lambda_l(iS)}{\cos\nu}\,\frac{S^2+(l+1)^2}{S^2-S_l^2}\right] \\
+ \sum_{k\geq 1} \frac{(-1)^k B_{2k}}{(2k)!} 
\frac{d^{2k-1}}{dS^{2k-1}}
\left\{\ln\left[\frac{\Lambda_l(iS)}{\cos\nu}\,\frac{S^2+(l+1)^2}{S^2-S_l^2}\right]\right\}_{S=S_l}
\end{multline}
where we recall that $S_l$ is the positive root of the function $\Lambda_l(iS)$ and the
$B_k$ are Bernoulli's numbers, $B_1=-1/2,B_2=1/6,\ldots$.
Since (\ref{eq:alter_thetal}) was obtained from Stirling's series, which is an {\sl asymptotic}
series, we expect that (\ref{eq:alter_thetal}) is also an asymptotic series.
We thus investigated numerically how many terms one has to keep in practice to have a good
accuracy. In Fig.\ref{fig:qga}a, it is found remarkably that the zeroth-order approximation,
consisting in omitting {\sl all} the terms in the sum over $k$ in (\ref{eq:alter_thetal}),
already gives in practice a sufficiently accurate approximation for the global scale $q_{\rm global}^{(l)}$.
In Fig.\ref{fig:qga}b, it is shown that the difference between the exact value of $\theta_l$, obtained from
the infinite product representation of $C_l(-S_l)$, and the zeroth-order approximation,  omitting all $k$ terms
in (\ref{eq:alter_thetal}), is non-zero but is very accurately accounted for
by the term $k=1$ in (\ref{eq:alter_thetal}).

From the zero-range theory by Efimov, it is expected that the geometric form Eq.~(\ref{eq:spectrum_analy}) of the spectrum
is asymptotically exact in the
limit of a large quantum number, $n\to +\infty$, see the constraint (\ref{eq:contrainte}). How large the values of $n$ should
be to reach this geometric behavior will be evaluated numerically in section \ref{sec:numerics} 
[in the mean time see the discussion that follows Eq.~(\ref{eq:exqglob})].
The advantage of the non-zero $R_*$ calculation is that it also gives the global energy scale in the spectrum, 
see the factor $q_{\rm global}^{(l)}$ in (\ref{eq:spectrum_analy}) explicitly given by (\ref{eq:global}).
As compared to the bosonic case
\cite{Mora_3corps,Mora_CRAS}, there is an additional knob here, which is the mass ratio $\alpha$.
How does the global energy scale vary with the mass ratio~?

\noindent {\sl For $\alpha\to \alpha_c^{(l)}$:} The behavior of the global energy scale close to the critical mass ratio
can be determined from the results of Appendix \ref{app:qrefc}, where it is shown that 
$\theta_l$ vanishes linearly with $S_l$:
\begin{multline}
\lim_{\alpha\to\alpha_c^{(l)}} \frac{\theta_l}{|s_l|} = -\psi(l+1)+3\psi(1)-\psi(-iv_0)-\psi(1-i v_0)
\\ +\sum_{n=1}^{+\infty} [\psi(-i u_n^c) + \psi(1-i u_n^c)-\psi(-i v_n)-\psi(1-i v_n)].
\label{eq:glob_lim}
\end{multline}
Here $u_n^c, n>0,$ is the value of the complex root $u_n$ of the function $S\to \Lambda_l(iS)$ at the critical mass
ratio, the $v_n=i(2n+l+1)$, $n\geq 0$, are the poles of $S\to \Lambda_l(iS)$, and the function $\psi(z)$ is the digamma function, that is
the logarithmic derivative of the $\Gamma$ function. As a consequence, the global energy scale has
a finite limit at the threshold for the Efimov effect! We give here a few  corresponding values 
obtained from the rapidly converging formula (\ref{eq:convergente}):
\bea
q_{\rm global}^{(l=1)}R_* \simeq 6.56577\cdot 10^{-2}, & 
q_{\rm global}^{(l=3)}R_* \simeq 6.12349\cdot 10^{-3} \nonumber \\
q_{\rm global}^{(l=5)}R_* \simeq 1.62809\cdot 10^{-3}, & 
q_{\rm global}^{(l=7)}R_* \simeq 6.48952\cdot 10^{-4}  \nonumber \\
& \label{eq:exqglob}
\eea
At this threshold, $|s_l|\to 0$ so we expect that Eq.~(\ref{eq:spectrum_analy}) becomes actually
exact for the quantum number $n\geq 1$, since $q_n^{(l)} R_*$ tends to zero in that limit, whereas 
(\ref{eq:spectrum_analy}) is clearly invalid for $n\leq -1$.
The case $n=0$ is dubious: Although $q_0^{(l)} R_*$ does not tend to zero
at the threshold, it assumes very small values, 
so may be Eq.~(\ref{eq:spectrum_analy}) still makes sense.
We can not however say more at this stage, 
and the question whether the quantum number $n=1$ 
corresponds or not to the ground trimer state (for a given
angular momentum $l$) will be answered in section \ref{sec:numerics}.

\begin{figure}[tb]
\includegraphics[width=8cm,clip=]{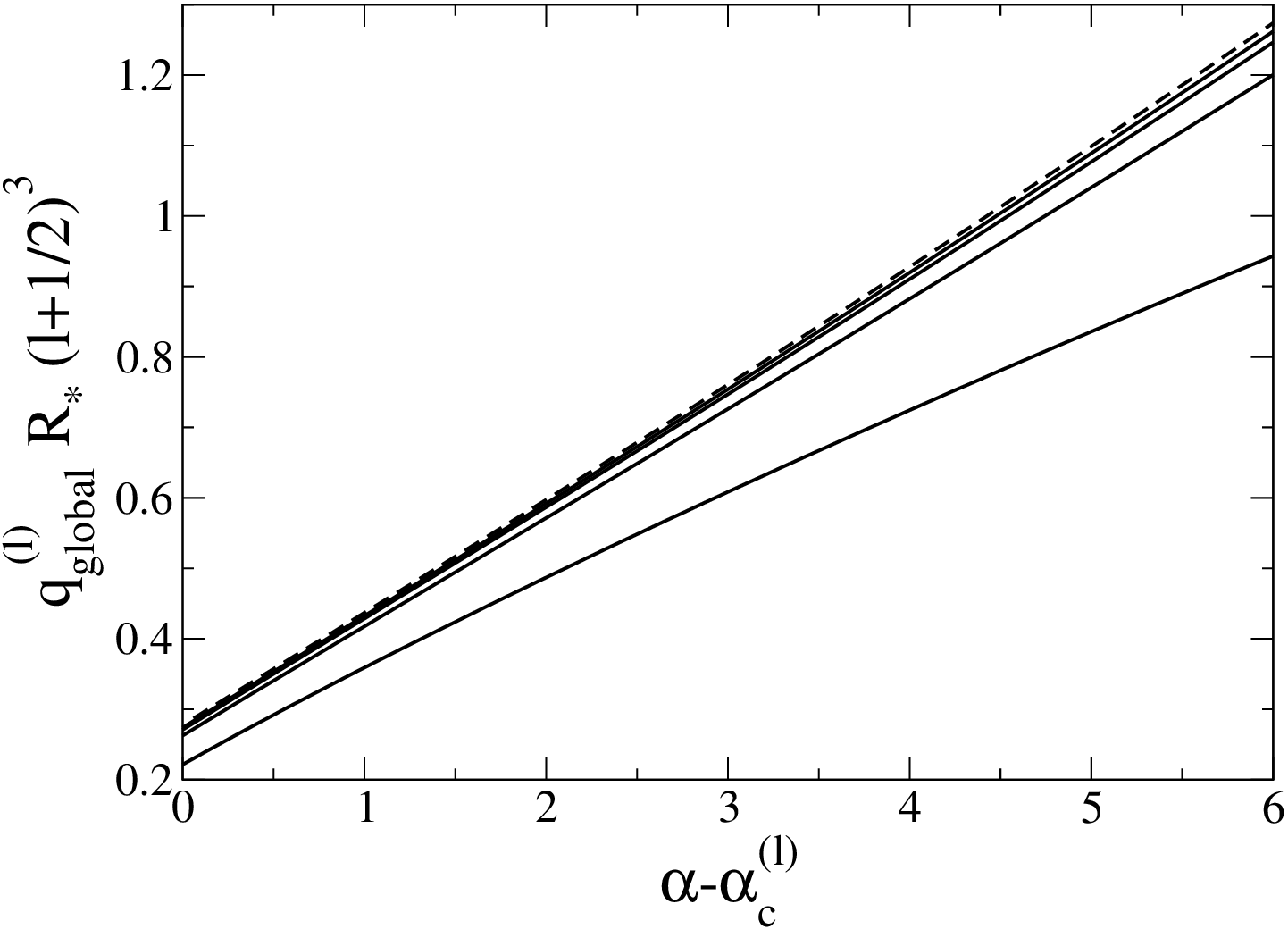}
\caption{Dependence, with the angular momentum $l$, of the global wavenumber scale $q_{\rm global}^{(l)}$ in the Efimov trimer spectrum,
as obtained from the exact result (\ref{eq:global}).
The figure shows the quantity $q_{\rm global}^{(l)}R_*$ multiplied by $(l+1/2)^3$, as a function
of $\alpha-\alpha_c^{(l)}$, for (odd) angular momenta equal to $l=1$, $l=3$, $l=5$ and $l=7$ from bottom
to top. This quantity
is observed to converge rapidly to a finite limit for increasing $l$. The dashed line represents
the analytical prediction (\ref{eq:grandl}) for that limit.
We recall that $E_{\rm global}=-\hbar^2 q_{\rm global}^2/(2\mu)$.}
\label{fig:qglob}
\end{figure}

\noindent {\sl For $l\to +\infty$:} 
Another interesting question is to determine how $q_{\rm global}^{(l)}$ depends on the angular momentum $l$
at a fixed distance of the mass ratio $\alpha$ from the critical value $\alpha_c^{(l)}$, that is roughly
at constant values of the Efimov exponent $s_l$.  
After a numerical evaluation of (\ref{eq:global}), as detailed in the note \cite{expli_poles},
we found that $q_{\rm global}^{(l)}$ drops rapidly for increasing $l$, roughly as $1/(\alpha_c^{(l)})^{3/2}$.
According to (\ref{eq:BO}) the critical mass ratio scales approximately as $(l+1/2)^2$, so that we expect
that $q_{\rm global}^{(l)}$ approximately scales as $1/(l+1/2)^3$, an approximation that becomes
rapidly excellent with increasing $l$ as soon as $l$ exceeds unity, see Fig.\ref{fig:qglob}.

The scaling of $q_{\rm global}^{(l)}$ as $1/l^3$ at a fixed distance from the critical mass ratio 
can be obtained analytically
for $l\to +\infty$ using the expression (\ref{eq:alter_thetal}) for the angle $\theta_l$ and
the expression Eq.~(\ref{eq:trans_hyper}) for $\Lambda_l(iS)$, as detailed in the Appendix
\ref{app:qref_grandl}:
\begin{multline}
\label{eq:grandl}
q_{\rm global}^{(l)} R_* \stackrel{\alpha-\alpha_c^{(l)}\ \mathrm{fixed}}{\underset{l\to +\infty}{\sim}}
\ \frac{(1+C)\, e^{3\gamma}}{l^3}\\
\times \exp\left\{\frac{\mathrm{Im}\, [\ln \Gamma(1+iS_l) + \ln\Gamma(1+2iS_l)]}{S_l}\right\},
\end{multline}
where $\gamma=0.577\ 215\ 664\ 9\ldots$ is Euler's constant and $s_l=iS_l$, $S_l>0$. 
In that limit, one can use the Born-Oppenheimer-type relation, $S_l^2= (\alpha-\alpha_c^{(l)})C^2/2$,
as shown by Eq.~(\ref{eq:justi_bo}).
The asymptotic result (\ref{eq:grandl}) is plotted as a dashed line in Fig.\ref{fig:qglob} and well reproduces the large-$l$
numerical results.

\noindent{\sl For $\alpha\to +\infty$:} The behavior of the global energy scale in the limit of an infinite
mass ratio (for a given $l$) is determined in the Appendix \ref{app:qrefasympt}. It is found that $q_{\rm global}^{(l)}$
has a finite limit, which remarkably is also independent of the angular momentum $l$:
\be
q_{\rm global}^{(l)}R_* \underset{\alpha\to +\infty}{\to}  2(1+C)\, e^J = 3.31582 \ldots
\label{eq:qglob_loin}
\ee
where $C$ is defined by (\ref{eq:C}) and $J$ is the integral
\be
J = \int_0^C dx\, \left[\frac{1}{C}\, \frac{1+x}{1-xe^x} - \frac{1}{C-x}\right]= 0.05630577\ldots
\label{eq:defJ}
\ee
As we shall see in subsection \ref{subsec:BO}, for $\alpha\to+\infty$, the low-lying part of the trimer spectrum 
is hydrogenoid rather than Efimovian, so that (\ref{eq:qglob_loin}) is relevant only for
trimers with diverging quantum number $n$.

\subsection{Born-Oppenheimer approximation and hydrogenoid trimer spectrum}
\label{subsec:BO}

As explained in subsection \ref{subsec:etae} the Born-Oppenheimer approach is a natural tool 
when the fermions become arbitrarily massive. The extra particle then mediates an attractive interaction
potential $\epsilon(r_{23})$ between the heavy fermions, with $\rr_{23}=\rr_2-\rr_3$ the relative coordinates
of the fermions. We calculate this interaction potential on a narrow Feshbach resonance.
Then from Schr\"odinger's equation for the relative wavefunction of the two fermions,
\be
E \psi(\rr_{23}) = -\frac{\hbar^2}{m} \Delta_{\rr_{23}} \psi(\rr_{23}) + \epsilon(r_{23}) \psi(\rr_{23})
\label{eq:bos}
\ee
which is in the odd $l$ sector due to fermionic antisymmetry,
we determine the low-energy trimer states in the limit $\alpha\to+\infty$.

We first fix the positions of the fermions to $\rr_2$ and $\rr_3$. Each fermion then acts on the extra particle
as a fixed scatterer of infinite scattering length, zero true range but finite effective range $r_e=-2 R_*$.
The effect of such a scatterer is then represented by modified contact conditions on the wavefunction
$\phi(\rr_1)$ of the extra particle in the so-called effective range approach \cite{Petrov_rstar,effec_range_app}.
For a bound state $\phi(\rr_1)$ of eigenenergy 
\be
\epsilon(r_{23})=  -\frac{\hbar^2 \kappa^2}{2M},
\ee
where the dependence of $\kappa>0$ on $r_{23}$ is for simplicity omitted in the writing, we impose the boundary conditions
\be
\phi(\rr_1) \underset{\rr_1\to\rr_n}{=} A_n \left[\frac{1}{|\rr_1-\rr_n|}-\frac{1}{a_{\rm eff}}\right] + O(|\rr_1-\rr_n|)
\label{eq:contact_modif}
\ee
in the vicinity of each scatterer $n=2,3$. The effective scattering length is energy dependent, as
$1/a_{\rm eff}=1/a-k^2 r_e/2$ for an incoming free wave of wavenumber $k$. 
Here the true scattering length $a$ is infinite, the effective range is $r_e=-2 R_*$
for a narrow Feshbach resonance, and $k=i\kappa$ is purely imaginary for a bound state, so that
$1/a_{\rm eff}=-\kappa^2 R_*$. In presence of the contact conditions, the extra-particle wavefunction
obeys Schr\"odinger's equation
\be
\epsilon(r_{23}) \phi(\rr_1) = -\frac{\hbar^2}{2M} \left[\Delta_{\rr_1} \phi(\rr_1) +\sum_{n=2}^{3} 4\pi A_n \delta(\rr_1-\rr_n)\right]
\ee
where the Dirac terms are due to the $1/|\rr_1-\rr_n|$ divergences. The general solution is expressed in terms of the
Green's function of the Laplacian at negative energy:
\be
\phi(\rr_1) = \sum_{n=2}^{3} A_n \frac{e^{-\kappa |\rr_1-\rr_n|}}{|\rr_1-\rr_n|}.
\label{eq:gen_sol}
\ee
The contact conditions (\ref{eq:contact_modif}) then impose $(1+\kappa R_*)A_{2,3} = A_{3,2} e^{-\kappa r_{23}}/(\kappa r_{23})$.
This $2\times 2$ system has a non-zero solution only for the {\sl symmetric} case $A_2=A_3$, where $\kappa$ solves
\be
1+ \kappa R_* = \frac{e^{-\kappa r_{23}}}{\kappa r_{23}}.
\label{eq:trans}
\ee
The wavefunction $\phi(\rr_1)$ is then symmetric under the exchange of $\rr_2$ and $\rr_3$.
As the Born-Oppenheimer ansatz for the total wavefunction is $\psi(\rr_{23})\phi(\rr_1;\rr_2,\rr_3)$,
where the parametric dependence of $\phi$ with the fermions positions is made explicit,
fermionic exchange symmetry indeed imposes $\psi(-\rr_{23})=-\psi(\rr_{23})$.

For a fixed $r_{23}$, Eq.~(\ref{eq:trans}) looks difficult to solve.
However, one can see that, for each positive $\kappa$,
it is solved by a single positive $r_{23}$
\cite{footnote_monotone}.
Furthermore, rewriting $\kappa$ in the left hand side as $(\kappa r_{23})/r_{23}$, $r_{23}/R_*$
may be expressed as an explicit function of $u=\kappa r_{23}$, where $u$ ranges from $0$
to the numerical constant $C$ defined in (\ref{eq:C}):
\be
\frac{r_{23}}{R_*} = \frac{u^2}{e^{-u}-u}.
\ee
This allows a straightforward plot and study of the Born-Oppenheimer potential $\epsilon(r_{23})$.
We reach in particular the useful limiting cases
\bea
\label{eq:vbo_loin}
\epsilon(r_{23})\!\!\! &\underset{r_{23}\to +\infty}{\sim}& \!\!\! -\frac{\hbar^2 C^2}{2M r_{23}^2} \\
\epsilon(r_{23})\!\!\! &\underset{r_{23}\to 0}{=}&\!\!\!   -\frac{\hbar^2}{2MR_*^2}
\left[\frac{R_*}{r_{23}} - \frac{2R_*^{1/2}}{r_{23}^{1/2}}
+ O(1)\right]\!.
\label{eq:vbo_pres}
\eea
The asymptotic behavior (\ref{eq:vbo_loin}) reproduces the Born-Oppenheimer potential (\ref{eq:eps0})
obtained for the usual Bethe-Peierls case $R_*=0$. For a large enough $\alpha$, and for each odd
values of the angular momentum $l$, it ensures that the spectrum of Eq.~(\ref{eq:bos}) is indeed Efimovian in the limit
of large quantum number $n$, that is for $E\to 0^-$. On the contrary, for fixed quantum numbers $n$ and $l$,
the bound states of (\ref{eq:bos}) for increasing $\alpha=m/M$ are increasingly localized in the low $r_{23}$
part Eq.~(\ref{eq:vbo_pres}) of the Born-Oppenheimer potential. This means that the low-energy part of the spectrum
is hydrogenoid, it becomes asymptotically equivalent for large $\alpha$ to the known spectrum of the hydrogen
atom if one takes for the electron mass $m_e=m/2$ and the CGS electron charge $e$ such that $e^2=\hbar^2/(2MR_*)$.
In terms of the wavenumber $q$ introduced in Eq.~(\ref{eq:defq}) we thus obtain the exact asymptotic result \cite{proviso}
\be
q_{n}^{(l)}R_* \underset{\alpha\to+\infty}{\sim} \left(\frac{\alpha}{8}\right)^{1/2} \frac{1}{n+l}.
\label{eq:hydro_asympt}
\ee
Here the integer quantum number $n$ start from $n=1$ in each odd angular momentum sector $l$, so that the
hydrogen spectrum reads $-m_e e^4/[2\hbar^2(n+l)^2]$ with this convention.

Since the eigenfunctions of the hydrogen atom are well known, it is possible to calculate the first correction 
to the hydrogenoid spectrum, treating the $1/r_{23}^{1/2}$ term in Eq.~(\ref{eq:vbo_pres}) to first order
in perturbation theory. The calculations are given in the Appendix \ref{app:hydrogen}, we present
here only the result:
\be
q_{n}^{(l)}R_* \!\!\underset{\alpha\to+\infty}{=}\!\! \frac{(\alpha/8)^{1/2}}{n+l}
\left[1 -
R_n^{(l)} 
\left(\frac{\pi(n+l)}{2\alpha}\right)^{1/2}\!\!\! + O\left(\frac{1}{\alpha}\right)\right],
\label{eq:hydro_cor}
\ee
where $R_n^{(l)}$ is a rational number given by
\begin{multline}
R_n^{(l)} = \frac{(n-1)!}{(n+2l)!}
\sum_{k=0}^{n-1}
\left[\frac{(2k)!}{(k!)^2}\right]^2 \frac{4^{-(n+2l+k)}}{(2k-1)^2}
\\ \times
\frac{[2(2l+1+n-k)]!}{(n-1-k)!(2l+1+n-k)!}.
\label{eq:Rnl}
\end{multline}
An interesting question is to know whether Eq.~(\ref{eq:hydro_cor}), that originates from the Born-Oppenheimer
approximation, is still exact \cite{not_ato}.
The discussion below Eq.~(\ref{eq:def_petit_delta}) allows to hope so.
We shall present numerical evidence in subsection \ref{subsec:hr} that this is indeed the case.

Another interesting aspect is to determine if the Born-Oppenheimer approximation, combined with a suitable
semi-classical (WKB type) approximation, is able to determine exactly the global scale $q_{\rm global}^{(l)}$
in the large mass ratio limit. 
For a fixed angular momentum $l$, using the technique presented in \cite{Migdal} 
and setting here $E=-\hbar^2 \mathcal{Q}^2/m$, we obtain the semi-classical quantization condition \cite{detsc}
\be
\int_{r_{\rm min}}^{r_{\rm max}} dr\, \left[\frac{\alpha}{2}\kappa^2(r)-\frac{(l+1/2)^2}{r^2}-\mathcal{Q}^2\right]^{1/2}
=(n-1/2) \pi
\label{eq:semicl}
\ee
where $r_{\rm min}$ and $r_{\rm max}$ are the lower and upper roots of the integrand, and the quantum number
$n$ is any integer $\geq 1$. In the large $n$ limit, $\mathcal{Q}$ tends to zero exponentially fast
(this is the Efimovian part of the spectrum), so that we can split the integral
in (\ref{eq:semicl}) in two intervals, the interval $r_{\rm min}<r<r_{\rm int}$ where one can make the approximation $\mathcal{Q}\simeq 0$,
and the interval $r_{\rm int}<r<r_{\rm max}$ where the Born-Oppenheimer potential
can be replaced by its asymptotic expression (\ref{eq:vbo_loin}). The intermediate value $r_{\rm int}$ simply
has to satisfy $R_* \ll r_{\rm int} \ll |s_{\rm BO}|/\mathcal{Q}$ where $s_{\rm BO}$ is given by Eq.~(\ref{eq:BO}). 
Then the result does not depend on the specific value
of $r_{\rm int}$ and one finds an approximation for the Efimovian spectrum for $n\to+\infty$:
\be
q_n^{(l)} R_* \approx  \frac{2^{3/2}|s_{\rm BO}|(C-u_{\rm min})(1+C)}{C^2(1+\alpha)^{1/2}}  e^{J_{\rm cl}}
e^{-(n-1/2)\pi/|s_{\rm BO}|},
\label{eq:bowkbspec}
\ee
where we have performed the change of variable  $u=\kappa(r) r$ in the integral over $r$ 
and one has $\alpha u_{\rm min}^2/2=(l+1/2)^2$.
The quantity $J_{\rm cl}$ is defined by the integral
\begin{multline}
J_{\rm cl} = -1+\int_{u_{\rm min}}^{C} \!\! du 
\left[\left(\frac{2}{u}+\frac{1+e^u}{1-ue^u}\right)
\left(\frac{u^2-u^2_{\rm min}}{C^2-u_{\rm min}^2}\right)^{1/2}\right.  \\
\left. -\frac{1}{C-u}\right].
\label{eq:gjcl}
\end{multline}
Whereas as expected \cite{detsc} this semi-classical result is disastrously bad for $\alpha$ 
close to the critical value $\alpha_c^{(l)}$
(it does not predict a finite $q_{\rm global}^{(l)}$ at the Efimovian threshold), it becomes increasingly accurate
for increasing $\alpha-\alpha_c^{(l)}$. Comparing to the exact numerical values
of Fig.\ref{fig:qga}a, we found that for $\alpha-\alpha_c^{(l)}>8$, the error on $q_{\rm global}^{(l)}$
is already less than $10\%$. In the limit $\alpha\to +\infty$,  the semi-classical result allows to recover
exactly the quantum result Eq.~(\ref{eq:qglob_loin}), since
$u_{\rm min}\to 0$ in that limit and $J_{\rm cl}$ then tends to $J$ of Eq.~(\ref{eq:defJ}).
Keeping the first correction linear in $u_{\rm min}$ in the semi-classical result we even get
the refined estimate
\be
q_{\rm global}^{(l)} R_* \underset{\alpha\to \infty}{=} 2 (1+C)\, e^J \left[1-\frac{\pi l}{|s_{\rm BO}|} + O(\ln\alpha/\alpha)\right].
\label{eq:refesti}
\ee
As this amounts to keeping the first $\alpha^{-1/2}$ term in a large $\alpha$ expansion, we can again hope
that the Born-Oppenheimer approximation (combined to the semi-classical one) gives the exact result in Eq.~(\ref{eq:refesti})
\cite{au_moins_l_grand}. 
In Fig.\ref{fig:qga}c we have calculated $J_{\rm cl}$ and hence the semi-classical value of $q_{\rm global}^{(l)}$ numerically to show
how it nicely interpolates between the large-$\alpha$ exact data of Fig.\ref{fig:qga}a (that still strongly depend
on $l$)
and the $l$-independent $\alpha\to +\infty$ limit of $q_{\rm global}^{(l)}$. Taking values of $\alpha$
as large as in Fig.\ref{fig:qga}c is straightforward in the semi-classical formula, but it would be a numerical challenge
for the exact expression Eq.~(\ref{eq:weier_Ft}) (not to mention real experiments).

\section{Numerical solution for trimer states}
\label{sec:numerics}

In this section, we proceed with the direct numerical solution of the integral equation
(\ref{eq:intl}), looking for the allowed bound state energies $E$ for various values of the angular momentum quantum
number $l$.  
The motivation is to look for trimer states that are not predicted (or not faithfully predicted) by the 
analytical results of section \ref{sec:arots}. First, in presence of the Efimov effect ($l$ odd,
$\alpha>\alpha_c^{(l)}$) the analytical formula (\ref{eq:spectrum_analy}) 
is guaranteed to be asymptotically exact in the large quantum number $n\to +\infty$ limit,
but the numerics can assess its accuracy for low values of $n$, $n\geq 1$, and can check
whether or not  $n=1$ in (\ref{eq:spectrum_analy}) corresponds to the ground state trimer for a
given angular momentum. Second, it is in principle possible that the narrow Feshbach resonance model
exhibits for $1/a=0$ non-Efimovian trimers, that would appear for a mass ratio lower than $\alpha_c^{(l)}$.
In single channel models, with real interaction potentials, such few-body bound
states were recently observed numerically \cite{Carlson,Blume} and their emergence was related to few-body resonances
\cite{Blume} that one may expect within the zero-range model when Eq.~(\ref{eq:lam_l_de_s=0}) has a real root $s$
between $0$ and $1$ \cite{WernerCastinsymetrie,Nishida}.
To be complete, let us mention that, for a {\sl finite} and positive value of the scattering length, we expect that there exist
a finite number of non-Efimovian $l=1$ trimer states for mass ratios below the critical mass ratio $\alpha_c^{(l=1)}$, as shown
in \cite{Kartav_discovery}. These interesting trimer states should have a vanishing energy right at the Feshbach resonance
($1/a=0$) so that they shall not show up in our numerical solution and their study is beyond the scope of the paper.

\subsection{Optimized numerical method}

The general method to numerically find the bound state spectrum is to approximate the operator appearing
in the momentum space integral equation by a matrix, after discretization and truncation of the momentum $k$,
and to perform a dichotomic or Newton search of the values of the energy $E<0$ such that the resulting matrix
has a zero eigenvalue. Here, we are in the particular case of an infinitely narrow Feshbach resonance
with an infinite scattering length, and a much more direct method can be used.
The kernel in (\ref{eq:intl}) does not indeed involve any interaction length, it exhibits $\hbar q$ as the only
momentum scale, where the wavenumber $q$ is the unknown since $E=-\hbar^2 q^2/(2\mu)$. We can thus rescale all wavenumbers by $q$,
setting $\check{k}=k/q, \check{K}=K/q$, and we introduce a reduced relative wavenumber
\be
\check{q}_{\rm rel}(\check{k}) \equiv \frac{q_{\rm rel}(q\check{k})}{q} = \left[1+\frac{1+2\alpha}{(1+\alpha)^2}\, \check{k}^2\right]^{1/2}.
\ee
For convenience, we also write the unknown function $f^{(l)}(k)$ as
\be
f^{(l)}(k=q\check{k})=\frac{\check{f}^{(l)}(\check{k})}{\check{k}\check{q}_{\rm rel}(\check{k})},
\label{eq:conve}
\ee
where the denominator shall ensure that the resulting integral operator is hermitian.
After multiplication of (\ref{eq:intl}) by $\check{k}/q_{\rm rel}(k)$, we obtain
\begin{multline}
-qR_* \check{f}^{(l)}(\check{k}) = \frac{1}{\check{q}_{\rm rel}(\check{k})} \check{f}^{(l)}(\check{k})  \\
+ \int_0^{+\infty}\!\!\frac{d\check{K}}{\pi}  \frac{\check{f}^{(l)}(\check{K})}{\check{q}_{\rm rel}(\check{k})\check{q}_{\rm rel}(\check{K})}
\int_{-1}^{1}\!\!\! du\, \frac{P_l(u) \check{k}\check{K}}{1+\check{k}^2+\check{K}^2+\frac{2\alpha}{1+\alpha}\check{k}\check{K} u}.
\label{eq:red1}
\end{multline}
Remarkably, the dimensionless quantity $-q R_*$ is simply the solution of an eigenvalue problem for a fixed
operator. After numerical discretization and truncation, one simply has to diagonalize {\sl once} a real symmetric matrix,
which is a well mastered numerical problem, and each {\sl negative} eigenvalue of that matrix will provide
a numerical approximation of the quantity $-q R_*$ for the bound states. We expect the numerical truncation
to be accurate if the maximal value $\check{k}_{\rm max}$ 
of $\check{k}$ in the numerical grid obeys, for each considered negative eigenvalue $-q R_*$:
\be
q \check{k}_{\rm max} R_* \gg 1,
\label{eq:cond_num}
\ee
so as to ensure that the effective range term in the two-body scattering amplitude is well taken into account.
For a given numerical diagonalization, the negative eigenvalues $-q R_*$ that are larger that $-1/\check{k}_{\rm max}$ thus can not
be trusted. In these estimates, we have dropped for simplicity a possible dependence of the criteria on
the mass ratio $\alpha$ \cite{en_fait}. 

In presence of the Efimov effect, the condition (\ref{eq:cond_num}) is quite severe, as $q R_*$ may
assume extremely small values. The way out is well known, one simply has to use a logarithmic scale,
with the change of variable $x=\ln\check{k}, X=\ln\check{K}$. To keep the hermiticity of the operator,
we reparametrize the unknown function
\be
\check{f}^{(l)}(\check{k}=e^x) = \frac{\check{F}^{(l)}(x)}{e^{x/2}}.
\label{eq:reparam}
\ee
Multiplying (\ref{eq:red1}) by $\check{k}^{1/2}$ and performing these changes, we finally obtain
the numerically useful form \cite{num_noyau_pas}:
\begin{widetext}
\begin{multline}
-q R_* \check{F}^{(l)}(x) = \frac{\check{F}^{(l)}(x)}{\left(1+e^{2x}\cos^2\nu\right)^{1/2}} 
+\int_{-\infty}^{+\infty}\!\! \frac{dX}{\pi} \frac{\check{F}^{(l)}(X)}{\left[\left(1+e^{2x}\cos^2\nu\right)\left(1+e^{2X}\cos^2\nu\right)\right]^{1/2}}
\int_{-1}^{1} \!\!\! du \ 
\frac{P_l(u)\ e^{3(x+X)/2}}{1+e^{2x}+e^{2X} + 2u e^{x+X}\sin \nu}
\label{eq:red2}
\end{multline}
\end{widetext}
where we recall that $\nu=\arcsin[\alpha/(1+\alpha)]$, see (\ref{eq:def_nu}).
Another interesting feature of the form (\ref{eq:red2}), that it shares with (\ref{eq:red1}),
is that the kernel remains bounded at large momenta.

\subsection{Efimovian results}

We now present results obtained from a numerical solution of the eigenvalue problem
(\ref{eq:red2}), for not too large values of the mass ratio so that the trimer spectrum
does not exhibit the hydrogenoid character predicted in subsection \ref{subsec:BO}. In the numerics, we took for the variable $x=\ln\check{k}$ a discretization
with a constant step $dx=0.09$ over an interval from $x_{\rm min}=\ln \left(10^{-2}\right)$ to
$x_{\rm max}=\ln \left(10^{15}\right)$. We explored the interval of mass ratio $\alpha=m/M$ from the small
value $1/200$ to the large value $200$.

We first explored the case of even angular momenta. According to the zero-range theory,
numerical observation of bound states in that case would reveal non-Efimovian trimers.
We went up to an angular momentum $l=12$ without finding any bound state: The minimal eigenvalues
$-q R_*$ found numerically were positive and of the order of $10^{-15}$ or larger.
We then explored odd angular momenta, looking for non-Efimovian bound states
for a mass ratio $\alpha < \alpha_c^{(l)}$ (and $\alpha < 200$). We went up to $l=13$
without finding any.

\begin{figure*}[htb]
\includegraphics[width=8cm,clip=]{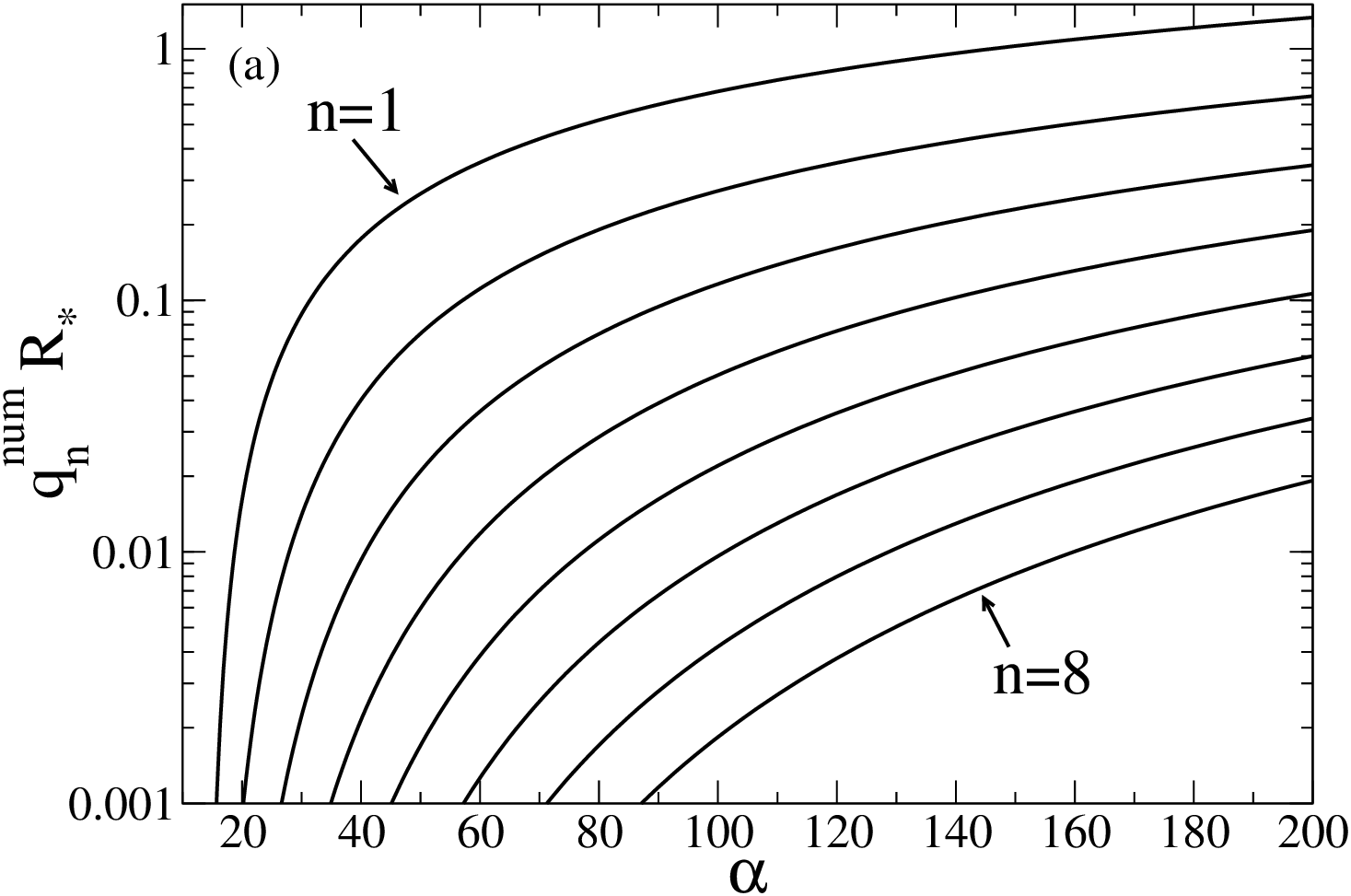}
\hspace{5mm}
\includegraphics[width=8cm,clip=]{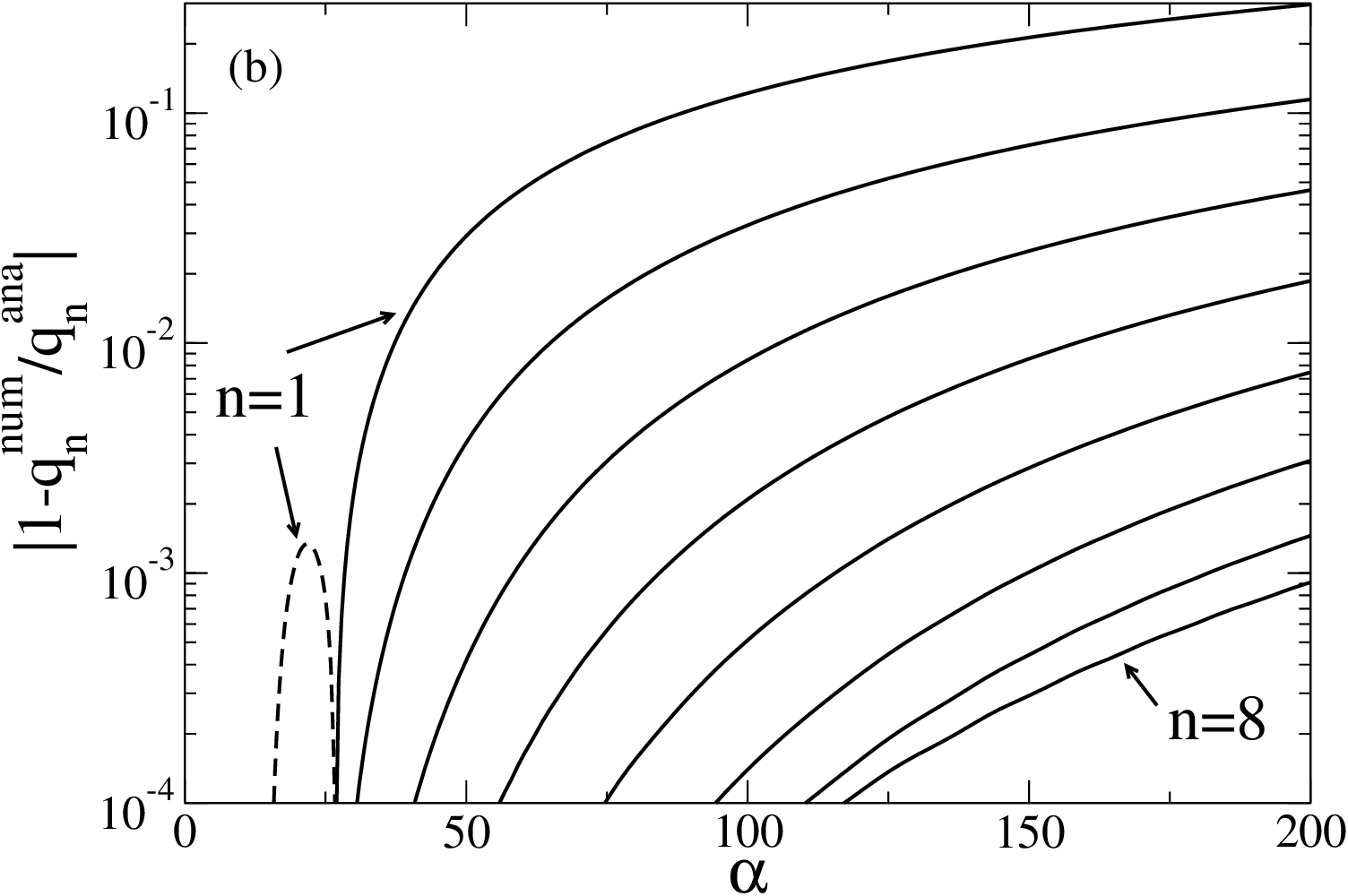}
\caption{Numerical study of the Efimovian part of the spectrum:
For an angular momentum $l=1$, quantities $q_n$ (labeled in descending order) from $n=1$ (ground trimer state) to $n=8$, as functions
of the mass ratio $\alpha=m/M$. We recall that the trimer energies are then $E_n=-\hbar^2 q_n^2/2\mu$. 
In (a) the numerical values are shown in log scale, restricting the figure
to experimentally accessible values $q R_*>10^{-3}$ (see text). In (b) the absolute value of the deviation from unity
of the ratio of the numerical $q_n$ to the analytical approximation
(\ref{eq:spectrum_analy}). The solid (respectively dashed) lines correspond to numerical values of $q_n$ larger (respectively
smaller) than the analytical ones.  
For both (a) and (b),
the vertical axis is in log scale, and the curves $n=1$ to $n=8$ are from top to bottom.
}
\label{fig:num1}
\end{figure*}

At this stage, it remained to explore Efimovian physics.
The numerical results for the most bound trimer states are shown for $l=1$ in Fig.\ref{fig:num1}.
In Fig.\ref{fig:num1}a we directly show the obtained values of $q R_*$ as functions of the mass ratio $\alpha$,
obviously in log scale for the vertical axis. This is useful to estimate which trimer states may be accessed
in an experiment. An experimental limitation is that the scattering length $a$ is not infinite, which will suppress
the too-weakly bound trimers, that is the trimer states with a too small value of $q$.
According to (\ref{eq:int_gen}) we see that if $1/|a| \ll q$, the term $1/a$ is small as compared to the term $q_{\rm rel}(k)$
for all values of $k$, so that the assumption $1/a=0$ should be a good approximation.
Assuming that producing in a controlled way a scattering length larger than $100$ $\mu$m (in absolute value)
becomes unrealistic, in particular for a narrow Feshbach resonance, we thus take
\be
|a^{\rm exp}| \lesssim 100\, \mu\mathrm{m}
\label{eq:aexp}
\ee
and we impose $q>10/|a| \approx 10/(100 \mu\mathrm{m})$. Since $R_*$ should be much larger than the
typical van der Waals length (typically of a few nanometers), we take $R_* > 10$~nm,
so that $q R_*> 10^{-3}$. Hence the interval of values on the vertical axis in Fig.\ref{fig:num1}a
\cite{magnetic_field_stabilisation}.

In the more theoretical Fig.\ref{fig:num1}b, we show the ratios of the numerical values
of $q R_*$ to the values obtained from the approximation (\ref{eq:spectrum_analy}), as functions of the mass ratio.
The numerical values are numbered as $n=1$ (ground state trimer state), $n=2,3, \ldots$, for $q$ in descending order (that is in ascending
order of the energies $E_n=-\hbar^2 q_n^2/2\mu$).
The numerical $q_n$ are then each divided by the approximate $q_n$ of (\ref{eq:spectrum_analy}). For $\alpha$ close to $\alpha_c^{(l=1)}$,
for example $\alpha \leq 20$, it is found that the ratio of numerical to analytical values of the $q$'s is extremely
close to unity: For the ground trimer $n=1$, the deviation is less than $10^{-3}$, and for $n=2$ and $n=3$ it is fluctuating between $\pm 10^{-5}$
which is probably due to numerical errors. Similar results are obtained for larger odd values of $l$
(not shown).

We have thus numerically obtained the important result that the quantum number $n=1$ in the analytical formula
(\ref{eq:spectrum_analy}) indeed corresponds to the ground trimer state (within each subspace of fixed $l$).
For $\alpha-\alpha_c^{(l)}$ not too large (away from the hydrogenoid regime), the binding energy of the ground
trimer state thus does not correspond to the naive expectation $\approx \hbar^2/(\mu R_*^2)$, it contains an extra factor
$\exp(-2\pi/|s_l|)$ which can be tiny. There is numerical evidence that this also holds for three bosons
resonantly interacting on a narrow Feshbach resonance \cite{observ_ludo}; the impossibility in the bosonic case
to connect the spectrum continuously to a $S_l=0$ limit (and to study the variation of the three-body parameter
in that limit) however makes the statement more subjective than in the present fermionic case.

\begin{figure}[htb]
\includegraphics[width=8cm,clip=]{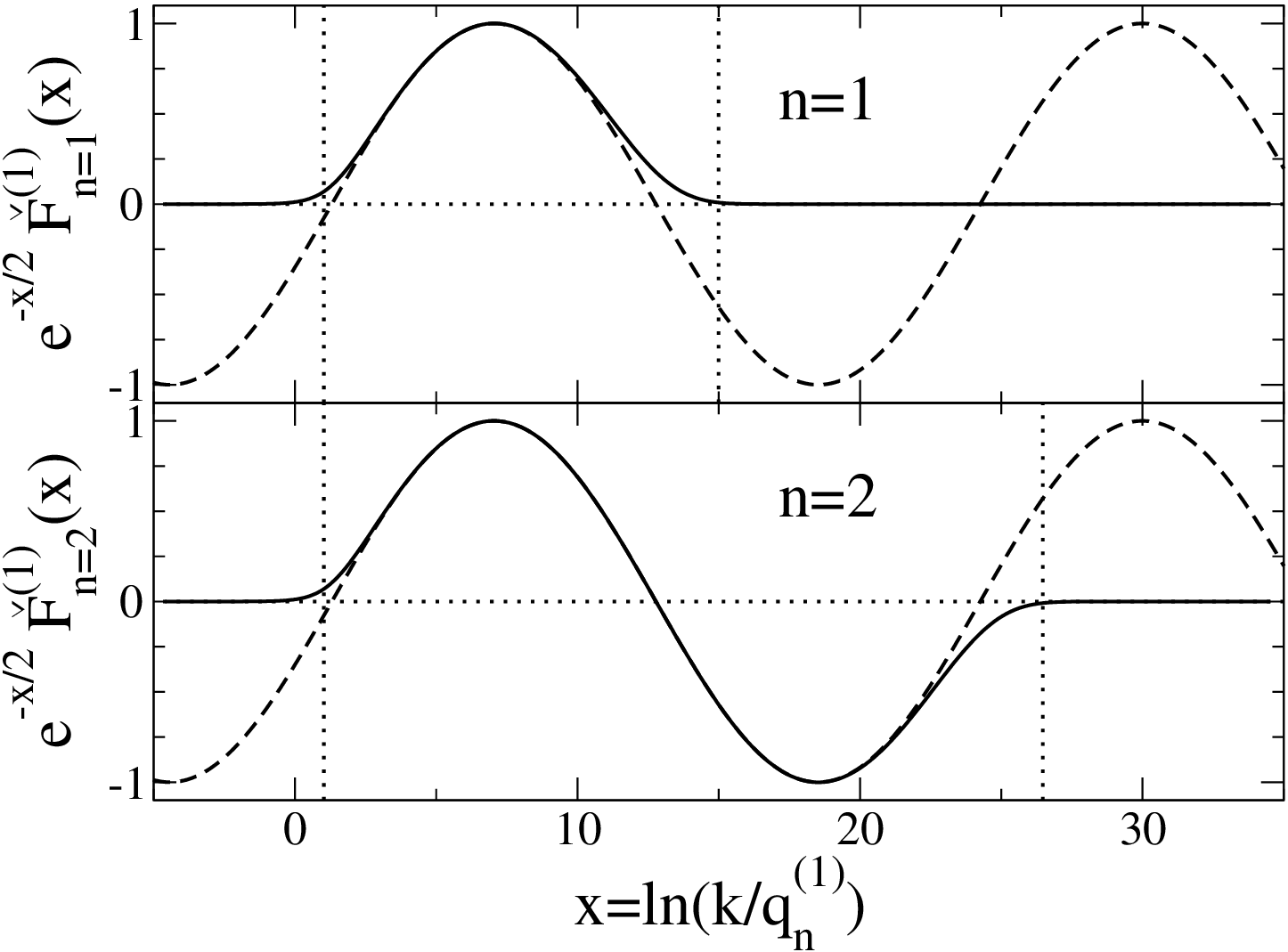}
\caption{For $l=1$ and a mass ratio $\alpha=14$, ground eigenvector ($n=1$) and first excited
eigenvector ($n=2$) of the eigenvalue problem Eq.~(\ref{eq:red2}), corresponding to the ground 
and first excited trimer states. Solid lines: Numerical results.
Dashed lines: Analytical form (\ref{eq:analytical_form}), 
meaningful over the matching interval $1/\cos\nu < k/q < 1/(qR_*\cos\nu)$
whose meaning is explained in Fig.\ref{fig:match} and whose borders are indicated by the vertical dotted lines.
The expression of $D(\kk)$ in terms of $\check{F}^{(1)}(x)$ and $Y_1^0(\kk)$ 
can be obtained from the text, with $x=\ln(k/q)$. 
An overall factor $\exp(-x/2)$ was applied to $\check{F}^{(1)}(x)$ for convenience, 
and the resulting functions are normalized to the maximal value of unity.
\label{fig:vecteur}
}
\end{figure}

To be complete, we have also calculated numerically the eigenvectors 
$x\to \check{F}^{(l)}(x)$ corresponding to the eigenvalues $-q_n^{(l)}R_*$, 
in the case $l=1$ for $\alpha=14$. According to Eqs.~(\ref{eq:conve},\ref{eq:reparam}) one
has $\check{F}^{(l)}(x)=e^{3x/2} (1+e^{2x}\cos^2\nu)^{1/2} f^{(l)}(qe^x)$.
For $k/Q \ll 1$, that is $e^x\cos\nu \ll 1$ since $Q=q/\cos\nu$,
we found that $\check{F}^{(l=1)}(x)\propto e^{5x/2}$ as deduced from
(\ref{eq:solun}) [multiplied by $i$]. For $k R_*\cos\nu \gg 1$, that is $e^x\cos\nu\gg e^{\pi
n/|s_1|}$, where $s_1=i S_1$ is the purely imaginary Efimov exponent for $l=1$,
we found that $\check{F}^{(l=1)}(x)\propto e^{-5x/2}$ as predicted
by \cite{plus_inf}. Finally, in the crucial intermediate
region $q/\cos\nu < k < 1/(R_*\cos\nu)$, which corresponds to the matching interval
of the $(E<0, R_*=0)$ and $(E=0, R_*>0)$ analytical solutions, see Fig.\ref{fig:match}, we compared the numerics to the analytical 
result deduced from Eq.~(\ref{eq:zr_grandk}) [multiplied by $i$]:
\be
\check{F}^{(l=1)}(x) \propto e^{x/2} \sin[|s_1|(x-x_0)]
\label{eq:analytical_form}
\ee
where $x_0=(\arctan |s_1|)/|s_1|-\ln(2\cos\nu)$.
From the numerics, see Fig.\ref{fig:vecteur}, we found that $x\to e^{-x/2} \check{F}^{(l=1)}(x)$
indeed exhibits half an oscillation of the sinus
for the ground trimer $n=1$ [the function remains positive everywhere],
and a full oscillation of the sinus for the first excited
trimer $n=2$ [the function changes sign once].

These nodal properties were expected from the fact noted in \cite{num_noyau_pas}
that the matrix elements of the kernel in (\ref{eq:red2}) are strictly negative
for $l$ odd, for all $x$ and $X$,  
whereas the diagonal element is strictly positive, for
all $x$. From a standard variational argument, that formulates
Eq.~(\ref{eq:red2}) in terms of the extremalization of an ``energy" functional
for a fixed norm, and that compares the ``energy" of
$x\to \check{F}^{(l)}(x)$ to the one of $x \to |\check{F}^{(l)}(x)|$, we conclude that
the function $\check{F}^{(l)}(x)$ for the ground trimer state has a constant
sign. As the other modes have to be orthogonal to the ground mode, this shows
that the ground trimer state is not degenerate, and that the excited
trimer states have a sign-changing function $\check{F}^{(l)}(x)$.
We have thus reached a fully consistent picture of the fact that
$n=1$ in Eq.~(\ref{eq:spectrum_analy}) is indeed the ground trimer state.

\subsection{Hydrogenoid results}
\label{subsec:hr}

\begin{figure}[htb]
\includegraphics[width=8cm,clip=]{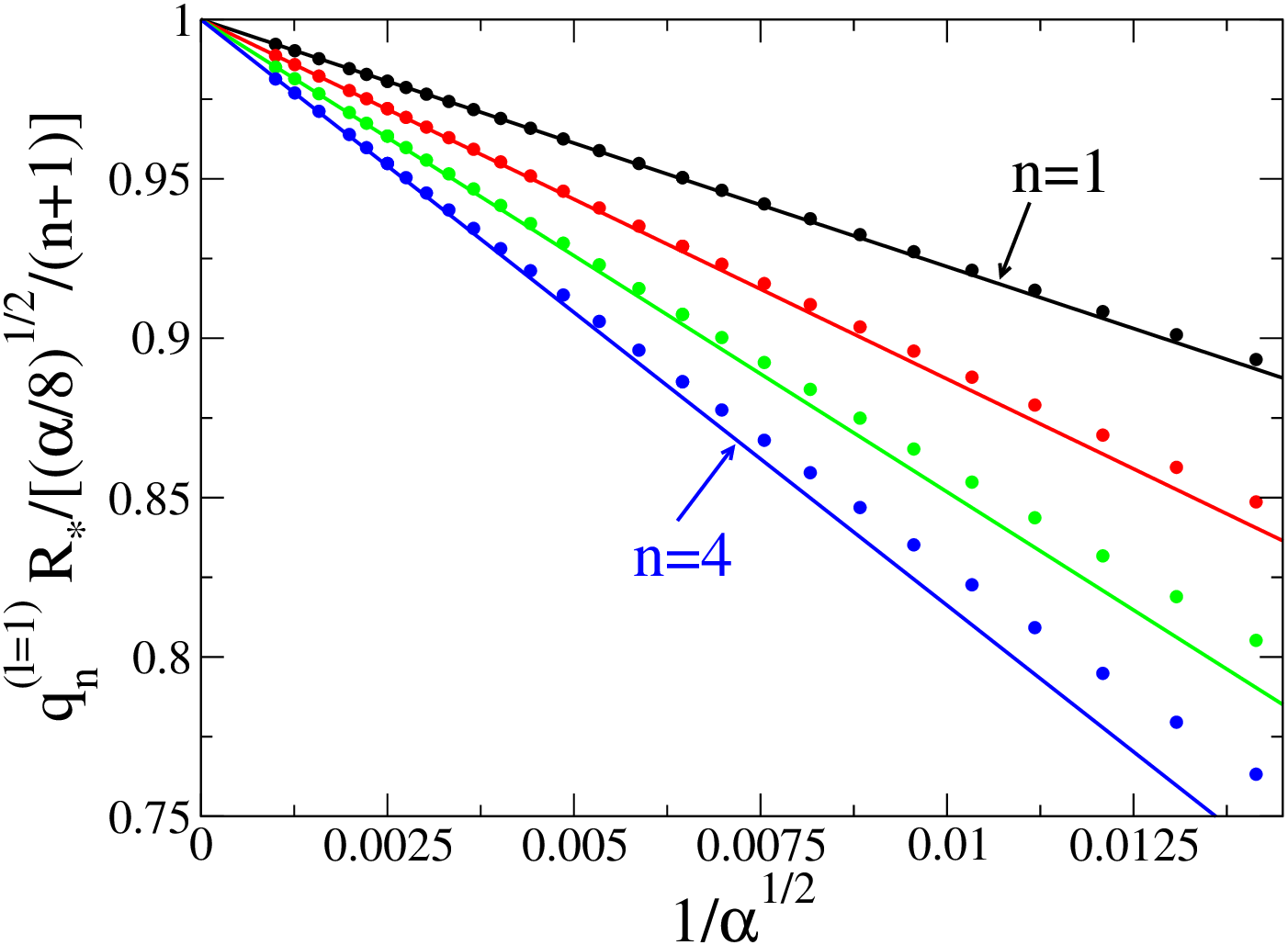}
\caption{(color online) Numerical study of the hydrogenoid part of the spectrum:
For an angular momentum $l=1$, wavenumbers $q_n^{(l)}$ for $n=1$ (ground trimer state), $n=2$, $n=3$ and 
$n=4$ (from top to bottom), divided
by the hydrogenoid asymptotic Born-Oppenheimer prediction (\ref{eq:hydro_asympt}), as functions of $1/\alpha^{1/2}$.
Disks: Numerical results. Straight lines: Asymptotic result (\ref{eq:hydro_cor}) including the first deviation
of the Born-Oppenheimer potential from the Coulomb form at short range. The color code is black for $n=1$,
red for $n=2$, green for $n=3$
and blue for $n=4$.}
\label{fig:hydro}
\end{figure}

We now explore numerically the trimer spectrum for extremely large values of the mass ratio $\alpha$.
In this limit, the low-energy trimers are not expected to be Efimovian anymore: According to the Born-Oppenheimer
approach of subsection \ref{subsec:BO}, the spectrum for fixed values of the quantum numbers $n$ and $l$
becomes hydrogenoid for $\alpha\to +\infty$. 
To test the asymptotic analytical formula (\ref{eq:hydro_cor}), we plotted in Fig.\ref{fig:hydro} 
the ratio of the wavenumber $q_n^{(l)}$ [such that $E_n^{(l)}=-\hbar^2 [q_n^{(l)}]^2/(2\mu)$] to the asymptotic
prediction (\ref{eq:hydro_asympt}) as a function of $1/\alpha^{1/2}$, for $l=1$ and a few values of $n$
(we recall that $n=1$ labels the ground trimer state for fixed $l$). The prediction Eq.~(\ref{eq:hydro_cor}) then corresponds
to straight lines, and we indeed observe that the numerical results approach these straight lines 
for diverging $\alpha$. This suggests that (\ref{eq:hydro_cor}), obtained in the Born-Oppenheimer approximation, is actually
asymptotically exact.
As expected, for increasing $n$, the trimers become spatially more extended, so that a larger value of $\alpha$
is required to make them hydrogenoid.

Some considered values of $\alpha$ in Fig.\ref{fig:hydro} are extremely large, up to $10^6$. This can not
be realized with atoms, as such a large mass ratio does not exist in the periodic table. Using an optical
lattice as suggested in \cite{idee_reseau} is possible if the trimer states have a spatial extension much larger
than the lattice spacing. This may require here unrealistically large values of $R_*$.
A futuristic alternative is to replace the fermionic atoms by large and round molecules, cooled to their internal
(vibrational and rotational) ground state, that have a vanishing total angular momentum
in that ground state, may be in the class of fullerenes \cite{fullerenes}, and that would exhibit a narrow Feshbach resonance
with the extra atom.

\section{Conclusion}
\label{sec:conclusion}

We have performed a detailed study
of the quantum three-body problem of two same-spin-state fermions of mass
$m$ interacting in free space with a distinguishable particle of mass $M$
on an infinitely  narrow Feshbach resonance, with a focus of the three-body
bound states (trimer states) of that system.  The interaction was assumed
to be tuned right on resonance, with an infinite $s$-wave scattering length
$a$, which makes it possible to obtain analytical results, since
the only length scale left in the problem is the so-called Feshbach
length $R_*$ \cite{Petrov_rstar}. The assumption $1/a=0$ 
also ensures that there is no two-body bound states.
This three-body problem however remains rich, richer than {\sl e.g.}\ the 
problem of three resonantly interacting bosons on a narrow
Feshbach resonance \cite{Petrov_rstar,Mora_3corps,Mora_CRAS}, 
because there is a tunable
parameter left, which is the mass ratio $\alpha=m/M$ of a fermion to
the extra particle.
The existence of this tunable parameter raises the following three
fundamental questions on the trimer states within each sector
of {\sl fixed} total angular momentum $l$.

First, does this system support trimer states for a mass ratio
$\alpha$ {\sl smaller} than the minimal mass ratio $\alpha_c$
required to activate the Efimov effect \cite{Efimov}~? Such non-Efimovian
trimer states may indeed emerge from three-body resonances recently discovered
numerically for a different interaction model \cite{Carlson,Blume},
with an Efimov exponent $s$ having a {\sl real} value between $0$ and $1$
\cite{WernerCastinsymetrie,Nishida}.
For the narrow Feshbach resonance, our numerical answer to this
question is negative.

Second, how does the Efimov trimer spectrum emerge when the mass ratio
$\alpha$ is varied across the critical value $\alpha_c$ (necessarily for $l$ odd)~? 
This is an intriguing
question, because there is no trimer state for $\alpha < \alpha_c$ 
(for fixed $l$) and there is an infinite number of trimer states
for $\alpha > \alpha_c$. We found that, for $\alpha$ tending to $\alpha_c$
from above, the {\sl whole} trimer spectrum, including the ground trimer state
 (and not simply the trimer
states with large quantum number $n$) forms a geometric sequence.
This of course can not be deduced from general zero-range Efimov's theory,
which guarantees the geometric nature of the spectrum only in the asymptotic
region of large quantum number $n$, and can not say anything about 
{\sl e.g.}\ the model-dependent ground trimer state.
We have also shown that the global energy scale $E_{\rm global}$ in that spectrum 
has a finite and non-zero limit for $\alpha\to \alpha_c$.
This simply means that, at the Efimovian threshold,
the energy of the ground state trimer 
vanishes as $E_{\rm global} \exp(-2\pi/|s|)$, the energy 
of the first excited trimer vanishes as $E_{\rm global} \exp(-4\pi/|s|)$,
and so on, where the modulus of the purely imaginary Efimov exponent 
$s$ vanishes as the square root of $\alpha-\alpha_c$. 
This constitutes a complete picture of the emergence of the Efimovian trimer states
when the mass ratio is varied across $\alpha_c$.
The dependence of $E_{\rm global}$ on the mass ratio $\alpha$
and on the angular momentum $l$ was further studied analytically, 
by an efficient and to our knowledge original 
expression of $E_{\rm global}$ in terms of an asymptotic
series, see Eqs.~(\ref{eq:global},\ref{eq:alter_thetal}).

Third, what is the nature of the most bound trimer states,
for example the ground state,
when the mass ratio $\alpha$ becomes significantly larger than
the critical value $\alpha_c$? As already mentioned,
Efimov's theory can not answer this model-dependent question.
For the narrow Feshbach resonance,
we found that the low energy part of the spectrum becomes asymptotically equivalent
to an hydrogenoid spectrum for a diverging mass ratio, that is
scaling as $E_0/(n+l)^2$, where $n\geq 1$, $l$ is odd, and
$E_0=-m\hbar^2/(4MR_*)^2$, and we calculated
the first deviation from this hydrogenoid spectrum for a finite $\alpha$.
As the hydrogenoid nature asymptotically takes over the Efimovian nature of the trimer
spectrum for $\alpha\to +\infty$ (except in a vicinity of the $E=0$ accumulation point where the spectrum
remains geometric), in
a continuous way when the mass ratio is varied, this constitutes a crossover
from an Efimovian to a hydrogenoid spectrum.

We have also discussed to which extent
all these predictions for the $2+1$ fermionic problem
on a narrow Feshbach resonance
may be addressed experimentally with cold atoms,
the variation of the mass ratio being obtained by combining
the discrete tuning provided by the choice of appropriate species for the fermions 
and the extra particle,
with an additional continuous fine tuning of the effective mass with
an optical lattice \cite{idee_reseau}. Reaching the large mass ratios required
to observe the hydrogenoid part of the spectrum is challenging, 
except for the futuristic alternative of replacing the fermionic
atoms with massive and round molecules.
On the contrary, the Efimovian effect in our system 
may be reachable experimentally for angular momentum
$l=1$, where the critical mass ratio is only $\alpha_c\simeq 13.607$,
provided that the energy ratio $\exp(2\pi/|s|)$ is not too large, that
is the mass ratio is far enough from the critical value $\alpha_c$
where $|s|=0$. From the bound (\ref{eq:aexp}) on the achievable
$s$-wave scattering length $a$ with a magnetic Feshbach resonance, the ground state trimer is directly 
observable for $\alpha > 15$, whereas 
directly observing at least {\sl two} Efimovian trimer states, 
to check the geometric nature of the spectrum, requires $\alpha> 20$.
On a Feshbach resonance as narrow as the one of ${}^{6}$Li with ${}^{40}$K, however, 
the bound (\ref{eq:aexp}) is probably too optimistic, if one does not implement
a magnetic field stabilization of metrologic quality \cite{magnetic_field_stabilisation}.

\acknowledgments
We warmly thank L. Pricoupenko and C. Mora for fruitful discussions at an early stage of this work, 
A. Sinatra for useful comments on the manuscript, 
and F. Werner for pointing out useful references. The group of Y.C. is a member of IFRAF.

\appendix

\section{Zero-range solution in position space}
\label{app:real_space}

In the case $R_*=0$, our three-body problem reduces to the zero-range infinite scattering
length problem that Efimov solved in real space \cite{Efimov} with the ansatz for the three-body
wavefunction:
\begin{multline}
\psi(\rr_1,\rr_2,\rr_3)= \frac{F(R)}{R^2}\left[\frac{\varphi(\alpha_{21})}{\sin(2\alpha_{21})}
Y_l^{m_l}\left(\rr_3-\frac{\alpha \rr_2+\rr_1}{1+\alpha}\right)\right. \\ 
\left.-\frac{\varphi(\alpha_{31})}{\sin(2\alpha_{31})}
Y_l^{m_l}\left(\rr_2-\frac{\alpha \rr_3+\rr_1}{1+\alpha}\right)
\right].
\label{eq:ansatz_psi}
\end{multline}
This contains two {\sl a priori} unknown functions of single variables,
$F$ and $\varphi$.  The various variables in this ansatz 
naturally appear in appropriately normalized Jacobi coordinates.
Using {\sl e.g} the convention of Appendix 3 in \cite{livre_unitaire}
(except for a permutation of the indices),
taking as in Eq.~(\ref{eq:psi3at}) particle $1$ as the extra particle of mass $m_1=M$, and $2$ and $3$ as the fermions
of mass $m_2=m_3=m$, we define the unnormalized Jacobi coordinates
$\yy_1=\rr_3-(m \rr_2+ M\rr_1)/(m+M)$ and $\yy_2=\rr_1-\rr_2$, the generalized reduced masses
$\mu_1^{-1}=m^{-1}+(m+M)^{-1}$ and $\mu_2^{-1}=M^{-1}+m^{-1}$, the hyperradius
$R\geq 0$ such that $\bar{m} R^2 =\sum_{i=1}^{3} m_i (\rr_i-\mathbf{C})^2$ where $\bar{m}$ is an
arbitrary unit of mass and $\mathbf{C}$ is the center of mass position of the three particles.
Then the conveniently normalized Jacobi coordinates are $\uu_i=(\mu_i/\bar{m})^{1/2} \yy_i$
with $i=1,2$, so that $R^2=u_1^2+u_2^2$. This also puts Schr\"odinger's equation in a reduced form,
\be
\sum_{i=1}^3 -\frac{\hbar^2}{2 m_i} \Delta_{\rr_i} = -\frac{\hbar^2}{2(2m+M)}\Delta_{\mathbf{C}}
+\sum_{i=1}^2 -\frac{\hbar^2}{2\bar{m}} \Delta_{\uu_i}.
\ee
The angles $\alpha_{21}$ and $\alpha_{31}$ belong to the interval $[0,\pi/2]$. The clever choice is
$\tan \alpha_{21}=u_2/u_1$ and the similar formula for $\alpha_{31}$ obtained
by exchanging the role of particles $2$ and $3$. Geometrically, this amounts to introducing
polar coordinates $(R,\alpha_{21})$ in the plane $(u_1,u_2)$. Explicitly this gives
\bea
\tan \alpha_{21} &=& \frac{|\rr_1-\rr_2|\cos\nu}
{|\rr_3-(\alpha \rr_2+\rr_1)/(1+\alpha)|} \\
\tan \alpha_{31} &=& \frac{|\rr_1-\rr_3|\cos\nu}
{|\rr_2-(\alpha \rr_3+\rr_1)/(1+\alpha)|}
\eea
where $\cos\nu$ is given by (\ref{eq:cosnu}).
Since the second (Faddeev) component in (\ref{eq:ansatz_psi}) is deduced from the first one
by a minus sign and the exchange of particles $2$ and $3$, which ensures the fermionic
antisymmetry of $\psi$, it suffices to calculate the action of $\Delta_{\uu_1}+\Delta_{\uu_2}$
on the first component. This is quite simple in spherical coordinates, since
$R$ and $\alpha_{21}$ depend only on the moduli $u_1$ and $u_2$, and the factor involving the spherical harmonic
function is simply $Y_l^{m_l}(\uu_1)$. Thanks to the factor $R^2 \sin (2\alpha_{21})= 2 u_1 u_2$ in the denominator,
one is left with the operator $\partial_{u_1}^2 + \partial_{u_2}^2$, that is the Laplacian
in the plane $(u_1,u_2)$, which has a simple expression in the polar coordinates $(R,\alpha_{21})$.
One finally finds that the ansatz (\ref{eq:ansatz_psi}) separated in hyperspherical coordinates
solves Schr\"odinger's equation if $\varphi$ solves the eigenvalue problem
\be
-\varphi''(\alpha_{21}) + \frac{l(l+1)}{\cos^2\alpha_{21}} \varphi(\alpha_{21})
= s^2 \varphi(\alpha_{21}).
\label{eq:diff_phi}
\ee
$s$ is {\sl a priori} unknown, but the general theory of {\rm e.g.}\ section 3.3 
in \cite{livre_unitaire} guarantees that it coincides with $s$ defined
in Fourier space by (\ref{eq:ansatz_puissance}).
As a consequence, the $R$ dependence of the ansatz also separates and the hyperradial 
function $F(R)$ is found to solve a Schr\"odinger equation for a fictitious particle
in two dimensions experiencing an effective $1/R^2$ potential:
\be
E F(R) = -\frac{\hbar^2}{2\bar{m}} \left[F''(R) + \frac{1}{R} F'(R)\right]
+\frac{\hbar^2 s^2}{2\bar{m} R^2} F(R).
\label{eq:hyper_radial}
\ee
The Efimov effect appears for $s^2 <0$.

To determine the eigenvalue $s^2$, one needs to specify the boundary conditions
for $\varphi$. For $u_1\to 0$, $\psi$ in general does not diverge. Since $\sin (2\alpha_{21})$
vanishes in the denominator, this imposes
\be
\varphi(\pi/2)=0.
\label{eq:cond1}
\ee
For $u_2\to 0$, on the contrary, $\psi$ diverges. More precisely, the Bethe-Peierls contact
conditions for an infinite scattering length impose that, when particles $1$ and $2$ approach
each other for a fixed position of their center of mass $\mathbf{C}_{12}$, that is
$\rr_2=\mathbf{C}_{12}-\rr/(1+\alpha)$ and $\rr_1=\mathbf{C}_{12}+\alpha \rr/(1+\alpha)$
with $r\to 0$, $\psi$ should behave as $A/r+O(r)$ where $A$ depends here on $\mathbf{C}_{12}-\rr_3$.
In other words, there should be no non-zero contribution behaving as $r^0$.
To calculate the $r^0$ contribution from the first Faddeev component, one has
to expand $\varphi(\alpha_{21})$ to first order in $\alpha_{21}$. In the second Faddeev
component one can directly set $\rr=0$, that is $\rr_1=\rr_2=\mathbf{C}_{12}$.
From the angular representation (\ref{eq:def_nu}) for the mass ratio,
we find $\arctan[(1+2\alpha)^{1/2}/\alpha]=\frac{\pi}{2}-\nu$ and
\be
\frac{1}{2} \varphi'(0) -(-1)^l \frac{\varphi(\frac{\pi}{2}-\nu)}{\sin(2\nu)}=0.
\ee
To try to recover the function $\Lambda_l(s)$, we heuristically multiply this
condition by $2 \cos\nu/\varphi'(0)$ to pull out a first additive term as in
(\ref{eq:lambda_expli}):
\be
\Lambda_l(s) \stackrel{?}{=} \cos\nu - (-1)^l \frac{\varphi(\frac{\pi}{2}-\nu)}{\varphi'(0)\sin\nu}.
\label{eq:heuristic}
\ee
Note that Eq.~(\ref{eq:diff_phi}) is independent of the mass ratio,
it occurs in the problem of three spin $1/2$ (same mass) fermions and even in the case of three
bosons.  The solution of (\ref{eq:diff_phi}) obeying (\ref{eq:cond1}) 
was given in these contexts in \cite{hypergeom}
in terms of a hypergeometric function. Reusing this solution
\begin{multline}
\varphi(\alpha_{21})= \cos^{l+1}\alpha_{21}\\  \times\,
{}_2 F_1\left(\frac{l+1+s}{2},\frac{l+1-s}{2},l+\frac{3}{2};\cos^2\alpha_{21}\right)
\end{multline}
leads to (\ref{eq:trans_hyper}) that was checked numerically to coincide with 
(\ref{eq:lambda_expli}), so that the question mark can be removed in (\ref{eq:heuristic}).
For completeness, we note that still another form of $\varphi(\alpha_{21})$, in terms of a finite sum,
was given in \cite{Werner3corps}.

\section{Efimovian spectrum global scale at threshold}
\label{app:qrefc}

At fixed angular momentum $l$, we evaluate the global scale $q_{\rm global}^{(l)}$ in the limit
$\alpha\to \alpha_c^{(l)}$. Since $s_l = i S_l$, $S_l>0$, vanishes in that limit,
we can simply expand the complex number $Z_l$ up to order $S_l$, neglecting terms $O(S_l^2)$.
The most difficult part is the bit $iS_l\, C_l(-S_l)$, that we rewrite using (\ref{eq:weier_Ft}),
$\Gamma(2iS_l) = \Gamma(1+2iS_l)/(2iS_l)$:
\begin{multline}
iS_l\, C_l(-S_l)= \frac{\Gamma(1+2iS_l)\Gamma(1-iv_0-iS_l)}{2\Gamma(1)\Gamma(-iv_0+iS_l)} \\
\times \prod_{n\in\mathbb{N}^*} \frac{\Gamma(-iu_n+iS_l)}{\Gamma(1-iu_n-iS_l)} \, \frac{\Gamma(1-i v_n-iS_l)}{\Gamma(-iv_n+iS_l)}.
\label{eq:recrit}
\end{multline}

We first need to figure out to which order the roots $u_n$ depend on $u_0=S_l$ (remember that the $v_n$'s do not depend
on the mass ratio). To zeroth order, right on the critical mass ratio,
$u_n=u_n^c$ by definition. Making apparent the dependence
of the function $\Lambda_l$ with $\alpha$, one has
\be
\Lambda_l(i u_n; \alpha)=0.
\label{eq:implicit}
\ee
It is clear on Eq.~(\ref{eq:rep_odd}) that $\Lambda_l$ is a regular function  of $\alpha$, so that 
$\Lambda_l(s;\alpha)$ around $\alpha_c^{(l)}$ varies to first order in $\alpha-\alpha_c^{(l)}$
and (\ref{eq:implicit}) simplifies to
\be
\Lambda_l(i u_n; \alpha_c^{(l)}) = O(\alpha-\alpha_c^{(l)}).
\label{eq:implicit2}
\ee
We also note that $\Lambda_l(s;\alpha)$ is an even function of $s$.
For $n=0$, $u_0=S_l$ and $u_0^c=0$; since $0$ is a double root of $s\to \Lambda_l(s;\alpha_c^{(l)})$,
one has to expand (\ref{eq:implicit2}) to second order in $u_n-u_n^c$ to get the leading contribution,
and $S_l^2$ varies linearly in $\alpha-\alpha_c^{(l)}$ as expected.
For $n>0$, $u_n^c\ne 0$ and $i u_n^c$ is a {\sl simple} root of $s\to \Lambda_l(s;\alpha_c^{(l)})$,
one has to expand (\ref{eq:implicit2}) to first order in $u_n-u_n^c$ to get the leading contribution,
so that
\be
u_n-u_n^c = O(\alpha-\alpha_c^{(l)})= O(S_l^2) \ \ \mbox{for}\ n>0.
\ee
To first order in $S_l$, we can thus replace the $u_n$'s in (\ref{eq:recrit}) by their values $u_n^c$ at threshold
(for $n>0$).

The last step is to expand the $\Gamma$ functions in (\ref{eq:recrit}) to first order
in $S_l$.  For any real positive number $x>0$, one has
\be
\frac{\Gamma(x+iS_l)}{\Gamma(x)} = 1 + i S_l \psi(x) + O(S_l^2)
\ee
where $\psi(z)=\Gamma'(z)/\Gamma(z)$ is the digamma function. One has in particular $\psi(1)=-\gamma$,
where $\gamma=0.577\ 215\ 664\ 9\ldots$ is Euler's constant, see {\sl e.g.}\ relation 8.362(1) in \cite{Gradstein}.
We recall that, by definition, $-i u_n$ for $n>0$ and $-i v_n$ for $n\geq 0$ are real positive. Finally,
the product over $n$ in (\ref{eq:global}) may be written as $\Gamma(l+1-iS_l)/\Gamma(1-i S_l)$
and expanded with the same technique.
All this leads to (\ref{eq:glob_lim}).

To reveal the convergence of the series in (\ref{eq:glob_lim}), it is useful to introduce
the function $\Phi(x)= \psi(x)+\psi(x+1)-2\ln x$. Then $\Phi(x)=O(1/x^2)$ for $x\to +\infty$
since $\psi(x+1)=\ln x+\frac{1}{2x}+O(1/x^2)$ according to relation 8.344 in \cite{Gradstein}.
Expressing $\psi(x)+\psi(x+1)$ as $\Phi(x) + 2\ln x$, for $x=-i u_n^c$ and for $x=-i v_n$,
and collecting all the logarithmic contributions as the logarithm of the product
$\prod_{n\geq 1} \frac{(u_n^c)^2}{v_n^2}$, that one can relate to the curvature in $s=0$
of the function $\Lambda_l(s;\alpha)$ thanks to (\ref{eq:weier_Lambda}), 
one obtains
\begin{multline}
\frac{\theta_l}{|s_l|} \underset{S_l\to 0^+}{\to}  -\psi(l+1)+3\psi(1)-\psi(-iv_0)-\psi(1-iv_0)
\\ +\ln\left[\frac{v_0^2}{2\cos\nu} \partial_s^2 \Lambda_l(0;\alpha_c^{(l)})\right]
+\sum_{n=1}^{+\infty} [\Phi(-i u_n^c) -\Phi(-i v_n)].
\label{eq:convergente}
\end{multline}
As one expects that $u_n/v_n\to 1$ for $n\to +\infty$, with $v_n=i(2n+l+1)$, 
from the supposed convergence of the infinite product
in (\ref{eq:weier_Lambda}), the sum in (\ref{eq:convergente}) is convergent.

In practice, it is found that the sum over $n\geq 1$ in (\ref{eq:convergente}) is so rapidly
convergent that it gives a very small contribution to the result. A useful, easy to evaluate
approximation is thus:
\begin{multline}
\lim_{\alpha\to\alpha_c^{(l)}} q_{\rm global}^{(l)} \simeq -\frac{(l+1)^2}{R_*\cos\nu}\, e^{-2\psi(l+1)-\psi(l+2)+3\psi(1)} \\
\times
\partial_s^2 \Lambda_l(0;\alpha_c^{(l)}).
\label{eq:easy}
\end{multline}
The relative error introduced by this approximation is at most $\simeq 5\times 10^{-4}$ for the values of $l$
(from $1$ to $11$) that we have considered.
Note that the approximation (\ref{eq:easy}) is equivalent to neglect all terms with $k\geq 1$ in
(\ref{eq:alter_thetal}) and to take the limit $\alpha\to \alpha_c^{(l)}$.

\section{Alternative representations of the function $C_l(S)$}
\label{app:alter}

To determine $q_{\rm global}^{(l)}$ of Eq.~(\ref{eq:global}), that is the phase $\theta_l$ of the complex number $Z_l$
in Eq.~(\ref{eq:Z_l}), in the large $l$ limit for fixed $\alpha-\alpha_c^{(l)}$ or in the large $\alpha$ limit
for $l$ fixed, the infinite product
form (\ref{eq:weier_Ft}) for the function $C_l(S)$ is inappropriate, even numerically.
We construct here more efficient representations, in the spirit having led to Eq.~(\ref{eq:convergente}).
Remarkably, the last form that we construct does not rely on the roots and poles of the function $\Lambda_l$.

We split the function $C_l(S)$ of Eq.~(\ref{eq:weier_Ft}) in parts whose phase is easy/difficult to evaluate.
Since $-i u_n$ is real for $n>0$, and $S$ is real, using $\Gamma(z)^*=\Gamma(z^*)$ we rewrite the factors of (\ref{eq:weier_Ft}) as
\be
\frac{\Gamma(-iS -iu_n)}{\Gamma(1+iS-iu_n)} = \frac{\Gamma(-iS -iu_n) \Gamma(1-iS-iu_n)}{|\Gamma(1+iS-iu_n)|^2}.
\ee
A similar rewriting can be performed on the factors involving $-iv_n$, $n>0$, so that,
apart from an infinite product that is real positive and does not contribute to the phase
of $C_l(S)$, we identify a hard part $D_l(S)$ that is the product of factors of the form $\Gamma(z)\Gamma(z+1)$,
with $z=-iu_n-iS$ in the numerator and $z=-i v_n-iS$ in the denominator.
We then set
\be
\Psi(z) = \frac{\Gamma(z)\Gamma(z+1)}{2\pi e^{2f(z)}} \ \ \mbox{with}\ \ \ f(z)=z\ln z-z.
\ee
Here we are in the case $\mathrm{Re}\, z>0$ so that the branch cut of the logarithm (on the real negative axis)
and the poles of $\Gamma(z)$ are out of reach.
Then
\be
D_l(S) = e^{\Sigma_l(S)} \prod_{n\in\mathbb{N}^*} \frac{\Psi(-iu_n-iS)}{\Psi(-iv_n-iS)}
\label{eq:dls}
\ee
with
\be
\Sigma_l(S) = 2\sum_{n\in\mathbb{N}^*} [f(-i u_n-iS)-f(-iv_n-iS)].
\ee
According to the relation 8.344 in \cite{Gradstein},  $\Psi(z)=1+O(1/z)$ at large $|z|$ so that one expects
that the infinite product in (\ref{eq:dls}) converges more rapidly than the original form (\ref{eq:weier_Ft}).
Furthermore, only the imaginary part of $\Sigma_l(S)$ is required, and it can be expressed as the integral
\be
\mbox{Im}\, \Sigma_l(S) = - \int_0^{S} dS' R_l(S') 
\label{eq:int_ims}
\ee
with
\be
R_l(S)=\ln\left[\frac{\Lambda_l(iS)}{\cos\nu}\,\frac{S^2+(l+1)^2}{S^2-S_l^2}\right].
\ee
Eq.~(\ref{eq:int_ims}) holds for $S=0$, since $-i u_n$ and $-i v_n$ are all real positive for $n>0$.
To check that it holds at non-zero $S$, one takes the derivative of (\ref{eq:int_ims}) with respect
to $S$, using $f'(z)=\ln z$ and $f(z)^*=f(z^*)$. One then recognizes the function $\Lambda_l(iS)$
from its Weierstrass representation (\ref{eq:weier_Lambda}), also using $v_0=i(l+1)$ and $u_0=S_l$.

It is possible to go further and to express the phase of the infinite product in (\ref{eq:dls})
in terms of derivatives of the function $R_l(S)$. Relation 8.344 in \cite{Gradstein} indeed gives
Stirling's representation of $\ln \Psi(z)$ as an {\sl asymptotic} series in $1/z$, and one also has
the $k$th derivative for $k\geq 1$:
\be
\frac{d^k}{dS^k} [\ln(-iu_n-iS)] = \frac{-i^k (k-1)! }{(-i u_n-iS)^k}
\ee
and similar relations obtained by taking the complex conjugate or by replacing $u_n$ with $v_n$.
Finally $D_l(S)=|D_l(S)|\exp[i\varphi_l(S)]$ with
\be
\varphi_l(S)= -\int_0^{S} dS' \, R_l(S') - \sum_{k\geq 1}\frac{(-1)^{k}B_{2k}}{(2k)!~} R_l^{(2k-1)}(S),
\label{eq:joliphi}
\ee
where the $B_{2k}$ are the Bernoulli numbers and $R_l^{(2k-1)}$ stands for the $(2k-1)^{\rm th}$ 
derivative of the function $R_l(S)$. The useful statement is then that 
\be
\frac{C_l(-S_l) e^{i\varphi_l(S_l)}}{\Gamma(2iS_l) \Gamma(l+1-iS_l) \Gamma(l+2-iS_l)}
\ \mbox{is real positive}
\label{eq:useful}
\ee
where we used the fact that $R_l(S)$ is an even function of $S$, and thus $\varphi_l(S)$
an odd function of $S$.
Minor transformations then lead to Eq.~(\ref{eq:alter_thetal}).
In short, these results originate from the lemma: For any $x>0$ and $S$ real,
\begin{multline}
\mbox{Im}[\ln\Gamma(x-iS)+\ln\Gamma(x+1-iS)]=
-\int_0^S dS' \ln(x^2+S'^{2})\\ -\sum_{k\geq 1} (-1)^k\frac{B_{2k}}{(2k)!} 
\frac{d^{2k-1}}{dS^{2k-1}}
\ln(x^2+S^2),
\end{multline}
where here again the series is only asymptotic.

\section{Efimovian spectrum global scale at large angular momenta}
\label{app:qref_grandl}

As we show here, asymptotically exact expressions of $q_{\rm global}^{(l)}$ 
for a diverging angular momentum $l$ can be obtained analytically.
This requires an asymptotic determination of the function $\Lambda_l(iS)$. To this end, the most
convenient starting point is the expression for $\Lambda_l$ in Eq.~(\ref{eq:trans_hyper}). Since
$\cos\nu$ tends to zero for $\alpha > \alpha_c^{(l)}$ in the large $l$ limit, we rewrite
this expression using relation 9.131(2) of \cite{Gradstein} that expresses an hypergeometric function
of the variable $z$ in terms of hypergeometric functions of the variable $1-z$. For $S$ real:
\begin{multline}
\frac{\Lambda_l(iS)}{\cos\nu} = 1+ (-1)^l\sin^l\nu  \left[
\left|\frac{\Gamma\left(\frac{l+1+iS}{2}\right)}{\Gamma\left(1+\frac{l+iS}{2}\right)}\right|^2
\frac{1}{2\cos\nu}
\right. \\ \times \left.
{}_2 F_1\left(\frac{l+1+iS}{2},\frac{l+1-iS}{2},\frac{1}{2};\cos^2\nu\right)
\right. \\ \left.
-\ {}_2 F_1\left(1+\frac{l+iS}{2},1+\frac{l-iS}{2},\frac{3}{2};\cos^2\nu\right) \right].
\label{eq:hyper_utile}
\end{multline}
For a fixed value of $S_l$, and thus considering $\alpha$ as a function 
of $S_l$, we expect the asymptotic expansion in the large $l$ limit:
\be
\cos \nu = \frac{a_1}{l}+\frac{a_2}{l^2}+\frac{a_3}{l^3}+\ldots
\label{eq:serie_cos}
\ee
Using (\ref{eq:hyper_utile}) for fixed $S$, the calculation of the leading coefficient $a_1$ is straightforward, 
expressing each ${}_2 F_1$
function in terms of its defining hypergeometric series, see relation (9.100) in \cite{Gradstein}, and taking the large
$l$ limit in each term of the series. For example, for any natural integer $k$:
\bea
\frac{(\frac{l+1}{2})^2\ldots(\frac{l+1}{2}+k-1)^2}{(\frac{1}{2})\ldots (\frac{1}{2}+k-1)}
\frac{\cos^{2k}\nu}{k!}\!\! &\underset{l\to+\infty}{\to}&\!\! \frac{a_1^{2k}}{(2k)!}, \\
\frac{(\frac{l+2}{2})^2\ldots(\frac{l+2}{2}+k-1)^2}{(\frac{3}{2})\ldots (\frac{3}{2}+k-1)}
\frac{\cos^{2k}\nu}{k!}\!\! &\underset{l\to+\infty}{\to}&\!\! \frac{a_1^{2k}}{(2k+1)!}.
\eea
The sum over $k$ then generates $\cosh$ and $\sinh$ functions of $a_1$. Also, the Gamma functions in
(\ref{eq:hyper_utile}) may be expanded using relation 8.344 in \cite{Gradstein}.
This lowest order calculation gives $\Lambda_l(iS_l)/\cos\nu=1-\exp(-a_1)/a_1+O(1/l)$.
Since $\Lambda_l(iS_l)$ vanishes (to all orders), one obtains $a_1=\exp(-a_1)$
so that 
\be
a_1=C
\ee
where $C$ was introduced in (\ref{eq:C}) in the Born-Oppenheimer context. This technique can be pushed in principle to
any order. We calculated  $a_2=-C/2$ and $a_3$. The fact that $S_l$ does not contribute to $a_2$ 
(and contributes to $a_3$ in the form of a term $S_l^2$) is due to the fact that
(\ref{eq:hyper_utile}) is an even function of $S$, and that it is always the ratio $S/l$ which appears
in the expansion. Turning the expansion (\ref{eq:serie_cos}) into an expansion
for the mass ratio, we obtain (\ref{eq:justi_bo}), (\ref{eq:alphac_grandl}) and (\ref{eq:defDelta}).

This large-$l$ expansion technique can even be extended to the case where $S_l/l=O(1)$ 
(which includes both the previous case of $S_l$ fixed and the new case $S_l/l$ fixed).
To leading order, one finds $S_l^2 \simeq (\alpha-\alpha_c^{(l)})C^2/2$ (as in the Born-Oppenheimer approximation) and
\be
\frac{\Lambda_l(iS)}{\cos\nu}\simeq 1-\frac{\exp\left[{-C\left(\frac{S^2+l^2}{S_l^2+l^2}\right)^{1/2}}\right]}
{C\left(\frac{S^2+l^2}{S_l^2+l^2}\right)^{1/2}}.
\ee
This shows that $\Lambda_l(iS)$ is a function of $S$ of width $\propto l$ in the large $l$ limit,
so that the derivatives in (\ref{eq:alter_thetal}) tends to zero in that limit, and the contribution is dominated
by the integral over $S$. Setting $x_l\equiv S_l/l$, assumed to be bounded as we said,
we then obtain an asymptotic expression for the global scale of the Efimovian spectrum:
\begin{multline}
\ln(q_{\rm global}^{(l)}R_*/2) \underset{l\to +\infty}{=}
\frac{\mathrm{Im}\, [\ln \Gamma(1+iS_l) + \ln\Gamma(1+2iS_l)]}{S_l}  \\
-3\ln l-\frac{1}{2} \ln(1+x_l^2) +
1-\frac{\arctan x_l}{x_l} 
+\int_0^{x_l}\!\! \frac{dx}{x_l}\, \ln \mathcal{F}(x) +O(\frac{1}{l})
\label{eq:gen_l_grand}
\end{multline}
with the function
\be
\label{eq:defcF}
\mathcal{F}(x)\equiv \frac{1}{x_l^2-x^2} \left[-1+\frac{e^{-C\left(\frac{1+x^2}{1+x_l^2}\right)^{1/2}}}{C\left(\frac{1+x^2}{1+x_l^2}\right)^{1/2}}\right].
\ee
In the case where $S_l$ has a fixed value, for $l\to +\infty$,
one has that $x_l\to 0$ and one may approximate $\mathcal{F}$
by keeping terms up to order $x_l^2$ and $x^2$ inside the square brackets of (\ref{eq:defcF}), so that
$\mathcal{F}(x)= (1+C)/2 +o(1)$. This gives (\ref{eq:grandl}).
In the case where $x_l$ fixed to a non-zero value, $S_l$ diverges for $l\to +\infty$ so that
the Gamma functions in (\ref{eq:gen_l_grand}) may be Stirling-expanded, leading to
\begin{multline}
\ln(q_{\rm global}^{(l)}R_*/2) \stackrel{x_l\ \mathrm{fixed}}{\underset{l\to+\infty}{\to}}
\ln\frac{x_l}{(1+x_l^2)^{1/2}} -\frac{\arctan x_l}{x_l}  \\
+\int_0^{x_l} \!\! \frac{dx}{x_l} \ln [(x_l^2-x^2)\mathcal{F}(x)].
\label{eq:slg}
\end{multline}
Furthermore, if $x_l\gg 1$, the first two terms in the right-hand side of (\ref{eq:slg}) tend to zero, and the integral
can be shown to approach $J+\ln(1+C)$, where $J$ is the integral (\ref{eq:defJ}). Remarkably,
one then recovers for $q_{\rm global}^{(l)}R_*$ the same estimate as in (\ref{eq:qglob_loin}),
which was obtained with a different limiting procedure ($S_l\to +\infty$ for $l$ fixed). This is may be
not surprising, since $S_l/l\gg 1$ in both cases, this is even obvious in a semi-classical picture,
see discussion below Eq.~(\ref{eq:gjcl}), where the large $\alpha$ limit is reached for
$\alpha C^2/2 \gg (l+1/2)^2$ (which implies $S_l\gg l$), irrespective of the fact that $l$ is large or not.

\section{Efimovian spectrum global scale at infinite mass ratio}
\label{app:qrefasympt}

The results (\ref{eq:useful}) or (\ref{eq:alter_thetal}) are quite useful to determine $q_{\rm global}^{(l)}$
for $\alpha\to +\infty$ for a fixed angular momentum $l$. 
One simply needs an asymptotic expansion of $\Lambda_l(iS)$
for $S$ large of the order of $S_l\to +\infty$, which implies that $S$ and $1/\cos\nu$ both
scale as $\alpha^{1/2}$. For $S\to +\infty$ it is apparent that the integral
over $\theta$ in Eq.~(\ref{eq:rep_odd}) is dominated by the contribution of a small interval ending in $\theta=\nu$,
since here $s=iS$.
Approximating $\sin(s\theta)/\sin(s\pi/2)\sim \exp[S(\theta-\pi/2)]$, we see that the small interval has a width 
scaling as $1/S$. We then Taylor-expand $P_l(\sin\theta/\sin\nu)$ around $\theta=\nu$ up to second order in $(\theta-\nu)$,
and we perform the integral over $\theta$ extending the lower bound of the integral to $-\infty$, which generates
an expansion in powers of $1/S$.
We can also consistently expand $\cos\nu$ and $\sin\nu$ up to second order in $\pi/2-\nu$, since $1/S$ and $\pi/2-\nu$
are of the same order. If one sets $\epsilon=\pi/2-\nu$, this gives
\begin{multline}
\label{eq:Lamas}
\frac{\Lambda_l(iS)}{\cos\nu} = 1 -\frac{e^{-\epsilon S}}{\epsilon S}\left[1+\frac{2}{3}\epsilon^2
-\frac{1}{2} l(l+1)  \left(\frac{\epsilon}{S} +\frac{1}{S^2}\right)
\right. \\ \left.  
+ O\left(\frac{1}{S^4}\right)\right]
\end{multline}
where we used $P_l(1)=1$ and $P_l'(1)=l(l+1)/2$.

A first application of Eq.~(\ref{eq:Lamas}) is an expansion of $S_l$ in powers of $\epsilon$.
Since $\Lambda_l(i S_l)=0$, one finds to leading order $S_l \epsilon =C$, 
where $C$ is given by (\ref{eq:C}). Going to next order gives a correction of order $\epsilon^2$ to $\epsilon S_l$.
Expressing $\epsilon$  as a power series in $1/\alpha$ from $\cos\epsilon=\alpha/(1+\alpha)$
gives (\ref{eq:Slasympt}).

A second application of Eq.~(\ref{eq:Lamas}) is the derivation of the infinite-mass-ratio limit of $q_{\rm global}^{(l)}$.
To leading order, $\Lambda_l(iS)/\cos\nu$ is a function of $\epsilon S$, and so is
\be
\label{eq:utilefinalement}
R_l(S) \simeq \ln \left[\left(1-\frac{e^{-x}}{x}\right)\frac{x^2}{x^2-C^2}\right]\Big|_{x=\epsilon S}.
\ee
This means that the $(2k-1)^{\rm th}$ derivatives of $R_l$ in Eq.~(\ref{eq:joliphi}) scale as $\epsilon^{2k-1}$
and are negligible. The integral over $S'$ in Eq.~(\ref{eq:joliphi}) is a leading contribution
that scales as $S_l$ as revealed by the change of variable $x=\epsilon S'$. Also
the denominator in (\ref{eq:useful}) contributes with a phase factor $\sim \exp(2i S_l \ln 2)$.
Using (\ref{eq:Z_l}) one finally obtains (\ref{eq:defJ}).

\section{First correction to the hydrogenoid spectrum}
\label{app:hydrogen}

Within the Born-Oppenheimer framework of subsection \ref{subsec:BO}, for the hydrogenoid part of the trimer spectrum,
we apply the first order perturbation theory to the $1/r_{23}^{1/2}$ term of Eq.~(\ref{eq:vbo_pres})
that we call here $\delta V$.
In terms of the Bohr radius $a_0=\hbar^2/(m_e e^2)=4R_*/\alpha$, the normalized hydrogenoid wavefunction is \cite{Introduzione}
\begin{multline}
\psi_n^{(l)}(\rr_{23})=\left(\frac{2}{(n+l)a_0}\right)^{3/2}\left[\frac{(n-1)!}{2(n+l)(n+2l)!}\right]^{1/2} 
\\ \times e^{-r_{23}/[(n+l)a_0]} \left[\frac{2r_{23}}{(n+l)a_0}\right]^l L_{n-1}^{2l+1}\left[\frac{2r_{23}}{(n+l)a_0}\right]
Y_l^{m_l}(\rr_{23}).
\end{multline}
Here $L_n^{\beta}$ is the usual Laguerre polynomial defined with the convention of \cite{Gradstein} (and not
with the one of \cite{Introduzione}).
After angular integration and the change of variable $u=2r_{23}/[(n+l)a_0]$ we obtain for the expectation
value of $\delta V$ in that wavefunction:
\begin{multline}
\langle \delta V\rangle = \frac{\hbar^2\alpha^{1/2}}{M R_*^2} \frac{(n-1)!}{[2(n+l)]^{3/2} (n+2l)!} \\
\times 
\int_0^{+\infty} du\, u^{2l+3/2} e^{-u} \left[L_{n-1}^{2l+1}(u)\right]^2.
\end{multline}
To evaluate this integral, we use the generating function technique of \cite{Introduzione}: We define
\be
I(x,y) = \int_0^{+\infty} du\, u^{\beta+\gamma} e^{-u} \varphi_\beta(u,x) \varphi_\beta(u,y)
\ee
where eventually we shall set $\gamma=1/2$ and $\beta=2l+1$, and where the generating function of the Laguerre
polynomials $L_n^\beta$ for fixed $\beta$ is given for $|z|<1$ by relation 8.975(1) in \cite{Gradstein}:
\be
\varphi_\beta(u,z)\equiv \sum_{n=0}^{+\infty} L_n^\beta(u) z^n = \frac{e^{-uz/(1-z)}}{(1-z)^{\beta+1}}.
\ee
On one hand, since $\varphi_\beta$ is the generating function,
\be
I(x,y) = \sum_{m,n=0}^{+\infty} x^m y^n \int_0^{+\infty} du\, u^{\beta+\gamma} e^{-u} L_m^\beta(u) L_n^\beta(u)
\ee
so we need the diagonal terms $n=m$ in this series expansion. On the other hand, from the explicit form
of $\varphi_\beta$:
\be
I(x,y) = \Gamma(1+\beta+\gamma) \frac{[(1-x)(1-y)]^\gamma}{(1-xy)^{1+\beta+\gamma}},
\ee
that it remains to expand in a series of $x$ and $y$ using three times $(1-X)^\nu = \sum_{q=0}^{+\infty}
X^q \Gamma(q-\nu)/[\Gamma(q+1)\Gamma(-\nu)]$, $\nu$ non-integer, to obtain for a non-integer $\gamma$:
\begin{multline}
\int_0^{+\infty}du\, u^{\beta+\gamma} e^{-u} \left[L_{n-1}^\beta(u)\right]^2= \\
\sum_{k=0}^{n-1}  \left(\frac{\Gamma(k-\gamma)}{\Gamma(k+1)\Gamma(-\gamma)}\right)^2
\frac{\Gamma(\beta+\gamma+n-k)}{\Gamma(n-k)}.
\label{eq:tbc}
\end{multline}
For $\gamma=1/2$, $\beta=2l+1$, expressing the Gamma function of integers and half-integers
in terms of factorials finally gives Eqs.~(\ref{eq:hydro_cor},\ref{eq:Rnl}) \cite{note_naive}.


\begin{thebibliography}{99}
\bibitem{footnote_clear}
Physically this is clear in a real space reasoning, since it turns the scattered
wave $\propto e^{ikr}/r$ into an exponentially decreasing function.
\bibitem{details}
One uses relation 9.132(1) in \cite{Gradstein} to map the hypergeometric function for a large and negative
argument $-k^2/Q^2$ to hypergeometric functions of small and positive argument $Q^2/(Q^2+k^2)$.
Second, the product of $\Gamma$ factors is simplified using the doubling formula 
8.335(1) of \cite{Gradstein} as $\Gamma(s)=2^{s-1} \Gamma(s/2)\Gamma[(1+s)/2]/\sqrt{\pi}$, 
then repeatedly the relation $\Gamma(z+1)= z \Gamma(z)$, and finally relation
8.332(2) of \cite{Gradstein} in the form $|\Gamma[(1+s)/2]|^2=\pi/\cos(\pi s/2)$.
\bibitem{Cp} Our function $C_l$ corresponds to the function $C_+$ in \cite{Mora_CRAS},
up to a Wick rotation, since a Barnes contour integral is used in \cite{Mora_CRAS}, rather than the more common
Fourier representation.
\bibitem{expli_poles}
Under inspection of (\ref{eq:rep_odd}), one finds that
the poles of $S\to \Lambda_l(iS)$, here $l$ odd, necessarily obey
$\sin (s\pi/2)=0$, imposing that $s$ is an even integer.
$s=0$ is not a pole due to the presence of the factor $\sin(s\theta)$ in the numerator of the integrand in (\ref{eq:rep_odd}).
Also $0< s < l$ can not be a pole: According to relation 1.391(1) in \cite{Gradstein}, for $s$ even integer,
$\sin(s\theta) = \cos\theta \, Q_s(\sin\theta)$ where $Q_s(X)$ is an odd polynomial of degree $s-1$.
Injecting this form in (\ref{eq:rep_odd}), making the change of variable $u=\sin\theta/\sin\nu$, and using
the fact that $P_l(u)$ and $Q_s(u\sin\nu)$ are odd function of $u$, one obtains the scalar
product between two polynomials, $\int_{-1}^{1} du \ P_l(u) Q_s(u\sin\nu)$. As the Legendre polynomials form an orthogonal
basis for this scalar product, the integral vanishes if the degree of $Q_s$ is less than the degree of $P_l$,
that is $s-1 < l$.
Numerically, we have found that the function $s\to \Lambda_l(s)$ is negative for $s=0$ (here $\alpha> \alpha_c^{(l)}$)
and is a decreasing function of $s$ over
the interval $[0,l+1[$. So the imaginary roots of $S\to \Lambda_l(iS)$, that is the real roots of $\Lambda_l(s)$,
are obtained by looking for changes of sign of the function $s\to \Lambda_l(s)$ 
with small steps (typically 100 steps) over each interval of $s$ between the two successive even integers $2n+l+1$ and $2(n+1)+l+1$,
$n\geq 0$.
A dichotomic search is then applied in case of sign change.
The infinite products in (\ref{eq:weier_Lambda},\ref{eq:weier_Ft}) are approximated
by keeping a maximal value of $n$, for example $n=1000$, leading to exactly the same number
of factors in the numerator and in the denominator. A check is to verify that this truncated product
in (\ref{eq:weier_Lambda}) accurately reconstructs the function $\Lambda_l(iS)/\cos\nu$ over {\sl e.g.}\ the interval $-5 < S < 5$
for $|S_l|\lesssim 1$, that is with an error less than $10^{-5}$, otherwise the maximal value of $n$
is increased (we went up to $n=2500$). We have also used the more rapidly convergent
representation (\ref{eq:dls},\ref{eq:int_ims}).
\bibitem{pas_poles}
We recall that the poles of $\Gamma(z)$ are the non-positive integers $n=0,-1,-2,\ldots$, that is $n\in \mathbb{Z}^-$,
with residues $(-1)^n/n!$.
As a particular consequence, the factor $\Gamma(1+iS+iS_l)$ in the denominator of Eq.~(\ref{eq:weier_Ft}) 
leads to the fact that $C_l(S)$ vanishes for $S\to -S_l + ip$, for any integer $p \geq 1$,
so that $\tilde{F}^{(l)}(S)$ has no pole there despite the division by the hyperbolic sine
in (\ref{eq:introC}).
Also, according to relation 8.326(1) in \cite{Gradstein}, $|\Gamma(x)/\Gamma(x-iy)|^2=\prod_{k=0}^{+\infty}
\left(1+\frac{y^2}{(x+k)^2}\right)$, for all $x\in \mathbb{R}\setminus\mathbb{Z}^{-}$ and $\forall y\in \mathbb{R}$,
the function $\Gamma(z)$ has no root in the complex plane.
\bibitem{plus_inf}
The same work can be gone for $x\to+\infty$. Closing the integration contour by a half-circle at infinity
in the upper half of the complex plane, one finds that the leading contribution to $F^{(l)}(x)$ originates
from the pole $z=v_0+i=(l+2)i$ (see Fig.\ref{fig:contour}) so that $D(\kk)$ 
vanishes as $1/k^{4+l}$ for $k\to +\infty$.
\bibitem{longue_expli}
One splits the functions $C_l$ and $\Lambda_l$ as
$C_l(S) = \frac{\Gamma(iS_l-iS)}{\Gamma(1+iS+iS_l)} \, C_l^{\rm other}(S)$
and $\Lambda_l(iS)=\cos\nu(S^2-S_l^2)$ $\times \Lambda_l^{\rm other}(iS)$.
Then from $\Gamma(z+1)=z\Gamma(z)$ one has $C_l^{\rm other}(S+i)=\Lambda_l^{\rm other}(iS) C_l^{\rm other}(S)$ for
all real $S$. A Weierstrass form for $C_l^{\rm other}(S)$ is deduced from (\ref{eq:weier_Ft}). Then
the property $[\Gamma(z)]^*=\Gamma(z^*)$ and the fact that $u_n$ and $v_n$ are purely imaginary
for $n>0$ lead to $[C_l^{\rm other}(z)]^*=C_l^{\rm other}(-z^*)$ in the complex plane. As a consequence,
\[
[C_l^{\rm other}(S_l)]^* = C_l^{\rm other}(-S_l).
\]
Expressing $C_l^{\rm other}$ in that equation in terms of the original function
$C_l$ (this requires taking a limit $S\to S_l$), 
together with the relation
$|\Gamma(2iS_l)|^2=\pi/[2 S_l \sinh (2\pi S_l)]$, {\sl cf.}\ relation 8.332(1) of \cite{Gradstein},
gives (\ref{eq:residues_Ft}).
Similarly one has
$\left[\tilde{F}^{(l)}(S)\right]^*=-\tilde{F}^{(l)}(-S^*)$ and $f^{(l)}(k)$ is purely imaginary
(for real $k$).
\bibitem{coherence}
This can also be obtained from the fact that $\mathcal{A}(\xx)$ in (\ref{eq:expr_acalli}) scales as $x^{-1\pm s}$
for $x\to 0$, and that the correctly normalized extra-particle
wavefunction (\ref{eq:gen_sol}) involves a factor $\kappa^{1/2}\propto 1/r_{23}^{1/2}$.
\bibitem{aide_epsilon}
To efficiently obtain (\ref{eq:defDelta})  we took $1/L=1/(l+1/2)$ as the small parameter,
so that $\cos\nu=(C/L)[1+\epsilon/L^2+O(1/L^3)]$ at the critical mass ratio. To calculate sums of the type
$S=\sum_{k\geq 0} P(k) C^{2k}/(2k)!$, where $P(k)$ is a polynomial, we expanded
$P(k)=\sum_{n\geq 0} c_n Q_n(k)$ on the basis of polynomials $Q_n(k)=(2k)(2k-1)\ldots (2k-n+1)$.
$Q_n(p/2)=0$  iff $0\leq p<n$, $p$ integer, so that the coefficients $c_n$ are straightforward
to express in terms of the $P(p/2)$'s. Then $S=\sum_{n\ \mathrm{even}} c_n C^n \cosh C + \sum_{n\ \mathrm{odd}} c_n C^n \sinh C$.
For $S=\sum_{k\geq 0} P(k) C^{2k+1}/(2k+1)!$, we defined $Q_n(k)=(2k+1)\ldots (2k+1-n+1)$,
with $Q_n[(p-1)/2]=0$ iff $0\leq p < n$, $p$ integer, so that
$S=\sum_{n\ \mathrm{even}} c_n C^n \sinh C + \sum_{n\ \mathrm{odd}} c_n C^n \cosh C$.
\bibitem{aide} 
The proportionality of Eqs.~(\ref{eq:zr_grandk},\ref{eq:asympt_rs}) implies that
$C_1 k^{iS_l}+\mbox{c.c.}= \lambda\times (C_2 k^{iS_l}+\mbox{c.c.})$, where
$C_1$ (respectively $C_2$) is the coefficient of $k^{iS_l}$ in the expression between curly brackets
in (\ref{eq:zr_grandk}) (respectively in (\ref{eq:asympt_rs})) and $\lambda$ is the proportionality factor.
Since the functions $k\to k^{\pm iS_l}$ are linearly independent, this is satisfied iff
$C_1 = \lambda C_2$ and $C_1^* = \lambda C_2^*$, that is $C_1^* C_2/(C_1 C_2^*)=1$.
This last equation may be naturally written is the form $X^{2 i S_l}=1$, which implies
that the quantity $X$ is an integer power of $e^{-\pi/S_l}$.
\bibitem{proviso}
For the zero range limit $b\to 0$ taken in section \ref{sec:model_equa} to be physically valid, $q_{n}^{(l)}b$ 
has to remain smaller than unity.
\bibitem{not_ato}
Note that the difference between the extra-particle mass $M$ and the reduced mass $\mu$ does not enter yet at this order.
\bibitem{detsc}
Starting from Eq.~(\ref{eq:bos}), one sets $\psi(\rr)=Y_l^{m_l}(\rr) f(r)/r^{1/2}$ so that
$f(r)$ obeys a 2D Schr\"odinger's equation. Then one sets $f(r)=F(x)$ with $x=\ln(r/R_*)$ 
so that $f''(r)+f'(r)/r=F''(x)/r^2$. After multiplication by $r^2$
one obtains $-F''(x)+W(x) F(x)=0$, to which one applies
the usual WKB approximation. Here $W(x)=(l+1/2)^2+ R_*^2 e^{2x} [\mathcal{Q}^2-\alpha \kappa^2(R_*e^x)/2]$,
where the $r$ dependence of $\kappa(r)$ is included in the writing.
The WKB approximation requires that $|\frac{d}{dx}[|W(x)|^{-1/2}]|\ll 1$
away from the turning points. For $(\mathcal{Q} R_*)^2 e^{2x}\lesssim |s_{\rm BO}|^2$, 
where $s_{\rm BO}$ is given by Eq.~(\ref{eq:BO}), this imposes
$|s_{\rm BO}|\gg 1$, that is one has to be far from the Efimovian threshold.
\bibitem{au_moins_l_grand}
Taking the large $x_l$ limit in Eq.~(\ref{eq:slg}) up to order $1/x_l$ included, we obtain exactly Eq.~(\ref{eq:refesti}),
knowing that $S_l$ and $|s_{\rm BO}|$ may be identified at this order.
This at least shows that the coefficient of the $1/|s_{\rm BO}|$ term in Eq.~(\ref{eq:refesti}) is exact in
the large $l$ limit.
Furthermore, one can extend the estimates (\ref{eq:Lamas}) and
(\ref{eq:utilefinalement}) to the interval $0 < S \ll 1/\epsilon$  by approximating $\Lambda_l(iS)$ by the limit $\Lambda_l^{\rm app}(iS)$
of Eq.~(\ref{eq:rep_odd}) for $\nu\to \pi/2$. Since $-S \Lambda_l^{\rm app}(iS) = \coth(\pi S/2) \prod_{n=0}^{(l-1)/2} \frac{S^2+(2n)^2}{S^2+(2n+1)^2}$,
$\int_0^{+\infty} dS\, \ln[-S \Lambda_l^{\rm app}(iS)]=-(l+1/2)\pi/2$, and
one finds from Eq.~(\ref{eq:alter_thetal}) that Eq.~(\ref{eq:refesti}) is exact for all $l$.
\bibitem{en_fait} 
From the expression of $\check{q}_{\rm rel}(\check{k})$, it is apparent that the actual condition
is $q R_* \check{k}_{\rm max} \cos\nu \gg 1.$
\bibitem{num_noyau_pas}
To evaluate in a numerically stable way the integral over $u$ in (\ref{eq:red2}),
we used that fact that $\int_{-1}^{1}\frac{du}{2} P_l(u)/(v-u) = Q_l(v)$ where $v>1$ and 
$Q_l(v)$ is a Legendre function of the second kind, see relation 7.224 in \cite{Gradstein}. 
We then used an open Fortran library to calculate that function, see {\sl e.g.}\  mlqnb.f90 in specfun, and
also a home-made routine using a high order $1/v$ expansion, with similar results.
Furthermore, for $x\gg 1$ and $X\gg 1$, one has $v\simeq\cosh(x-X)/\sin\nu$, so that one faces
the function $y\to Q_l[\cosh y/\sin \nu]$, which behaves for $|y|\ll 1$ and for $|\nu-\pi/2|\ll 1$
as $\frac{1}{2}\ln [4/(y^2 +\cos^2\nu)]$. We thus checked that our discretization
step for $x$ (and $X$) verifies $dx=d(\ln\check{k}) < \cos\nu$.
Note that, from the usual definition 
$P_l(u)=(2^{-l}/l!)\frac{d^l}{du^l}[(u^2-1)^l]$, one can show 
(by $l$ integrations
by part) that $Q_l(v)>0$ for $v>1$, for $l$ of any parity.
The kernel in (\ref{eq:red2}) 
has thus positive matrix elements for $l$ even, and negative
for $l$ odd.
\bibitem{magnetic_field_stabilisation}
We estimate here the issue of magnetic field stabilization for a narrow Feshbach resonance such
as the one of ${}^6$Li-${}^{40}$K. The $1/a=0$ assumption
imposes the filtering $q R_* > 10 R_*/|a|$ on our results, as we said. The finite
experimental value of the scattering length is, according to Eq.~(\ref{eq:adb}),
$a\approx a_{\rm bg} \Delta B/\delta B$ where $\delta B$ is the experimental control
on the magnetic field. For the modest choice $10 R_*/|a| = 10^{-1}$, which 
amounts to restricting our Fig.\ref{fig:num1}a to $q R_*>10^{-1}$,
and taking typical ${}^6$Li-${}^{40}$K Feshbach resonance parameters,
$R_* \approx 100$nm, $\Delta B \approx 0.1$mT, $a_{\rm bg}\approx 3$nm \cite{Levinsen,Wille,Tiecke},
we reach the condition $\delta B\approx 30$nT=$0.3$mG, which requires magnetic shielding
but remains accessible \cite{Reinhard}.
\bibitem{note_naive}
The naive calculation of the integral
(\ref{eq:tbc}) after expansion of the polynomial in terms of the monomials $u^q$ gives a sum of $n^2$ terms 
that suffers the sign problem.
\end{thebibliography}
\end{document}